\documentclass[12pt]{article}
\pdfoutput=1
\usepackage{ulem,color,euscript,latexsym,epsfig,amssymb,amsmath,cite,slashed,bbm,slashed,dsfont}
\usepackage{latexsym}
\usepackage{amsmath} 
\usepackage[usenames,dvipsnames]{xcolor}
\usepackage{ulem}
\usepackage{cite}
\usepackage{graphicx}
\usepackage{hyperref}
\usepackage{subcaption}

\newcommand{\bea}{\begin{eqnarray}}
\newcommand{\eea}{\end{eqnarray}}
\newcommand{\tr}{\mathrm{tr\,}}

\newcommand{\nn} {\nonumber}

\newcommand{\Tr}{{\rm Tr}}
\newcommand{\pa}{\partial}
\newcommand{\mm}{\mathcal}

   \def\d{\delta}    \def\n{\nu}  \def\s{\sigma}      
\def\G{\Gamma}  \def\D{\Delta}

\include{macros}
\catcode`@=12

\numberwithin{equation}{section}

\begin{document}

\begin{titlepage}

\begin{flushright}
\end{flushright}
\bigskip
\def\thefootnote{\fnsymbol{footnote}}

\begin{center}
\vskip -10pt
{\Large
{\bf
A Study of Quantum Field Theories \\\vskip 15pt in AdS at Finite Coupling}
}
\end{center}

\bigskip
\begin{center}
{\large 
Dean Carmi$^\Psi$, Lorenzo Di Pietro$^\sigma$, Shota Komatsu$^\phi$}

\end{center}

\renewcommand{\thefootnote}{\arabic{footnote}}

\begin{center}
\vspace{0.2cm}
$^\Psi$ {Walter Burke Institute for Theoretical Physics, Caltech, Pasadena, CA 91125, USA\\}
$^\sigma$ {Perimeter Institute for Theoretical Physics, Waterloo, ON N2L2Y5, Canada
\\}
$^\phi$ {School of Natural Sciences, Institute for Advanced Study, Princeton, NJ 08540, USA}

\vskip 5pt

\end{center}

\noindent
\begin{center} {\bf Abstract} \end{center}
\noindent
 
We study the $O(N)$ and Gross-Neveu models at large $N$ on AdS$_{d+1}$ background. Thanks to the isometries of AdS, the observables in these theories are constrained by the SO$(d,2)$ conformal group even in the presence of mass deformations, as was discussed by Callan and Wilczek \cite{Callan:1989em}, and provide an interesting two-parameter family of quantities which interpolate between the S-matrices in flat space and the correlators in CFT with a boundary. For the actual computation, we judiciously use the spectral representation to resum loop diagrams in the bulk. After the resummation, the AdS $4$-particle scattering amplitude is given in terms of a single unknown function of the spectral parameter. We then ``bootstrap'' the unknown function by requiring the absence of double-trace operators in the boundary OPE. Our results are at leading nontrivial order in $\frac{1}{N}$, and include the full dependence on the quartic coupling, the mass parameters, and the AdS radius. In the bosonic $O(N)$ model we study both the massive phase and the symmetry-breaking phase, which exists even in AdS$_2$ evading Coleman's theorem, and identify the AdS analogue of a resonance in flat space. We then propose that symmetry breaking in AdS implies the existence of a conformal manifold in the boundary conformal theory. We also provide evidence for the existence of a critical point with bulk conformal symmetry, matching existing results and finding new ones for the conformal boundary conditions of the critical theories. For the Gross-Neveu model we find a bound state, which interpolates between the familiar bound state in flat space and the displacement operator at the critical point.

\vspace{1.6 cm}
\vfill

\end{titlepage}

\setcounter{footnote}{0}

\tableofcontents
\newpage


\section{Introduction}
Developing non-perturbative approaches to strongly interacting quantum field theories is undoubtedly an important subject in theoretical physics. Among various approaches attempted over the years, the one which is gaining renewed interest and has been producing numerous breathtaking results in recent years is the application of the {\it bootstrap} method. The basic idea of bootstrap is to start with a minimal set of assumptions and constrain the observables by imposing the self-consistency conditions. It was proposed initially in the study of scattering amplitudes in the late 60's \cite{Eden:1966dnq} in order to describe strong interaction. Although the idea was abandoned to a large extent soon after the advent of quantum chromodynamics, it was revived decades later in a different guise in the study of two-dimensional conformal field theories \cite{Belavin:1984vu}. Rather recently the approach was also successfully applied to conformal field theories in higher dimensions \cite{Rattazzi:2008pe}, most notably to the three-dimensional Ising model \cite{ElShowk:2012ht}, with the help of numerical implementation and has been a subject of active research since then. Motivated by this success, the original idea of the S-matrix bootstrap was also revisited, resulting in several interesting outcomes \cite{Caron-Huot:2016icg,Paulos:2016but,Paulos:2017fhb, Doroud:2018szp, He:2018uxa,Cordova:2018uop,Paulos:2018fym}.

Another, perhaps more elementary approach to non-perturbative physics is the resummation of diagrams\footnote{When combined with other techniques such as Borel resummation and conformal mapping, the standard perturbation theory can also be a powerful tool for studying nonperturbative physics. This was recently demonstrated for $\lambda \phi^{4}$ theory in two dimensions in an impressive work \cite{Serone:2018gjo}.} based on the Schwinger-Dyson equation. Although the method is tractable only in specific situations such as the large-$N$ limit, it has an advantage that  it allows us to compute various observables explicitly as functions of the coupling constants and study the renormalization group flows and the critical points analytically.

The main goal of this article is to shed new lights on the conformal and the S-matrix bootstraps and the relation between them, by analyzing large-$N$ field theories using a combination of Schwinger-Dyson techniques and the idea of bootstrap. More specifically, we consider the $O(N)$ vector model and the Gross-Neveu model in $(d+1)$-dimensional anti-de-Sitter space (AdS$_{d+1}$).

There are mainly three motivations for studying these theories. Firstly it helps to make connections between the correlation functions in conformal field theories (CFTs) and the S-matrix in flat space: On the one hand, thanks to the isometries of the AdS spacetime, the observables in quantum field theories in AdS$_{d+1}$ are constrained by the $d$-dimensional conformal group SO$(d,2)$ even in the presence of mass deformation. One can therefore study them using the standard techniques of conformal field theories such as the operator product expansion and crossing symmetry. On the other hand, by taking an appropriate limit, one can extract from them the flat-space S-matrix. 

\begin{figure}[t]
\centering
\begin{minipage}{0.3\hsize}
\centering
\includegraphics[clip,height=5cm]{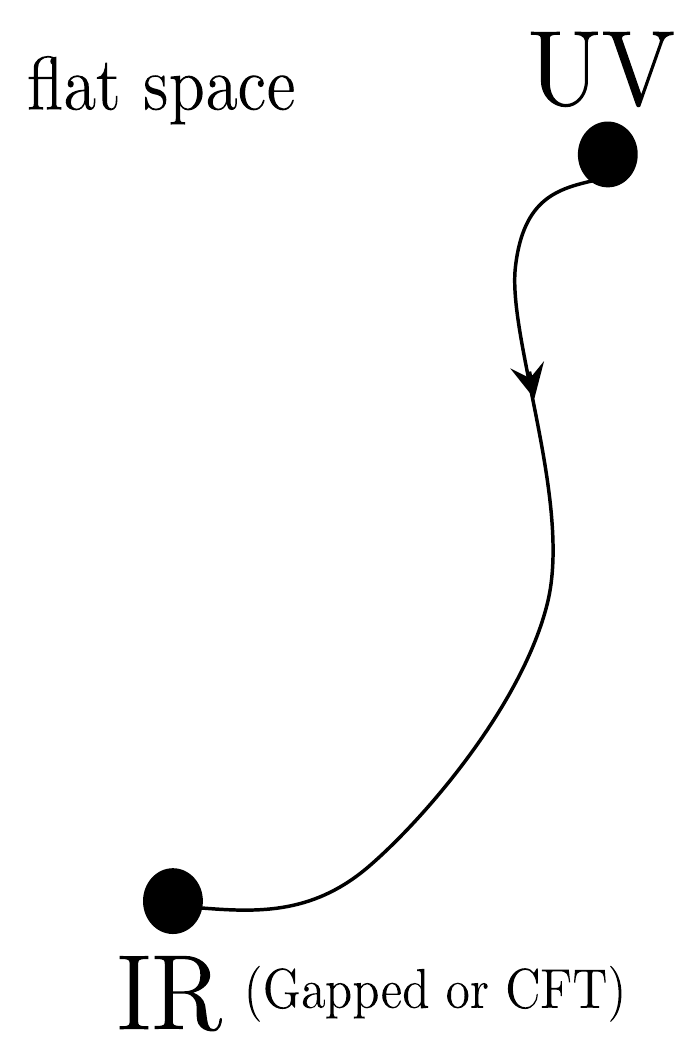}\\
(a)
\end{minipage}
\begin{minipage}{0.3\hsize}
\centering
\includegraphics[clip,height=5cm]{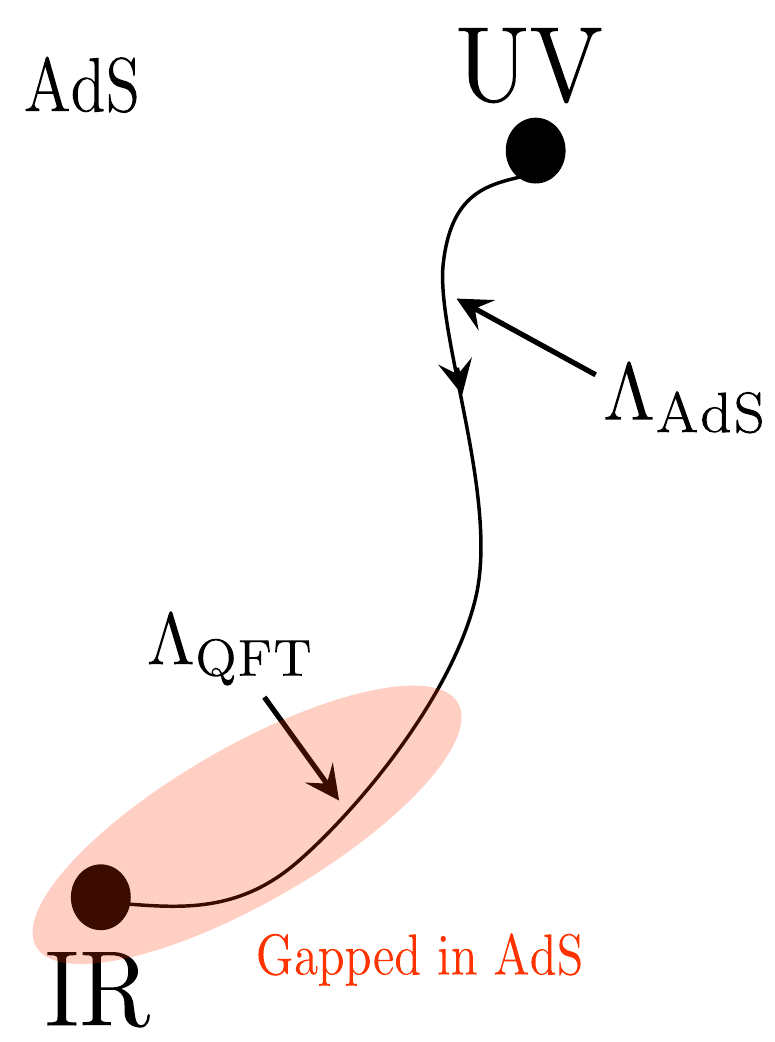}\\
(b)
\end{minipage}
\begin{minipage}{0.35\hsize}
\centering
\includegraphics[clip,height=5cm]{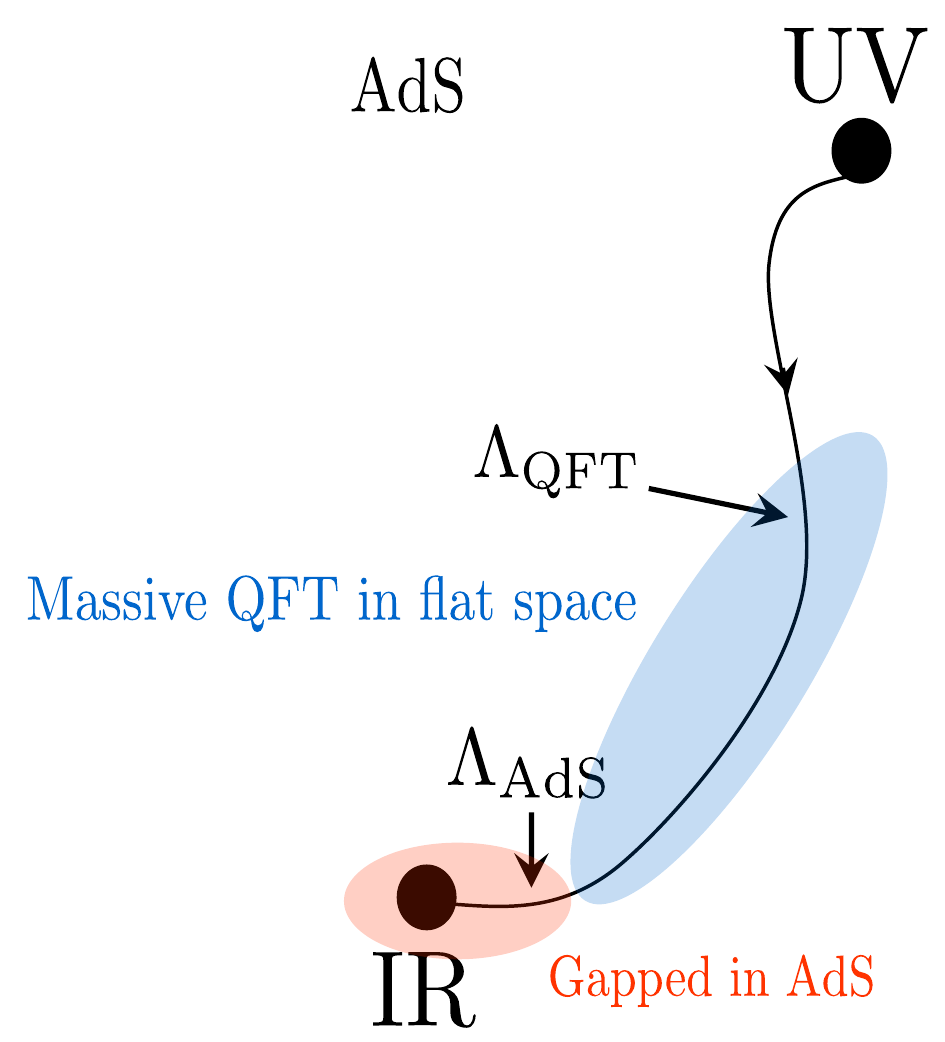}\\
(c)
\end{minipage}
\caption{Schematic pictures of RG flows of massive QFTs in flat space and AdS. (a) The RG flow in flat space. In flat space, the theory starts from the UV fixed point and flows either to a gapped phase or to a fixed point governed by CFT. (b) The RG flow in AdS for $\Lambda_{\rm AdS}\gg \Lambda_{\rm QFT}$. In this case, the theory starts seeing the AdS curvature as soon as it flows away from the UV fixed point, and simply flows to the gapped phase in AdS. (c) The RG flow in AdS for $\Lambda_{\rm AdS}\ll \Lambda_{\rm QFT}$. When $\Lambda_{\rm AdS}$ is small enough, the theory does not see the AdS curvature until it reaches the deep IR. Therefore there is a wide range of scales in which the physics can be well-approximated by the massive QFT in flat space.}\label{fig:flowsflat}
\end{figure}
To appreciate this point further, it is helpful to briefly discuss the physics of a massive quantum field theory in AdS (see also figure \ref{fig:flowsflat}). In flat space, the theory starts from the UV fixed point and flows either to a gapped phase or to an IR fixed point. Also in AdS, the UV limit is described by the same flat-space UV fixed point since the AdS curvature becomes negligible in the UV limit. However, the details of how it flows to the IR depend on the relative magnitude of the AdS scale $\Lambda_{\rm AdS}=1/L_{\rm AdS}$ and the mass scale of the theory $\Lambda_{\rm QFT}$. If the AdS scale is much larger than the mass scale of the theory ($\Lambda_{\rm AdS}\gg\Lambda_{\rm QFT}$), the theory starts seeing the AdS curvature as soon as it flows away from the UV fixed point. Therefore there is no scale in which the theory is governed by nontrivial flat-space physics and it flows simply to a gapped phase in AdS. In asymptotically-free theories, this also implies that the physics is controlled by perturbation theory in AdS since the effective coupling constant at the AdS scale is small. This advantage was first pointed out by Callan and Wilczek \cite{Callan:1989em} and further discussions on the case of Yang-Mills theory in AdS were given in \cite{Aharony:2012jf}\footnote{Other recent studies of quantum field theory on rigid AdS background are \cite{Aharony:2010ay,Aharony:2015zea, Aharony:2015hix,Doyon:2004fv}.}.
On the other hand, if the AdS scale is much smaller $(\Lambda_{\rm AdS}\ll \Lambda_{\rm QFT})$, the theory does not see the AdS curvature until it reaches the deep IR and therefore it can be well-described by a massive QFT in flat space for a wide range of scales\footnote{We can see all these features explicitly in the $O(N)$ vector model. See section \ref{sec:Corr}}. Thus by considering a theory in AdS and sending $\Lambda_{\rm AdS}$ to zero, one can compute the observables in flat space, in particular the S-matrix, from the observables in AdS. This allows us to analyze the physics of massive quantum field theories in flat space using the powerful techniques of conformal field theories. Such an idea was already employed in \cite{Paulos:2016fap} in which they studied the S-matrix in flat space by analyzing the conformal bootstrap numerically and taking the flat-space limit\footnote{Some of the relations between the S-matrix bootstrap in flat space and the conformal bootstrap were clarified recently with the help of the analytic functional bootstrap \cite{Mazac:2016qev,Mazac:2018mdx}.}. In this paper, we will see another use of this idea which is more analytical; namely we compute resummed loop diagrams by imposing the consistency of the operator product expansion in the boundary conformal theory and take the flat-space limit to reproduce the S-matrix. For details, see section \ref{sec:Corr}. Besides being an efficient computational tool, the connection between the flat-space S-matrix and the conformal correlators may also be used as a way to understand general analytic properties of the nonperturbative S-matrix in flat space. Unlike the correlators in conformal field theories whose analytic properties are well under control thanks to the conformal block expansion, not much is known about the analytic properties of the S-matrix in flat space at the non-perturbative level. Given such a situation, it would be interesting to see what the correlators in AdS, which are constrained by the conformal symmetry, can tell us about the properties of the S-matrix upon taking the flat-space limit. Although we are still quite far from satisfactory understanding of the analyticity of the flat-space S-matrix, we will see in this paper that several important features of the flat-space S-matrix, namely resonances and bound states, can be reproduced from the correlators of large-$N$ vector models in AdS.

The second reason for studying these theories is to better understand non-perturbative solutions to the crossing equation in CFT. As a result of intensive studies in the last couple of years, various results for CFTs in higher dimensions were obtained not just numerically but also analytically. However, most of the analytical results so far concern infinitesimal corrections to generalized-free-field CFTs\footnote{In generalized-free-field CFTs, correlation functions are simply given by a product of two-point functions and the operator spectrum consists of ``multi-trace operators''.}. For instance, it was shown in \cite{Komargodski:2012ek,Fitzpatrick:2012yx} that every CFT contains a universal large-spin sector in which the operator spectrum asymptotes to that of the generalized-free-field CFT, and infinitesimal corrections to it are governed by the low-twist operators of the theory\footnote{There have been remarkable development in the large spin expansion in the last few years. See for instance \cite{Alday:2015eya,Alday:2015ewa,Alday:2016njk,Simmons-Duffin:2016wlq,Caron-Huot:2017vep,Lemos:2017vnx} for other important developments.}. Although it is quite remarkable that one can make such a universal statement about general CFTs, it is indispensable to develop better understanding of the CFT data away from such a universal sector since the operator spectrum, or even the number of operators, in a general CFT is quite different from that of generalized free fields. From this perspective, it would be desirable to have examples of solutions to the crossing equation which exhibit a non-perturbative reorganization of the spectrum. The large-$N$ vector models that we study in this paper precisely provide such an example: Both in the O($N$) vector model and the Gross-Neveu model, we show that the dimensions of the operators receive finite shifts from those of the generalized free field. Furthermore, in the Gross-Neveu model we show the existence of an extra operator which corresponds to a bound state in flat space, while in the $O(N)$ vector model in the symmetry-breaking phase we find a distinctive pattern of the anomalous dimensions of the operators which can be thought of as the AdS analogue of the resonance phenomenon (see figure \ref{fig:phaseshiftano}). Both of these phenomena can only be seen after the resummation of diagrams in AdS and cannot be deduced from the existing analytical results in the literature.

 Thirdly, it encompasses the study of conformal boundary conditions in flat space. When the theory in AdS is tuned to be at a critical point, we obtain a conformal field theory in AdS. In such a special case, one can view the theory as describing a conformal field theory with a boundary (BCFT) since the AdS spacetime is conformally equivalent to the flat half-space ($\mathbb{R}_{+}\times \mathbb{R}^{d}$). Using this correspondence, one can extract BCFT data directly from the observables in AdS. We will show this explicitly by reproducing the existing results in the literature and also finding new ones about the conformal boundary conditions of the critical $O(N)$ and Gross-Neveu models.  
 
 The rest of the paper is organized as follows: In section \ref{sec:generalities}, we review the generalities of the $O(N)$ model, including the Hubbard-Stratonovich trick and the large-$N$ effective action. We then discuss in section \ref{sec:Phases} the phases of the $O(N)$ model in AdS$_2$ and AdS$_3$ by analyzing the effective potential, and show the existence of the symmetry-breaking vacua and of the gapped vacuum. In particular, we discuss the parameter regions in which the two phases coexist owing to the Breitenlohner-Freedman bound. In section \ref{sec:Corr}, we compute the two-point functions of $\sigma$ and the AdS 4-particle scattering amplitude of $O(N)$ vector fields $\phi^i$ in the gapped phase. We do so by first resumming Witten diagrams and expressing the final result in terms of a single unknown function. We then bootstrap the unknown function by requiring the consistency of the OPE of the boundary conformal theory. We also provide results in the symmetry-breaking phase, and in that context we describe the AdS analogue of a resonance in flat-space. We furthermore propose a relation between symmetry breaking in AdS and the existence of a conformal manifold in the boundary conformal theory.
 
 In section \ref{sec:Critical}, we discuss in detail the case when the bulk theory becomes conformal. We first analyze the structure of the OPE expansion in the bulk and propose a diagnosis for bulk conformality in AdS background. We then apply this idea to find the critical point of the O($N$) model, and extract BCFT data from the previously-computed correlators. In section \ref{sec:fermion}, we perform similar analyses in the Gross-Neveu model. In particular, we show the existence of a bound state in AdS, we analyze the critical point and we compute some data of the associated BCFT. Finally we conclude and comment on future directions in section \ref{sec:conclusion}.

\section{Generalities of the $O(N)$ model}\label{sec:generalities}
\subsection{Review of the $O(N)$ Model}
The Lagrangian of the $O(N)$ model is
\begin{equation}
\mathcal{L} = \frac12 (\partial \phi^i)^2 + \frac{m^2}{2} (\phi^i)^2 + \frac{\lambda}{2 N} ((\phi^i)^2)^2~,
\end{equation}
where $i = 1, \dots, N$ and summation over $i$ is implicit.

The model admits a $1/N$ expansion with $\lambda$ fixed \cite{Coleman:1974jh, Moshe:2003xn}, as we will now review. A convenient way to obtain the large-$N$ expansion is by introducing a Hubbard-Stratonovich (HS) auxiliary field $\sigma$ 
\begin{equation}\label{eq:HSLag}
\mathcal{L} = \frac12 (\partial \phi^i)^2 + \frac{m^2}{2} (\phi^i)^2 - \frac{1}{2 \lambda} \sigma^2 + \frac{1}{\sqrt{N}}  \,\sigma (\phi^i)^2 ~.
\end{equation}
The integration contour for $\sigma$ runs on the imaginary axis. Note that the equation of motion simply sets
\begin{equation}
\sigma = \frac{\lambda}{\sqrt{N}}(\phi^i)^2~,
\end{equation}
hence $\sigma$ is identified with the composite operator $(\phi^i)^2$ inside correlation functions.

\begin{figure}
\centering
\includegraphics[clip, height=2cm]{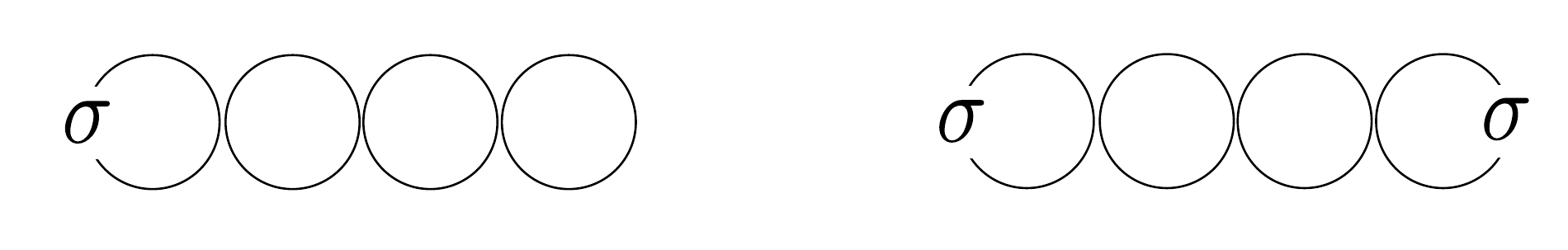}
\caption{Examples of 1PI diagrams for the one- and two-point functions of $\sigma$ which are not suppressed by powers of $N$. The black lines are the propagators of $\phi^{i}$, and we also depicted how O($N$) indices are contracted.\label{fig:notsuppressed}}
\end{figure}

At large $N$ and fixed $\lambda$ there are still loop corrections coming from the Lagrangian \eqref{eq:HSLag} that are not suppressed by inverse powers of $N$. To see this, consider the 1PI 1-point and 2-point correlation functions of $\sigma$ at 1 loop (see figure \ref{fig:notsuppressed}): they are given by a closed loop of the $\phi^i$ field with either 1 or 2 external lines of $\sigma$, giving a contribution of order $\mathcal{O}(\sqrt{N})$ and $\mathcal{O}(1)$, respectively. As a consequence, higher-loop connected diagrams built out these 1PI diagrams will also fail to be suppressed by inverse powers of $N$. In order to take these contributions into account, there are two standard ways to proceed: one is to write the Schwinger-Dyson equation and explicitly resum the diagrams. The other is to consider the path integral of $\sigma$ and $\phi^{i}$, determine the saddle point and compute the fluctuations around it. Both approaches yield the same integral equation in the end and are physically equivalent. In this paper, we adopt the second approach since it is more algorithmic and helps to consider different phases of the theory. 
In practice, the path integral calculation amounts to considering a modified Lagrangian, obtained by adding to \eqref{eq:HSLag} the generating functional of all 1-loop 1PI correlators of $\sigma$. The latter is simply computed by summing the bubbles of $\phi^i$ with an arbitrary number of insertions of the $\sigma$ field, giving
\begin{equation}\label{eq:Gamma}
\Gamma[\sigma]_{\rm{1\,loop}} = \frac{N}{2} \, \tr \log\left(-\square + m^2 + \frac{2}{\sqrt{N}}\sigma \right)~.
\end{equation}
Hence the modified, non-local Lagrangian is
\begin{equation}
\mathcal{L}_{\rm eff} = \frac12 (\partial \phi^i)^2 + \frac{m^2}{2} (\phi^i)^2 - \frac{1}{2 \lambda} \sigma^2 + \frac{1}{\sqrt{N}}  \,\sigma (\phi^i)^2 + \frac{N}{2} \, \tr \log\left(-\square + m^2 + \frac{2}{\sqrt{N}}\sigma \right)~.\label{eq:LagEff}
\end{equation}
Using this Lagrangian \eqref{eq:LagEff}, we can derive the following results: 
\begin{itemize}
\item[(i)]{The $\mathcal{O}(\sqrt{N})$ 1-point function of $\sigma$ is determined simply by minimizing the potential. Expanding the fields around constant values
\begin{align}
\sigma & = \sqrt{N} \Sigma + \delta\sigma~, \label{eq:expS}\\
\phi^i & = \sqrt{N} \Phi^i + \delta\phi^i~, \label{eq:expC}
\end{align}
the potential is given by the constant term in the Lagrangian, namely
\begin{equation}
V(M^2, \Phi^i) = N\left(- \frac{(M^2 - m^2)^2}{8\lambda} + \frac{M^2}{2} (\Phi^i)^2 + \frac12 \tr\log\left(-\square + M^2\right)\right)~,\label{eq:effpot}
\end{equation}
where we are using the shifted variable $M^2 = m^2 + 2\Sigma$. We find that the equation for the vacuum at leading order at large $N$ are
\begin{align}
0 & =  \frac{\partial V}{\partial \Phi^i} =N M^2 \Phi^i  ~, \label{eq:vacC} \\
0 & = \frac{\partial V}{\partial M^2}=\frac{N}{2}\left( \frac{m^2 - M^2}{2 \lambda}  + (\Phi^i)^2 + \tr \frac{1}{-\square + M^2} \right)~; \label{eq:vacS}
\end{align}
} 
\item[(ii)]{Expanding to second order in $\delta\sigma$ around the minimum, we can determine the full $\mathcal{O}(1)$ propagator, which resums the $\phi^i$ bubble diagrams. We can formally write it as follows: denote with $B(x,y)$ the bubble diagram, i.e.
\begin{equation}
B(x,y) = \left[\left(\frac{1}{-\square + M^2}\right)(x,y)\right]^2~,
\end{equation}
where $M^2$ is obtained by solving the equations for the vacuum \eqref{eq:vacC}-\eqref{eq:vacS} above. Then we have
\begin{equation}
\langle \delta\sigma(x) \delta\sigma(y)\rangle = -\left[\frac{1}{\lambda}\mathds{1} + 2 \, B \right]^{-1}(x, y)~.\label{eq:Sprop}
\end{equation}
Here the symbol $\mathds{1}$ is the identity operator and the inverse $[\ast]^{-1}$ is the operator inverse. In other words we should view the two-point function and the bubble as integral operators acting on functions of spacetime $f(x)$ through convolution, e.g. $B[f](x)\equiv \int d^{d+1}y\sqrt{g(y)}B(x,y)f(y)$ ;
}
\item[(iii)]{We can compute observables in the large-$N$ expansion using ordinary Feynman diagrams, with the Feynman rules induced by \eqref{eq:LagEff}, namely: the propagator for $\delta\sigma$ is the resummed one in eq. \eqref{eq:Sprop}, the propagator for $\delta\phi^i$ is just a free-field massive propagator with mass-squared $M^2$, there is a cubic vertex between $\delta\sigma$ and two $\delta\phi^i$'s of order $1/\sqrt{N}$, and there are self-interactions of $\delta\sigma$ induced by the 1-loop 1PI effective action. In this perturbation theory we do not include diagrams containing as a subdiagram any 1PI 1-loop n-point function of $\delta\sigma$, because those are already accounted for by the the full propagator and self-interactions of $\delta\sigma$.}
\end{itemize}

The approach described here is valid also on any curved background, if we interpret $\square$ as the scalar Laplacian on the background. In this general case we include a possible quadratic coupling to curvature in the definition of $m^2$. Depending on the number $d+1$ of spacetime dimensions, some couplings can have UV divergences and will have to be appropriately renormalized. In the following we will work in $2\leq d+1 < 4$, where the theory is super-renormalizable and at most a renormalization of the vacuum energy and of the parameter $m^2$ is needed. An important modification to the vacuum equation arises if the Euclidean spacetime has a finite volume \cite{Hartnoll:2005yc}, but we do not need to consider it because Euclidean AdS has infinite volume. On the other hand, in AdS we have to specify boundary conditions.  


\subsection{Boundary Conditions on AdS Background} 

We need to prescribe boundary conditions for the fields $\phi^i$ at the boundary of AdS. Working at large $N$ and finite $\lambda$, the theory is not a free-field theory in the usual sense, and therefore one seems to face the hard problem of understanding boundary conditions for interacting fields. Luckily, the virtue of the HS trick is that at leading order at large $N$ the interaction is just encoded in the non-trivial propagator of the auxiliary field $\sigma$, whereas the $\phi^i$ are decoupled from $\sigma$ and free. They only start interacting via the exchange of $\sigma$ at subleading order in $1/N$, and that interaction can be treated perturbatively, so it does not require us to understand interacting boundary conditions. 

More precisely, what we just wrote applies to the fluctuations $\delta\phi^i$ and $\delta\sigma$ around the vacuum configuration. However, while it is true that the path integral for the fluctuations $\delta\phi^i$ is gaussian at leading order at large $N$, even at this order the coupling $\lambda$ enters through the effective mass-squared $M^2$, which is determined dynamically by the VEV of $\sigma$. 

Let us briefly remind what are the possible boundary conditions on AdS$_{d+1}$ for a given mass-squared $M^2$ and AdS radius $L$, as discussed in \cite{Klebanov:1999tb}. Define $\Delta_+$, $\Delta_-$, with $\Delta_+ = d -\Delta_- \geq \Delta_-$, to be the two solutions to the quadratic equation
\begin{equation}
\Delta(\Delta-d) = L^2M^2~,
\end{equation}
for the variable $\Delta$. Choosing Poincar\'e coordinates $(z, \vec{x})$, where $z > 0$ and $\vec{x}\in \mathbb{R}^d$, with metric
\begin{equation}
d s^2 = L^2\frac{dz^2 + (d\vec{x})^2}{z^2}
\end{equation}
the solution to the Klein-Gordon equation behaves near the boundary at $z = 0$ as
\begin{equation}
\delta\phi^i(z, \vec{x}) \underset{z\to 0}{\longrightarrow}  z^{\Delta_+}(A_+^i(\vec{x})+\mathcal{O}(z^2) ) +  z^{\Delta_-}(A_-^i(\vec{x})+\mathcal{O}(z^2) )~.
\end{equation}
We restrict the functions $\delta\phi^i$ in the path integral to have the same $z$-dependence near the boundary as either the $+$ or $-$ mode. This defines a boundary condition that preserves the isometries of AdS and the $O(N)$ symmetry. In addition we need to require that the Euclidean action is a finite function of the $\delta\phi^i$'s, i.e. that the allowed modes are normalizable. For $L^2M^2 \geq -\frac{d^2}{4}+1$ only the $+$ mode is  normalizable and gives a valid boundary condition, while for $- \frac{d^2}{4} \leq L^2M^2 <  -\frac{d^2}{4}+1$ both the $+$ and $-$ modes are normalizable, hence there are two possible boundary conditions. If the boundary condition $\pm$ is chosen, the boundary conformal theory contains an operator in the vector representation of $O(N)$ of scaling dimension $\Delta_\pm$. 

Note that the case with $+$ boundary condition is continuously connected to the $O(N)$ model in flat space, via the flat-space limit $L \to \infty$ with $M^2$ and $\lambda$ fixed. On the other hand, the case with $-$ boundary condition is an intrinsically curved-space regime. In the following we will concentrate on the case of $+$ boundary condition.


\section{Phases of the $O(N)$ Model on AdS}\label{sec:Phases}
We now move on to the study of phases of the $O(N)$ model on AdS. As was pointed out initially by Callan and Wilczek, the AdS spacetime acts as a symmetry-preserving IR regulator\footnote{See also for \cite{Kiritsis:1994yv} the application of a similar idea to superstring.}. This fact leads to several important differences from the flat-space analysis, such as the coexistence of the gapped vacuum and the symmetry-preserving vacuum, and the existence of the symmetry breaking vacua even in two dimensions evading the famous Coleman-Mermin-Wagner theorem in flat space \cite{Mermin:1966fe, Coleman:1973ci}. 

\subsection{Effective Potential on AdS$_{d+1}$}
To obtain the effective potential we need to compute the trace in eq. \eqref{eq:effpot}. Calculations of functional determinants in AdS have appeared in \cite{Burgess:1984ti, Inami:1985wu, Camporesi:1993mz, Gubser:2002zh, Hartman:2006dy, Giombi:2013fka}. Here we will employ the spectral representation of the bulk-to-bulk propagator, reviewed in the appendix \ref{app:SpRep}. With the $+$ boundary condition
\begin{equation}
\left(\frac{1}{-\square + M^2}\right)(x,y) = \frac{1}{L^{d-1}}\int_{-\infty}^{+\infty} d\nu \,\frac{1}{\nu^2 + \left(\Delta -\frac{d}{2}\right)^2} \, \Omega_\nu(x, y)~,\label{eq:srprop}
\end{equation}
where the conformal dimension $\Delta$ is related to the mass $M$ by $\Delta=\Delta_{+}=\tfrac{d}{2}+\sqrt{\tfrac{d}{2}^2+M^2}$ and $\Omega_{\nu}$ is the harmonic function in AdS
 \begin{align}
-\square_{x} \Omega_{\nu}(x,y)=\left(\frac{d^2}{4}+\nu^2\right)\Omega_{\nu}(x,y)\,.
 \end{align}
Using the result \eqref{eq:coincpt} for the coincident point limit of the harmonic function, we obtain\footnote{Here the $\tr$ is normalized dividing by the volume of the Euclidean spacetime.}
\begin{equation}
\tr\frac{1}{-\square + M^2} =   \frac{ \Gamma\left(\frac{d}{2}\right)}{4 \pi^{\frac{d+2}{2}}\Gamma(d)L^{d-1}}\int_{-\infty}^{\infty} d\nu\frac{\Gamma\left(\frac{d}{2} \pm i \nu\right)}{(\nu^2 + \frac{d^2}{4} + L^2 M^2)\Gamma(\pm i \nu)}~,\label{eq:scaltadpole}
\end{equation}
where $\Gamma(z\pm a)\equiv\Gamma(z+a)\Gamma(z-a)$. 

This integral is UV divergent for $d\geq1$. The UV divergence can be reabsorbed in a renormalization of the parameter $m^2/\lambda$, which in fact by power counting is expected to be UV divergent in $d\geq 1$. We can evaluate the integral using dimensional regularization, i.e. taking $d < 1$ and then analytically continuing the result in $d$. For $d<1$ the integral can be evaluated by closing the contour at infinity in the complex $\nu$ plane, giving
\begin{equation}
\tr\frac{1}{-\square + M^2} =  \frac{\Gamma\left(\frac{d}{2}\pm \sqrt{\frac{d^2}{4} + L^2 M^2}\right)\sin\left(\pi \left(\frac{d}{2} - \sqrt{\frac{d^2}{4} + L^2 M^2}\right)\right)}{(4\pi)^{\frac{d+1}{2}}\Gamma(\frac{d+1}{2})\cos(\frac{\pi d}{2}) L^{d-1}}~.\label{eq:trprop}
\end{equation}
This dimensionally-regularized result is finite for integer $d\geq 1$ and even, corresponding to power-law divergences in a cutoff regularization, while it has poles for integer $d\geq 1$ and odd, corresponding to logarithmic divergences.

The trace just computed determines the derivatives of the effective potential w.r.t. $M^2$, see eq. \eqref{eq:vacS}. Knowing the derivative is sufficient to determine the vacuum. The constant in the integration over $M^2$ is a UV divergent vacuum energy. 

To ensure stability, we will only consider $\lambda > 0$, while the mass-squared parameter $m^2$ can take either sign. It is convenient to work with dimensionless quantities by setting $L = 1$ and measuring the other parameters $m^2$ and $\lambda$, as well as the variables $M^2$ and $\Phi^i$ that the effective potential depends on, in units of $L$. In these conventions, the flat-space limit is obtained by sending all the quantities with positive mass dimension to infinity, at a relative rate fixed by their units, e.g. $m^2 /\lambda^{\frac{2}{3-d}}$ and $m^2/M^2$ are kept fixed in the limit.


\subsection{AdS$_3$}

\begin{figure}[t]
\centering
\begin{subfigure}{.45\textwidth}
  \centering
  \includegraphics[width=1\linewidth]{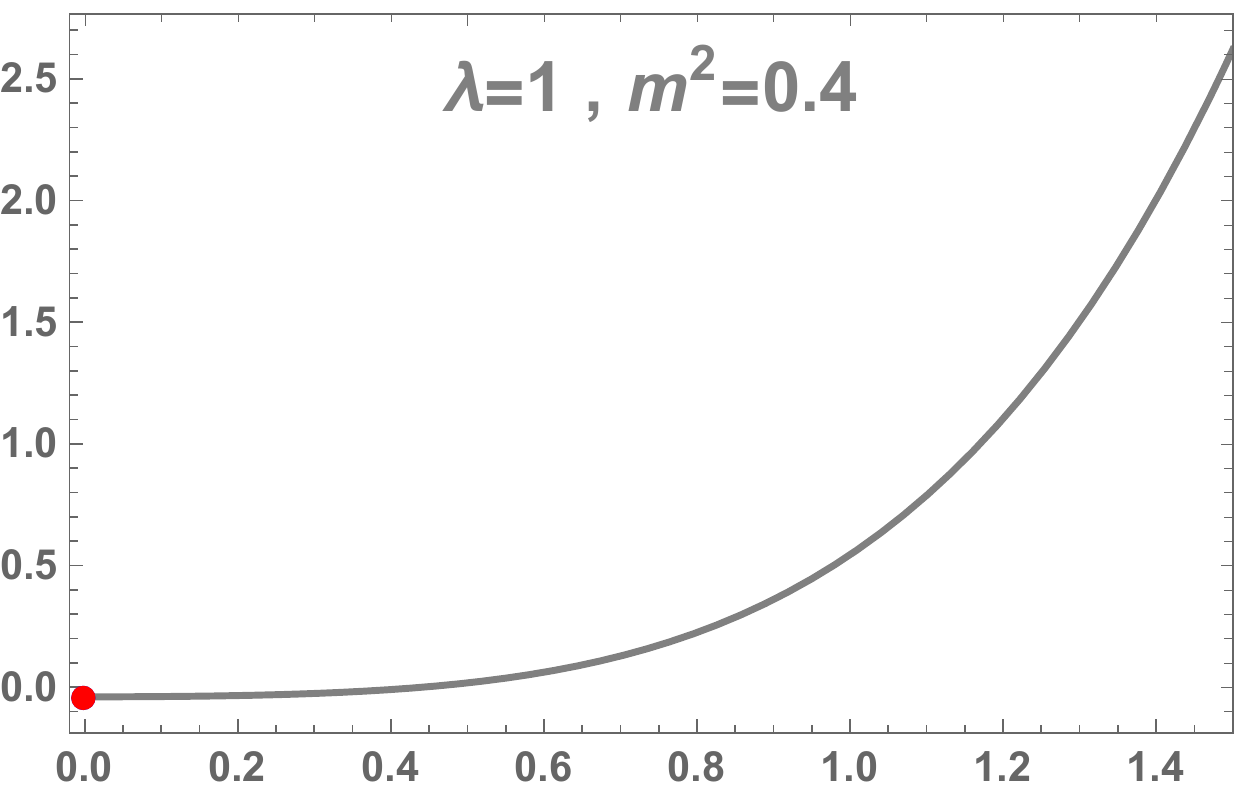}
  \label{fig:sub1}
\end{subfigure}%
\begin{subfigure}{.45\textwidth}
  \centering
  \includegraphics[width=1\linewidth]{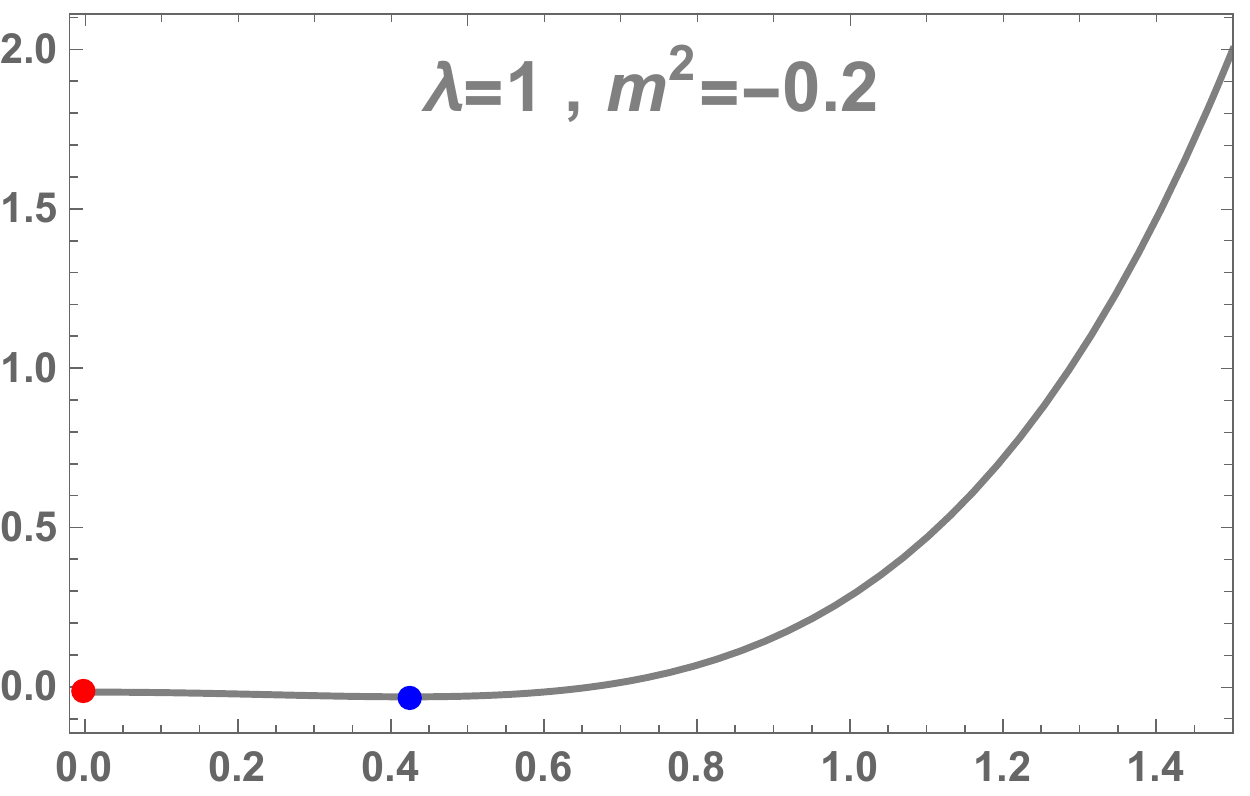}
  \label{fig:sub2}
\end{subfigure}
\begin{subfigure}{.45\textwidth}
  \centering
  \includegraphics[width=1\linewidth]{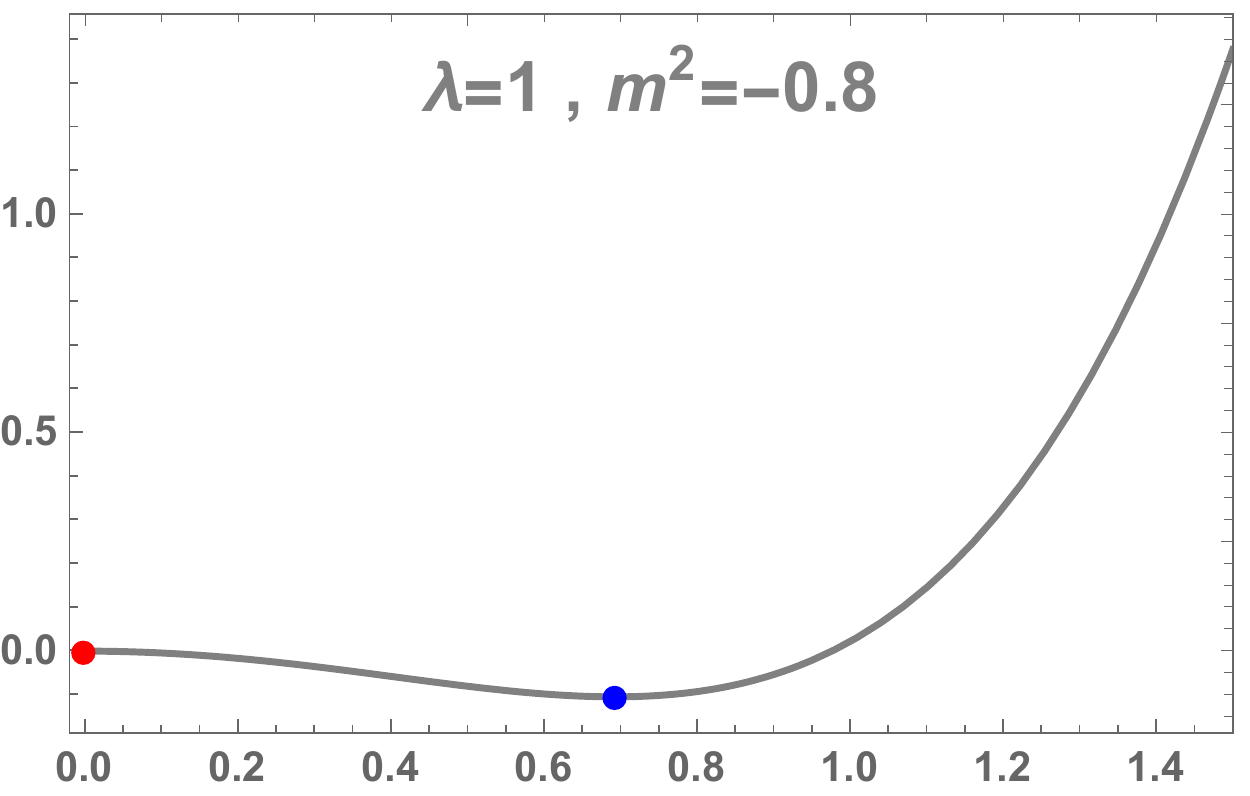}
  \label{fig:sub3}
\end{subfigure}
\begin{subfigure}{.45\textwidth}
  \centering
  \includegraphics[width=1\linewidth]{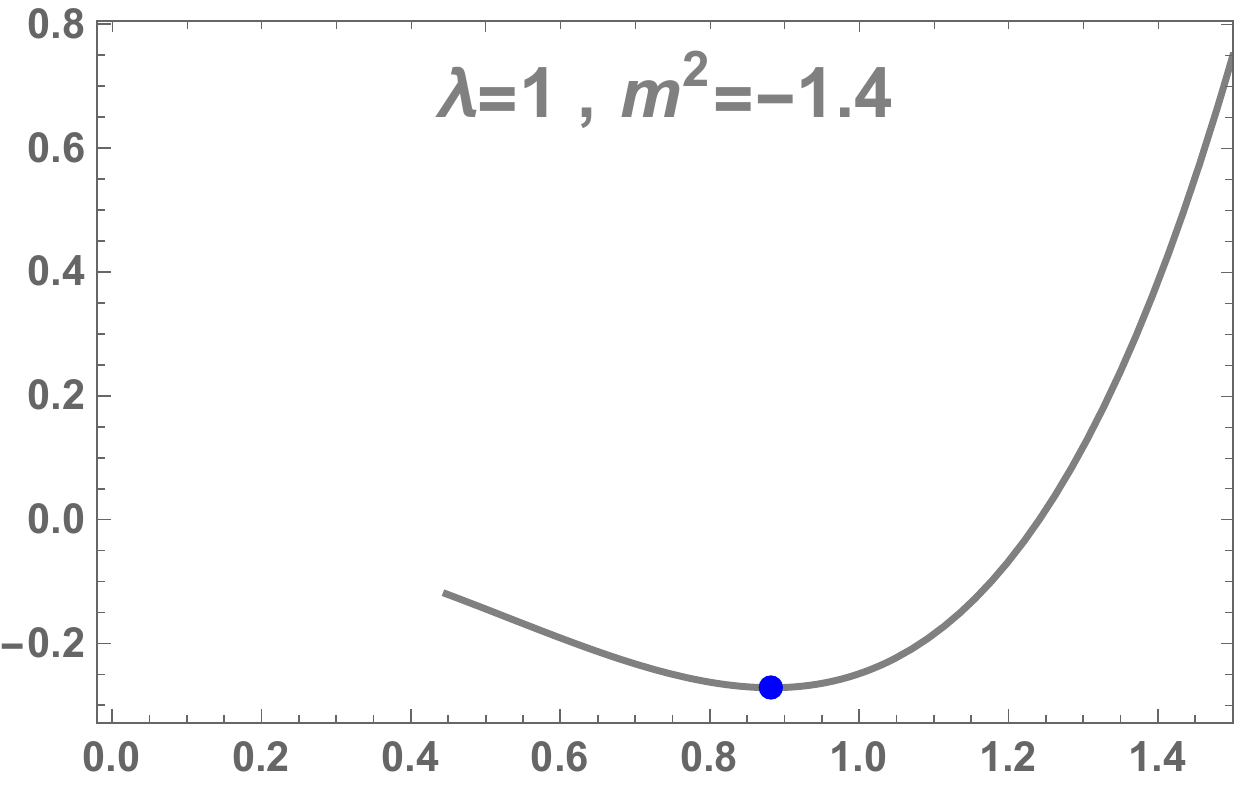}
  \label{fig:sub4}
\end{subfigure}
\caption{The large-$N$ effective potential $\frac{1}{N \lambda^3}V(\Phi^i)$ as a function of $|\Phi|$ in $d=2$ (i.e. in AdS$_3$), for $\lambda =1$ and various values of $m^2$. Dimensionful quantities are expressed in units of the AdS radius $L$. The position of the symmetry-preserving (-breaking) vacuum, when it exists, is indicated with a red (blue) dot. The line interrupts when there is no real solution to eq. \eqref{eq:minSAdS}, i.e. when $\frac{m^2+1}{2} + \lambda(\Phi^i)^2 < 0$.}
\label{fig:potd2}
\end{figure}

Plugging $d=2$ in eq. \eqref{eq:trprop} the expression simplifies to
\begin{equation}
\tr\frac{1}{-\square + M^2} =-\frac{\sqrt{1+ M^2}}{4\pi}~.
\end{equation}
Hence the effective potential, up to a constant, is
\begin{equation}
\frac{V(M^2,\Phi^i)}{N} = -\frac{(M^2-m^2)^2}{8\lambda} +\frac12 M^2 (\Phi^i)^2 -\frac{(1+  M^2)^\frac32}{12 \pi}~,
\end{equation}
and the equations for the vacuum are
\begin{align}
0 &= \frac{2}{N}\,\partial_{M^2} V =  \frac{m^2-M^2}{2 \lambda} + (\Phi^i)^2 - \frac{\sqrt{1 +  M^2}}{4\pi} \label{eq:minSAdS} \\
0 &= \frac{1}{N}\partial_{\Phi^i} V =  \Phi^i \,M^2 \label{eq:minCAdS}~.
\end{align} 
The solution of eq. \eqref{eq:minSAdS} is
\begin{equation}
M^2_{\Phi^i} = - 1+\frac{\lambda^2}{16\pi^2}\left(-1 + \sqrt{1+\frac{16\pi^2}{\lambda^2}\left(m^2 +2 \lambda (\Phi^i)^2 +1\right)}\right)^2~.\label{eq:TofrAdS}
\end{equation}
Plugging this solution back in the effective potential, we obtain a function $V(\Phi^i) \equiv V(M^2_{\Phi^i},\Phi^i)$ of a single variable $|\Phi|\equiv \sqrt{(\Phi^i)^2}$, with $\partial_{\Phi^i} V(\Phi^i) = \Phi^i M^2_{\Phi^i}$. In fig. \ref{fig:potd2} we show the plot of this function for $\lambda =1$ and various values of $m^2$. Dots represents the position of stable vacua. We stress that $m^2$ is a renormalized mass-squared parameter, whose value depends on the scheme, and we recall that we are adopting dimensional regularization as explained in the previous subsection.

For $m^2 > - 1$ there is a symmetry-preserving vacuum at $\Phi^i =0$. This vacuum is stable because the effective mass-squared of the fluctuations is $M^2 =M^2_{\Phi^i = 0}>- 1$, above the Breitenlohner-Freedman (BF) bound \cite{Breitenlohner:1982jf}. This fact---the mass squared can be slightly negative without affecting the stability of the vacuum---is in marked contrast to the analysis in flat space, and leads to the coexistence of the vacua which we discuss in the next paragraph. Note that from eq. \eqref{eq:TofrAdS} that $M^2_{\Phi^i = 0}$ is real and above the BF bound also in the range $ - 1 - \frac{\lambda^2}{16\pi^2}\leq m^2 < - 1$, but in this range $\sqrt{1 + M^2_{\Phi^i = 0}}$ would be negative, therefore the solution is not acceptable.
 
For $m^2< \frac{\lambda}{2 \pi}$ the points at 
\begin{equation}
(\Phi^i)^2 =\frac{1}{2\lambda} \left(\frac{\lambda}{2 \pi} -m^2\right)\equiv |\Phi|^2 > 0~,
\end{equation}
are symmetry-breaking vacua. At any of these points $M^2_{\Phi^i}= 0$, giving $N-1$ Goldstone bosons. The radial mode in the classical limit $\lambda \ll 1$ has a non-zero mass-squared $m^2_\rho=4 |\Phi|^2 \lambda$, while at the quantum level it mixes with $\delta \sigma$, as we will see in more details in section \ref{subsec:corrsymbreak}.  

The symmetry-breaking vacua and the symmetry-preserving vacuum coexist in the region of parameters 
\begin{equation}
-1 < m^2 \leq \frac{\lambda}{2\pi}~,\label{eq:coex}
\end{equation}
and the two solutions coincide at $m^2=\frac{\lambda}{2 \pi}$.\footnote{While the values at the extrema of this window are scheme-dependent, their difference is not.} The symmetry-breaking vacuum is always energetically favored in this range. A similar situation happens in flat space when there is a first-order phase transition. In such a case, the vacuum with larger energy is meta-stable meaning that it can decay to the true vacuum via a bubble nucleation. There are a few important differences in AdS background:\footnote{Since here we are considering a theory without dynamical gravity, the classic result of \cite{Coleman:1980aw} cannot be applied directly.} there is an additional potential due to the curvature that can hinder the expansion of the bubble towards the boundary, and the evolution of the bubble is periodic in real time. It is therefore possible that both of these two vacua can play the role of the ``true vacuum'' of the theory in this parameter range. Further studies are needed to clarify this point.  


\subsection{AdS$_2$}\label{eq:AdS2}
Expanding eq. \eqref{eq:trprop} around $d=1$ we find
\begin{align}
\tr\frac{1}{-\square + M^2} &= \frac{1}{4\pi}\left(-\frac{2}{d-1} + \log(4\pi )-\gamma\right)\nonumber \\& -\frac{1}{2\pi}\left( \psi\left(\tfrac12 + \sqrt{\tfrac 14 + M^2 }\right)\right) + \mathcal{O}(d-1)~,
\end{align}
where $\gamma$ is the Euler-Mascheroni constant, and $\psi (x)=d\log \Gamma(x)/dx$ is the digamma function. We reabsorbe the pole, together with the $M^2$-independent terms $\frac{1}{4\pi}(\log(4\pi)-\gamma)$, in a renormalization of the parameter $m^2/\lambda$, namely
\begin{equation}
\left(\frac{m^2}{\lambda}\right)_{\rm bare} = \mu^{1-d} \left(\frac{m^2}{\lambda}\right)_{\rm ren}\left(1-\frac{1}{2\pi}\left(-\frac{2}{d-1} + \log(4\pi )-\gamma\right)+ \mathcal{O}(d-1)\right)~.
\end{equation}
Here $\mu$ is the scale introduced by dimensional regularization. Therefore we have
\begin{equation}
\tr\frac{1}{-\square + M^2}|_{\rm ren} = -\frac{1}{2\pi}  \psi\left(\tfrac12 + \sqrt{\tfrac 14 +  M^2 }\right) + \frac{1}{4\pi}\log(\mu^2)~,
\end{equation}
which gives the following effective potential up to a constant
\begin{align}
\frac{V(M^2,\Phi^i)}{N} & =  -\frac{(M^2-m^2)^2}{8\lambda} +\frac12 M^2 (\Phi^i)^2  \nonumber\\ & -\frac{1}{4\pi}\int_0^{M^2} dz \, \psi\left(\tfrac12 + \sqrt{\tfrac 14 + z }\right) + \frac{M^2}{8\pi}\log(\mu^2)~.
\end{align}
The vacuum equations are
\begin{align}
0 &= \frac{2}{N}\,\partial_{M^2} V =  \frac{m^2-M^2}{2 \lambda} + (\Phi^i)^2 -\frac{1}{2\pi}   \psi\left(\tfrac12 + \sqrt{\tfrac 14 +  M^2 }\right) + \frac{1}{4\pi}\log(\mu^2)~, \label{eq:minSAdSd1} \\
0 &= \frac{1}{N}\partial_{\Phi^i} V =  \Phi^i \,M^2 \label{eq:minCAdSd1}~.
\end{align} 
Here we see an important difference from the flat-space case. On $\mathbb{R}^2$, the digamma function in eq. \eqref{eq:minSAdSd1} is replaced by its flat-space limit, giving $-\frac{1}{4\pi} \log(M^2/\mu^2)$. Therefore, there is no solution of the vacuum equations with $M^2 = 0$ and $\Phi^i \neq 0$, i.e. there is no symmetry breaking \cite{Coleman:1974jh}. On the other hand, on AdS$_2$ there is no singularity as $M^2$ goes to $0$ and in fact we can find symmetry-breaking solutions. In the corresponding vacua we have $N-1$ fields with $M^2=0$, i.e. the Goldstone bosons. In the scheme we are adopting the symmetry-breaking solutions exist for 
\begin{equation}
\frac{-m^2}{\lambda} \geq \frac{1}{\pi} \gamma +\log\mu\,.
\end{equation} 
The existence of symmetry-breaking vacua for the $O(N)$ model at large $N$ on AdS$_2$ background was first observed in \cite{Inami:1985dj}. 

The educated reader might object to our assertion of the existence of the Goldstone phenomenon in two dimensions. The Coleman-Mermin-Wagner theorem \cite{Mermin:1966fe, Coleman:1973ci} ---which states that a continuous symmetry cannot be spontaneously broken in two dimensions--- is evaded here because the curvature of the background cures the IR singularity in the propagator of the massless scalars. Note that we recover the absence of symmetry breaking in the flat-space limit, because $\mu $ goes to $\infty$ and the lower bound on $|m^2|/\lambda$ goes to $+\infty$. It is also worth stressing that while the large-$N$ limit sometimes gives rise to symmetry breaking and phase transitions even in situations in which they are impossible at finite $N$ (e.g. the spontaneous breaking of the continuous axial symmetry in the Nambu-Jona-Lasinio (NJL) model in two dimensions \cite{Gross:1974jv}, or the confinement-deconfinement transition for gauge theories on a compact spatial manifolds \cite{Witten:1998zw, Aharony:2003sx}), this is not what is happening here, as we clearly see from the fact that the same large-$N$ theory in flat space does not have symmetry breaking. To understand this better, it is useful to compare the large-$N$ $O(N)$ model in $\mathbb{R}^2$ with the large-$N$ NJL model in $\mathbb{R}^2$. In the first example the spontaneous breaking of the symmetry would give rise to a number of Goldstone bosons that grows with $N$, hence taking the large-$N$ limit is not helpful for taming the IR singularity in the massless propagator. By contrast, in the second example there is only one Goldstone boson, whose loops are suppressed at large $N$, hence the IR singularity is only visible at subleading orders in $1/N$. Having understood the reason for the absence of symmetry breaking in flat space, we conclude that it is really the curvature of the background that is responsible for the existence of symmetry breaking in the $O(N)$ model on AdS$_2$. 

\begin{figure}[t]
\centering
\begin{subfigure}{.45\textwidth}
  \centering
  \includegraphics[width=1\linewidth]{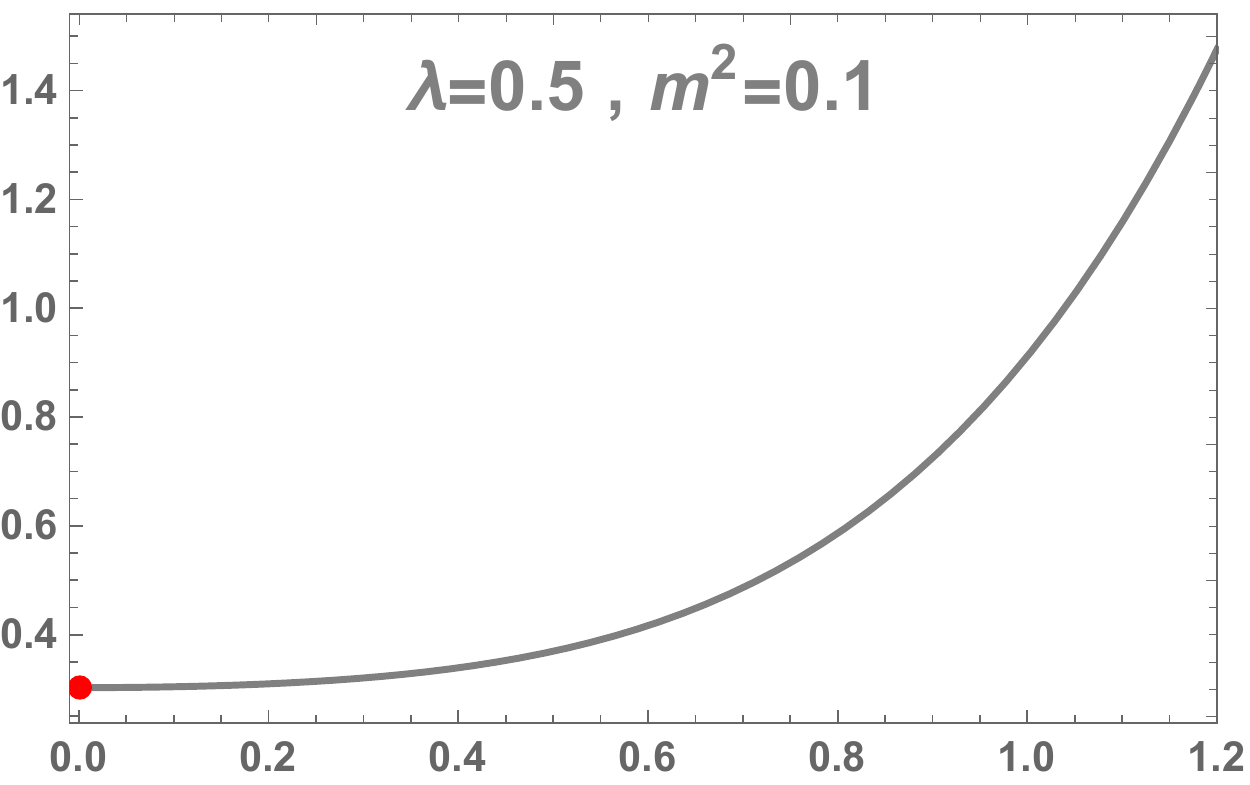}
  \label{fig:sub1}
\end{subfigure}
\begin{subfigure}{.45\textwidth}
  \centering
  \includegraphics[width=1\linewidth]{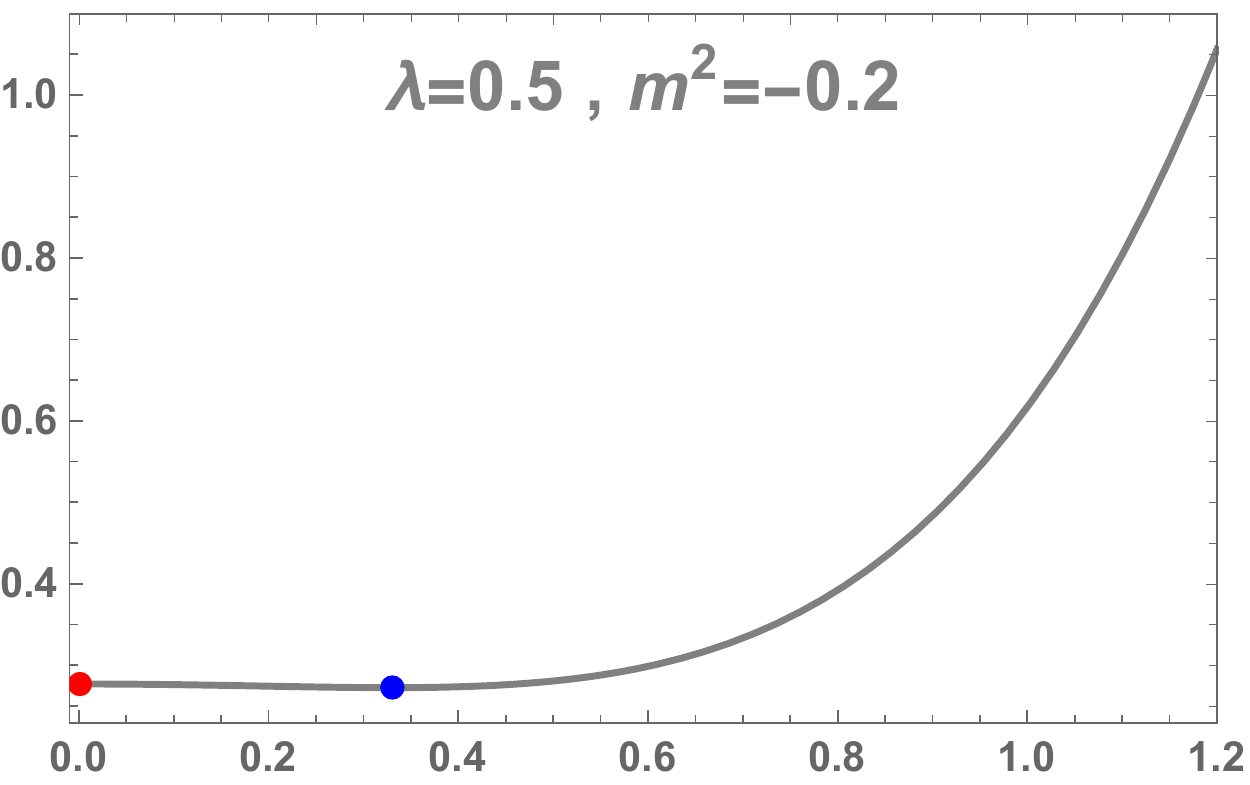}
  \label{fig:sub2}
\end{subfigure}
\begin{subfigure}{.45\textwidth}
  \centering
  \includegraphics[width=1\linewidth]{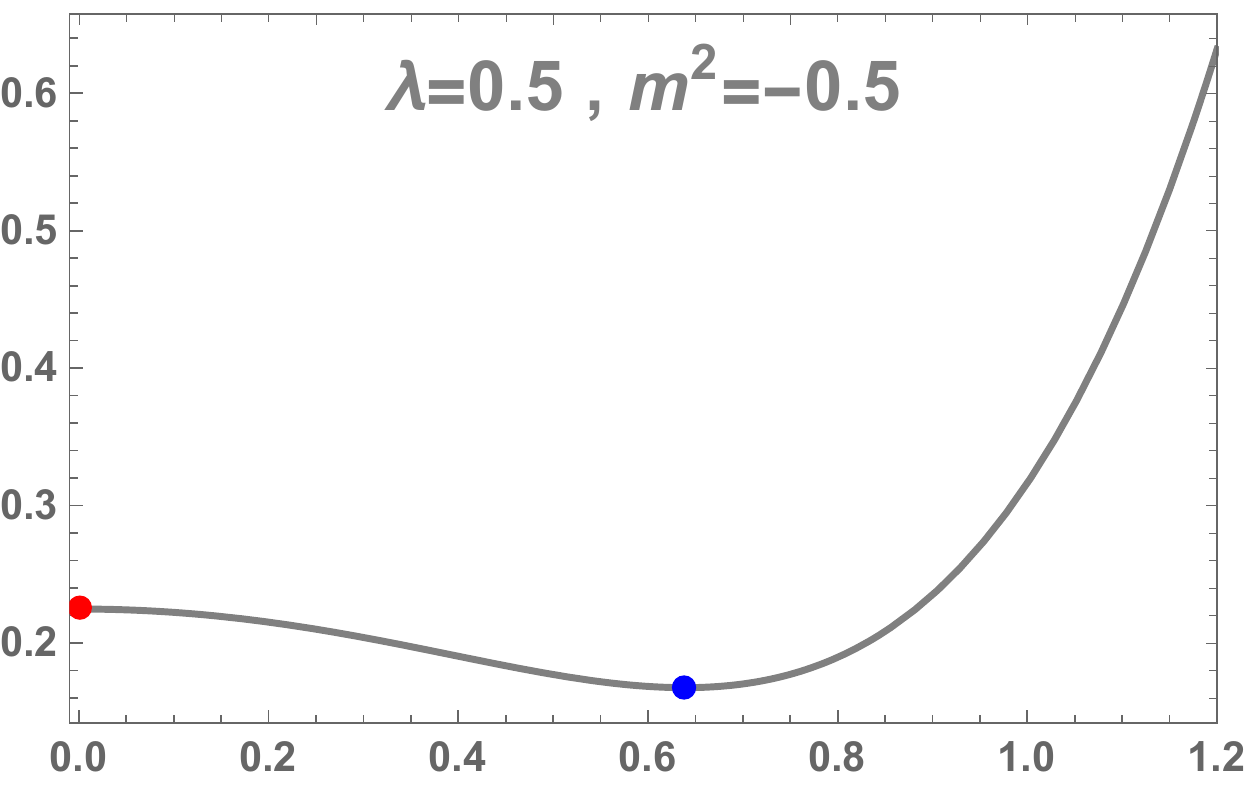}
  \label{fig:sub3}
\end{subfigure}
\begin{subfigure}{.45\textwidth}
  \centering
  \includegraphics[width=1\linewidth]{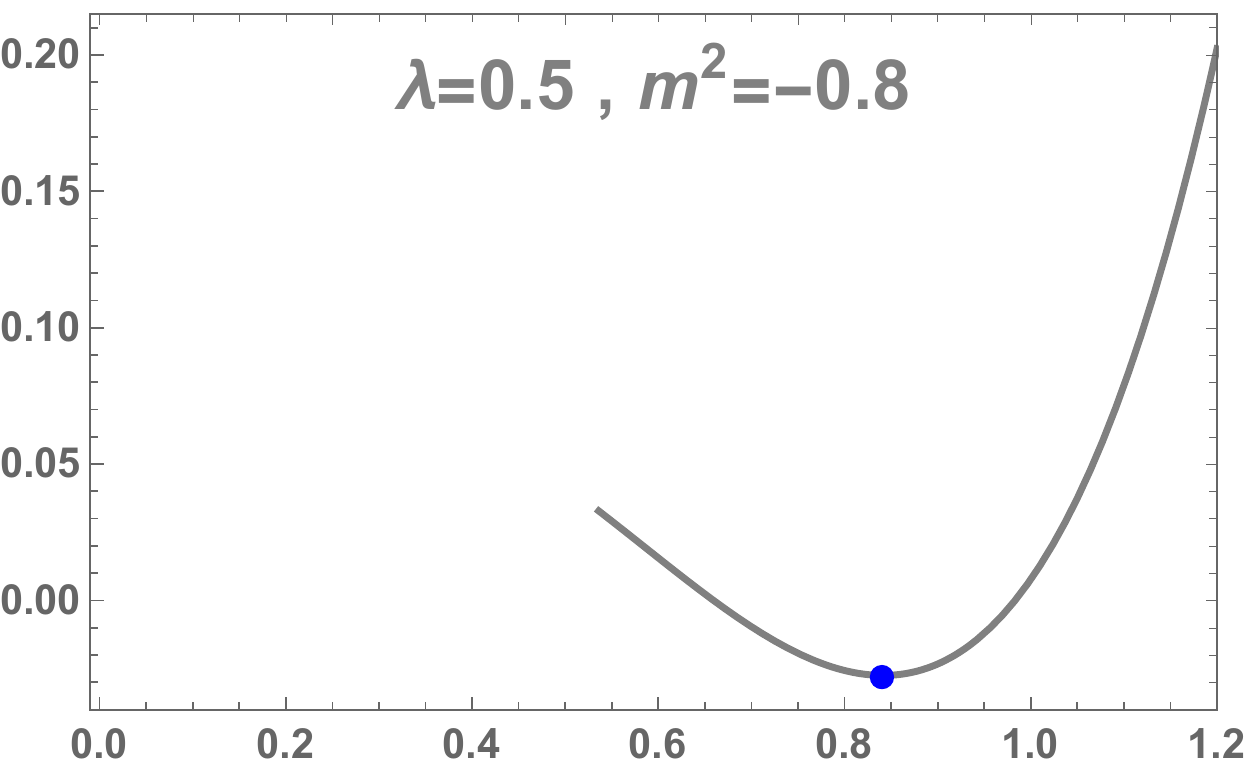}
  \label{fig:sub4}
\end{subfigure}
\caption{The large-$N$ effective potential $\frac{1}{N \lambda}V(\Phi^i)$ as a function of $|\Phi|$ in $d=1$ (i.e. in AdS$_2$), for $\lambda = 0.5$ and various values of $m^2$. Dimensionful quantities are expressed in units of the AdS radius $L$. The position of the symmetry-preserving (-breaking) vacuum, when it exists, is indicated with a red (blue) dot. The line interrupts when there is no solution to eq. \eqref{eq:minSAdSd1}, i.e. when $\frac{m^2+\frac14}{2} + \lambda(\Phi^i)^2 +\frac{\lambda}{2\pi}(\gamma + \log(4\mu))< 0$.}
\label{fig:potd1}
\end{figure}

Let us next discuss the symmetry-preserving vacuum. Note that $M^2 + \frac{\lambda}{\pi} \psi\left(\tfrac12 + \sqrt{\tfrac 14 + M^2 }\right)$ increases monotonically to $+\infty$ in the range $M^2 > -\frac{1}{4 }$ above the BF bound. Hence we can always find a stable symmetry-preserving solution for $M^2$ by setting $\Phi^i = 0$ and solving \eqref{eq:minSAdSd1}, as long as $m^2  >  - \frac{1}{4} - \frac{\lambda}{\pi}(\gamma + \log(4\mu)) $. 

Similarly to the AdS$_3$ case, the symmetry-breaking and symmetry-preserving vacua coexist in the range
\begin{equation}
 -\frac{1}{4 } - \frac{\lambda}{\pi}(\gamma + \log(4 \mu ))  < m^2 \leq -\frac{\lambda}{\pi}(\gamma + \log(\mu))~.
\end{equation}
In fig. \ref{fig:potd1} we show the plot of the effective potential, re-expressed as a function of a single variable $|\Phi|$ by plugging the (numerical) solution of eq. \eqref{eq:minSAdSd1} for $M^2$, for $\lambda = 0.5$. and various values of $m^2$.

Let us finally point out an interesting analogy with Yang-Mills theory in AdS$_4$, which was studied in \cite{Aharony:2012jf}. At sufficiently low energies, the $O(N)$ model in the symmetry-breaking vacua can be well-described by a large-$N$ non-linear sigma model, which on $\mathbb{R}^2$ is asymptotically free and generates a mass dynamically, consistently with Coleman's theorem. On the other hand, in AdS$_2$ the symmetry-breaking vacuum is stable in the IR and the theory contains massless Goldstone bosons if the coupling $\lambda/ |m^2|$ is sufficiently weak (or equivalently if the AdS radius is sufficiently small), as was shown in the analysis of this section. Similarly, the Yang-Mills theory in $\mathbb{R}^4$ is asymptotically free and gapped due to confinement, while in AdS$_4$ it can also be in the deconfined phase and contain light, weakly-coupled gluons if the AdS radius is sufficiently small \cite{Aharony:2012jf}. Therefore, from the point of view of the non-linear sigma model, the symmetry-breaking phase of the $O(N)$ model can be thought of as an analogue of the deconfined phase, and the phase transition between the gapped phase and the symmetry-breaking phase is in analogy with the confinement-deconfinement transition of the Yang-Mills theory on AdS$_4$. It would be interesting to study more in  details this transition of the large-$N$ non-linear sigma model, and we hope to return to this problem in the near future.


\section{Correlators of the $O(N)$ Model in AdS}\label{sec:Corr}

We now compute correlation functions of the $O(N)$ vector model in AdS. The computation proceeds in three steps: As a first step, we derive the spectral representation for the bulk two-point functions of the $\sigma$ field. The result is expressed in terms of a single unknown function which physically describes the bubble integral in AdS. Then in the second step, we contract this two-point function with bulk-boundary propagators and construct the boundary four-point functions of the fundamental fields $\phi^{i}$. Lastly in the third step, we require the consistency of the OPE expansion of the boundary conformal theory---in particular the absence of the double-trace operators---and then determine the unknown function.  The idea of determining the correlator by imposing the absence of the double-trace operators is similar to the so-called Mellin/Polyakov bootstrap \cite{Gopakumar:2016wkt,Gopakumar:2016cpb,Gopakumar:2018xqi}, but in our case the large-$N$ limit (in the bulk) helps to simplify the analysis greatly and allows us to obtain more powerful results. 
After determining the correlator, we  analyze it in several parameter regimes and discuss the physics it describes. Our primary focus is on the correlators in the gapped phase, but we also provide several interesting results for the symmetry-breaking phase. In particular, in the symmetry-breaking phase we find a distinctive pattern of the anomalous dimensions of the double-trace operators which can be interpreted as the AdS analogue of a particle resonance in flat space.


\subsection{``Bootstrapping'' Correlators in the Gapped Phase}\label{sec:Corr41}
Let us perform the computation following the strategy outlined above.
\subsubsection{Computation at large N}
\begin{figure}
\centering
\includegraphics[clip,height=1cm]{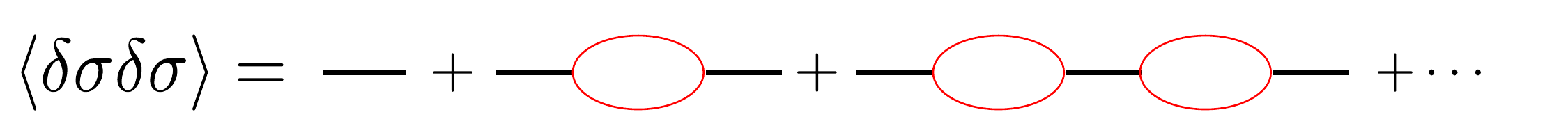}
\caption{The resummation of the two-point function of $\delta\sigma$. The thick black lines are the bare propagators of $\sigma$ while the red curves are propagators of $\phi^{i}$. \label{fig:geometricseries}}
\end{figure}

\paragraph{Step 1: Spectral representation of the bulk two-point function} The first step is to compute the bulk two-point function of $\sigma$, namely $\langle\delta \sigma(x) \delta \sigma(y)\rangle$. Its formal expression can be straightforwardly derived from the quadratic fluctuations in the effective action \eqref{eq:LagEff} as discussed in section \ref{sec:generalities}. The result reads
\begin{align}\label{eq:delsigxy}
\langle \delta\sigma(x) \delta\sigma(y)\rangle = -\left[\frac{1}{\lambda}\mathds{1} + 2 \, B \right]^{-1}(x, y)~,
\end{align}
where $B(x,y)$ is a product of two bulk-to-bulk propagators
\begin{align}\label{eq:productofbtob}
B(x,y) = \left[\left(\frac{1}{-\square + M^2}\right)(x,y)\right]^2~.
\end{align}
Physically the function $B(x,y)$ describes the bubble diagram in the bulk and the expression \eqref{eq:delsigxy} is the sum of the geometric series of bubble diagrams (see figure \ref{fig:geometricseries}),
\begin{align}
-\left[\frac{1}{\lambda}\mathds{1} + 2 \, B \right]^{-1}(x, y) =-\left[\lambda -2 \lambda^2 B +4\lambda^{3} B\star B -8\lambda^{4}B\star B\star B+\cdots \right]\,,
\end{align}
where $B\star B$ denotes the convolution integral $\int d^{d+1}z\sqrt{g (z)}   B(x,z) B(z,y)$. To perform this sum explicitly, we need to express $B(x,y)$ in the basis in which its action is diagonal. In flat space, this can be achieved simply by the Fourier-transformation since $B(x,y)$ in flat space depends only on the difference of the coordinates.

The analogue of the Fourier transform in AdS is called the {\it spectral representation}, which we review in Appendix \ref{app:SpRep}. We already employed the spectral representation of the propagator for the calculation of the effective potential in section \ref{sec:Phases}. Here we will use that more generally it can be defined for any bi-local function $F(x,y)$ in AdS, that only depends on the distance between the two points. Namely, any such function can be expanded in the basis of harmonic functions
\begin{equation}
F(x,y) = \int^{\infty}_{-\infty}d\nu \tilde{F}(\nu)\Omega_{\nu}(x,y)~.\label{eq:specrecall}
\end{equation}
The spectral representation shares one important property with the standard Fourier transform: It converts convolutions into products. Namely we have
\begin{align}\label{eq:convolutiontoproduct}
F\star G(x,y)=\int d\nu\, \tilde{F}(\nu) \tilde{G}(\nu) \Omega_{\nu} (x,y)
\end{align}
where $\tilde{F}(\nu)$ and $\tilde{G}(\nu)$ are the spectral representations of $F(x,y)$ and $G(x,y)$.

 Now, unlike a single bulk-to-bulk propagator, the spectral representation of the product of two bulk-to-bulk propagators \eqref{eq:productofbtob} takes a complicated form in general. For the time being, we do not need its explicit form and we will just treat it as an unknown function $\tilde{B} (\nu)$,
 \begin{align}
 B(x,y) = \left[\left(\frac{1}{-\square + M^2}\right)(x,y)\right]^2=\int_{-\infty}^{\infty} d\nu \,\tilde{B}(\nu )\Omega_{\nu}(x,y)\,.
 \end{align}
From this expression, we can immediately derive the expression for the two-point function of $\sigma$ using the property \eqref{eq:convolutiontoproduct} as
\begin{align}\label{eq:spectralforsigma}
\langle\delta \sigma (x)\delta\sigma (y)\rangle=-\int_{-\infty}^{\infty} d\nu \,\frac{1}{\lambda^{-1}+2 \tilde{B} (\nu)}\Omega_{\nu}(x,y)\,.
\end{align}
\paragraph{Step 2: Computing the four-point functions of $\phi^{i}$}
The next step is to construct the four-point function of $\phi^{i}$. For this purpose, it is convenient to introduce the embedding coordinates of AdS and of the boundary CFT. The embedding coordinates of a point on the boundary of AdS are given by
\begin{align}
P=\left(\frac{1+\vec{x}^{\,2}}{2},\frac{1-\vec{x}^{\,2}}{2},\vec{x}\right)\,,
\end{align}
where $P$ satisfies $P^{I}P_{I}=0$ and the signature\footnote{Throughout this section, we consider Euclidean AdS. If one wants to obtain the result for the original (Lorentzian) AdS, one can simply analytically continue the final results.} is $(-,+,+,\cdots)$. The inner product of two different $P$'s is related to the distance between two points,
\begin{align}
P_{12}\equiv -2 P_1\cdot P_2 = |\vec{x}_1-\vec{x}_2|^2
\end{align}
On the other hand, the embedding coordinates for a point in the bulk are given by
\begin{align}
X=\left(\frac{1}{2}\left(z+\frac{1}{z}+\frac{\vec{x}^{\,2}}{z}\right),\frac{1}{2}\left(-z +\frac{1}{z}-\frac{\vec{x}^{\,2}}{z}\right), \frac{\vec{x}}{z}\right)\,,
\end{align}
where $z$ and $\vec{x}$ are the Poincar\'e coordinates of AdS,
\begin{align}
ds^2=\frac{dz^2 +(d\vec{x})^2}{z^2}\,.
\end{align}
Using both $P$ and $X$, one can express the bulk-to-boundary propagator of the scalar field (with dimension $\Delta$) in the following simple way,
\begin{align}
K_{\Delta}(P,X)=\frac{\sqrt{\mathcal{C}_{\Delta}}}{(-2 P\cdot X)^{\Delta}}\,,
\end{align}
with
\begin{align}\label{eq:CDeltaDef}
\mathcal{C}_{\Delta}=\frac{\Gamma (\Delta)}{2\pi ^{d/2}\Gamma (\Delta-\frac{d}{2}+1)}\,.
\end{align}

Using the embedding coordinates, one can express the four-point function of $\phi^{i}$ at the leading order in the $1/N$ expansion, which is simply given by mean-field theory
\begin{align}
\langle \phi^{i}(P_1)\phi^{j}(P_2)\phi^{k}(P_3)\phi^{l}(P_4)\rangle|_{\mathcal{O}(1)}= \frac{\delta^{ij}\delta^{kl}}{(P_{12})^{\Delta}(P_{34})^{\Delta}}+\frac{\delta^{ik}\delta^{jl}}{(P_{13})^{\Delta}(P_{24})^{\Delta}}+\frac{\delta^{il}\delta^{jk}}{(P_{14})^{\Delta}(P_{23})^{\Delta}}\,.\nonumber
\end{align}
At the next order, the four-point function can be computed by contracting the bulk two-point function of $\sigma$ with the bulk-to-boundary propagators using the vertex $\sigma (\phi^{i})^2/\sqrt{N}$ in the effective action \eqref{eq:LagEff}. The result reads
\begin{equation}
\begin{aligned}
\langle \phi^{i}(P_1)\phi^{j}(P_2)\phi^{k}(P_3)\phi^{l}(P_4)\rangle|_{\mathcal{O}(1/N)}= &\frac{\delta^{ij}\delta^{kl} g_{12|34}+\delta^{ik}\delta^{jl} g_{13|24}+\delta^{il}\delta^{jk} g_{14|23}}{N}\,,\label{eq:scalarfourpfun}
\end{aligned}
\end{equation}
with\footnote{The prefactor $4$ is a standard combinatorial factor for the Feynman diagram.}
\begin{align}
\begin{aligned}
g_{12|34}& =4\int d X_1 dX_2 \,\,\langle\delta\sigma (X_1)\delta\sigma(X_2)\rangle\\
&\qquad \times K_{\Delta}(P_1,X_1)K_{\Delta}(P_2,X_1) K_{\Delta}(P_3,X_2)K_{\Delta}(P_4,X_2)\,.
\end{aligned}
\end{align}
To evaluate this integral, we use the spectral representation for $\langle\delta\sigma (X_1)\delta\sigma(X_2)\rangle$ in eq. \eqref{eq:spectralforsigma} and the following formula for the integral of the harmonic function, which we prove in Appendix \ref{ap:split}:
 \begin{align}\label{eq:integralweneedtocompute}
\begin{aligned}
\int dX_1 dX_2\, \Omega_{\nu}(X_1,X_2)K_{\Delta}(P_1,X_1)K_{\Delta}(P_2,X_1)K_{\Delta}(P_3,X_2)K_{\Delta}(P_4,X_2)=\\
\frac{1}{(P_{12})^{\Delta}(P_{34})^{\Delta}}\frac{\Gamma_{\Delta-\frac{d}{4}-\frac{i\nu}{2}}^{2}\Gamma_{\Delta-\frac{d}{4}+\frac{i\nu}{2}}^{2}}{64\pi^{\frac{d}{2}+1}\Gamma_{\Delta}^2\Gamma_{1-\frac{d}{2}+\Delta}^2}\left[\frac{\Gamma_{\frac{d}{4}+\frac{i\nu}{2}}^{4}\mathcal{K}_{\frac{d}{2}+i\nu}(z,\bar{z})}{\Gamma_{\frac{d}{2}+i\nu}\Gamma_{i\nu}}+\frac{\Gamma_{\frac{d}{4}-\frac{i\nu}{2}}^{4}\mathcal{K}_{\frac{d}{2}-i\nu}(z,\bar{z})}{\Gamma_{\frac{d}{2}-i\nu}\Gamma_{-i\nu}}\right]\,.\end{aligned}
\end{align}
Here $z$ and $\bar{z}$ are the conformal cross ratios of the boundary points defined by
\begin{align}\label{eq:crossratiodef}
\frac{P_{12}P_{34}}{P_{13}P_{24}}=z\bar{z}\,,\qquad \frac{P_{14}P_{23}}{P_{13}P_{24}}=(1-z)(1-\bar{z})\,,
\end{align}
and $\mathcal{K}_{\Delta}$ is the scalar conformal block in $d$ dimensions.
As a result, we obtain\footnote{To arrive at the formula, we used the invariance of the integrand under $\nu\to-\nu$ and combined the contributions from two terms in the second line of \eqref{eq:integralweneedtocompute} into one.}
\begin{align}\label{eq:g1234explicit}
g_{12|34}=-\frac{1}{(P_{12})^{\Delta}(P_{34})^{\Delta}} \int \frac{d\nu}{2\pi} \frac{1}{\lambda^{-1}+2\tilde{B}(\nu)}\frac{\Gamma_{\Delta-\frac{d+2i\nu}{4}}^2\Gamma_{\Delta-\frac{d-2i\nu}{4}}^2\Gamma_{\frac{d+2i\nu}{4}}^4}{4\pi^{\frac{d}{2}}\Gamma_{\Delta}^2\Gamma_{1-\frac{d}{2}+\Delta}^2\Gamma_{i\nu}\Gamma_{\frac{d}{2}+i\nu}}\mathcal{K}_{\frac{d}{2}+i\nu}(z,\bar{z}) \,,
\end{align}
where we used the abbreviations $\Gamma_{x}\equiv \Gamma (x)$. To generate the OPE expansion, one simply needs to close the contour on the lower-half plane and read off the residues at the poles.

\paragraph{Step 3: Bootstrapping the bubble function}
Now that we derived the expression for $g_{12|34}$ \eqref{eq:g1234explicit}, the remaining task is to determine the unknown function $\tilde{B}(\nu)$. As mentioned above, the function $\tilde{B}(\nu)$ comes from a scalar bubble diagram in AdS and one can in principle compute it directly using Witten diagrams. For AdS$_3$, this was carried out explicitly in \cite{Giombi:2017hpr} by using the split representation\footnote{For recent developments on the computation of loop diagrams using the split representation, see \cite{Yuan:2017vgp,Yuan:2018qva}. See also alternative approaches \cite{Cardona:2017tsw,Bertan:2018khc,Bertan:2018afl} which do not rely on the split representation.} of the bulk-to-bulk propagators. However the computation is rather involved and it seems especially hard to obtain an explicit expression for even-dimensional AdS, such as AdS$_2$. Below, we present an alternative method which does not involve the evaluation of the diagram at all. Instead, we just impose the consistency of the OPE expansion of the boundary conformal theory and ``bootstrap'' the bubble function. By doing so, we succeed in obtaining an explicit expression valid in any dimensions\footnote{The expression for the bubble diagram in any dimension as a sum over double-trace propagators was derived using the orthogonality of bulk-to-bulk propagators in \cite{Fitzpatrick:2010zm}, and extended to the case of different masses in the two propagators in \cite{Fitzpatrick:2011hu}. Using harmonic analysis in AdS, similar expressions in Mellin space were derived in \cite{Penedones:2010ue,Fitzpatrick:2011dm}. For odd-dimensional AdS, this result was reproduced from the analytic conformal bootstrap in \cite{Aharony:2016dwx}. We also make use of harmonic analysis and the bootstrap idea but our method seems much simpler than these analyses, and can be readily generalized to other cases, for instance to fermions. See subsection \ref{subsec:GNcompcorr}.}.

To see this, let us project the four-point function of $\phi^i$ to the $O(N)$ singlet sector in the s-channel (namely $12\to 34$ channel). This can be achieved by contracting the correlator against a tensor $\delta_{ij}\delta_{kl}/N^2$ and the result reads
\begin{align}
\begin{aligned}
&\frac{1}{N^2}\langle \phi^{i}(P_1)\phi^{i}(P_2)\phi^{k}(P_3)\phi^{k}(P_4)\rangle=\\
&\frac{1}{(P_{12})^{\Delta}(P_{34})^{\Delta}}+\frac{1}{N}\left[\frac{1}{(P_{13})^{\Delta}(P_{24})^{\Delta}}+\frac{1}{(P_{14})^{\Delta}(P_{23})^{\Delta}}+g_{12|34}\right] +\mathcal{O}(1/N^2)\,.
\end{aligned}
\end{align}
As shown above, the projection to the singlet sector suppresses the contribution of the $t$- and $u$-channel diagrams and therefore diagrams which naively come from different orders of the $1/N$ expansion contribute at the same order in the large $N$ expansion. This fact has an important bearing on the OPE expansion of this correlator as we see below. 

In terms of the s-channel OPE, the leading $\mathcal{O}(1)$ term simply represents a contribution of the identity operator. At $\mathcal{O}(1/N)$, there are two kinds of contributions: The first contribution comes from the generalized-free-field correlators (the first two terms inside the square bracket), and can be decomposed into a sum of the double-trace conformal blocks, namely
\begin{align}\label{eq:GFFOPE}
\frac{1}{(P_{13})^{\Delta}(P_{24})^{\Delta}}+\frac{1}{(P_{14})^{\Delta}(P_{23})^{\Delta}}=\frac{1}{(P_{12})^{\Delta}(P_{34})^{\Delta}} \sum_{\substack{\ell,n\\\ell:\text{ even}}}2c_{n,\ell}^2\mathcal{K}_{2\Delta+2n+\ell,\ell}(z,\bar{z})\,,
\end{align} 
 with\footnote{For the derivation of the OPE coefficients for the double-trace operators, see \cite{Fitzpatrick:2011dm}.}
 \begin{align}
 \begin{aligned}\label{eq:GFF3ptcoef}
 c_{n,\ell}^{2}=\frac{(-1)^{\ell}\left[(\Delta-\frac{d}{2}+1)_n(\Delta)_{\ell+n}\right]^2}{\ell ! n! (\ell+\frac{d}{2})_n(2\Delta+n-d+1)_{n}(2\Delta+2n+\ell-1)_{\ell}(2\Delta+n+\ell-\frac{d}{2})_n}\,,
 \end{aligned}
 \end{align}
 where $(a)_b$ is the Pochhammer symbol, $(a)_b=\Gamma(a+b)/\Gamma(a)$. The OPE expansion of the second contribution $g_{12|34}$ can be read off from the spectral parameter integral \eqref{eq:g1234explicit} by closing the contour on the lower-half plane. As is clear from the structure of the integrand, there are two sets of poles:
 \begin{enumerate}
 \item The poles at $\frac{d}{2}+i\nu=2\Delta +2n$ $(n\in \mathbb{N}_{\geq 0})$, which come from the factor $\Gamma_{\Delta-\frac{d+2i\nu}{4}}^2$.
 \item The poles coming from
 \begin{align}
 \lambda^{-1}+2\tilde{B}(\nu)=0\,.
 \end{align}
 \end{enumerate} 
 The first set of poles is precisely at the position of the scalar double-trace primaries $\phi^{i}\square^{n}\phi^i$ while the second set of poles are at generic positions which depend on the coupling constant $\lambda$.
 When the coupling constant $\lambda$ vanishes, we do not have the interacting diagram (namely $g_{12|34}=0$) and therefore the OPE expansion only yields the double-trace primaries \eqref{eq:GFFOPE}, which correspond to freely moving two-particle states. Now, once we turn on the coupling, the two-particle states are no longer free and we expect that (at least some of) their energies get slightly shifted\footnote{If we just consider a few perturbative Witten diagrams, they typically yield infinitesimal corrections to the dimensions of the operators which manifest themselves as logarithmic terms in the OPE expansion. By contrast, here we are {\it resumming} the diagrams and therefore we expect that the dimensions of operators receive finite shifts.}. However, as we saw above, even at finite $\lambda$ there are poles precisely at the locations of the non-interacting double-trace primaries. The only way to make it consistent with the physical intuition is to require that {\it the first set of poles in $g_{12|34}$ precisely cancels the corresponding generalized-free-field contribution \eqref{eq:GFFOPE}}. This leads to the relation
 \begin{align}
 \frac{1}{\lambda^{-1}+2\tilde{B}(\nu)}\frac{\Gamma_{\Delta-\frac{d+2i\nu}{4}}^2\Gamma_{\Delta-\frac{d-2i\nu}{4}}^2\Gamma_{\frac{d+2i\nu}{4}}^4}{4\pi^{\frac{d}{2}}\Gamma_{\Delta}^2\Gamma_{1-\frac{d}{2}+\Delta}^2\Gamma_{i\nu}\Gamma_{\frac{d}{2}+i\nu}}\quad \overset{\frac{d}{2}+i\nu\, \sim \,2\Delta+2n}{\sim} \quad \frac{-2c_{n,0}^2}{\frac{d}{2}+i\nu -(2\Delta+2n)}\,.
 \end{align}
 Now, on the left-hand side of this relation, we have $\Gamma_{\Delta-\frac{d+2i\nu}{4}}^2$ which produces double poles at $\frac{d}{2}+i\nu=2\Delta+2n$ while the right-hand side is just a simple pole. This means that the function $\tilde{B}(\nu)$ must have simple poles\footnote{Note that simple poles in $\tilde{B}(\nu)$ correspond to simple zeros of $1/(\lambda^{-1}+2\tilde{B}(\nu))$.} with appropriate residues at these points. Working out the residues, we conclude that the singularity of $2\tilde{B}(\nu)$ is given by
 \begin{align}
 2\tilde{B}(\nu)\overset{\frac{d}{2}+i\nu\, \sim \,2\Delta+2n}{\sim}-\frac{1}{\frac{d}{2}+i\nu-(2\Delta+2n)}\frac{(\frac{d}{2})_n\Gamma_{\Delta+n}\Gamma_{\Delta+n-\frac{d}{2}+\frac{1}{2}}\Gamma_{2\Delta+n-\frac{d}{2}}}{(4\pi)^{\frac{d}{2}}\Gamma_{n+1}\Gamma_{\Delta+n+\frac{1}{2}}\Gamma_{\Delta+n-\frac{d}{2}+1}\Gamma_{2\Delta-d+n+1}}\,.
 \end{align}
 Since the bubble function $\tilde{B}(\nu)$ must be symmetric under the shadow transform $\nu\to -\nu$, there are also poles on the upper half plane with the same residues.
 
 Now, to fully determine $\tilde{B}(\nu)$, we need two further inputs; the knowledge about the existence of other poles and the behavior at infinity.
Let us first discuss the existence of other poles. If there were other poles, they would contribute already to the $\mathcal{O}(\lambda)$ correction of the four-point function (namely a simple one-loop diagram in AdS) and predict the existence of new operators. This would mean that the operator spectrum changes discontinuously once we turn on the coupling. However, on general grounds, we do not expect that to happen perturbatively.\footnote{We expect that such a change of spectrum happens only when there appears a bound state. However, a bound state is not something that one can see perturbatively; it can only be seen once one resums diagrams (as we will see in section \ref{sec:fermion}).} Therefore we conclude that there are no other poles. Let us next discuss the behavior at infinity. Physically the limit $\nu\to \infty$ corresponds to the high energy limit. Since the curvature of AdS becomes negligible in the high energy limit, one can determine the asymptotics of $\tilde{B}(\nu)$ from the high energy limit of the flat-space scattering amplitude. This leads to the following asymptotics of $\tilde{B}(\nu)$ in AdS$_{d+1}$:
\begin{align}
\tilde{B}(\nu) \sim 1/\nu^{3-d} \qquad (\nu\to \infty)\,.
\end{align}
Using these inputs,  we can determine $\tilde{B}(\nu)$ uniquely to be\footnote{Note that we rewrote the expression into a manifestly shadow-symmetric form.}
 \begin{align}
 \tilde{B}(\nu)=-\sum_{n=0}^{\infty}\frac{2\Delta+2n-\frac{d}{2}}{\nu^2+(2\Delta+2n-\frac{d}{2})^2}\frac{(\frac{d}{2})_n\Gamma_{\Delta+n}\Gamma_{\Delta+n-\frac{d}{2}+\frac{1}{2}}\Gamma_{2\Delta+n-\frac{d}{2}}}{(4\pi)^{\frac{d}{2}}\Gamma_{n+1}\Gamma_{\Delta+n+\frac{1}{2}}\Gamma_{\Delta+n-\frac{d}{2}+1}\Gamma_{2\Delta-d+n+1}}\,. \label{eq:bubblesum}
 \end{align}
We find perfect agreement with the expression for the bubble as a sum of double-trace exchanges found in position space in \cite{Fitzpatrick:2010zm}, upon translating it to the spectral representation. Moreover, given the simplicity of the propagator in the spectral representation, the sum in \eqref{eq:bubblesum} can be performed explicitly (with the help of Mathematica) and we finally get 
 \begin{align}\label{eq:finalBnu}
 \begin{aligned}
 &\tilde{B}(\nu)=\frac{\Gamma_{\Delta}\Gamma_{\Delta-\frac{d}{2}+\frac{1}{2}}\Gamma_{2\Delta-\frac{d}{2}}}{4(4\pi)^{\frac{d}{2}}}\\
 &\times \left(\Gamma_{\Delta-\frac{d+2i\nu}{4}}\,{}_5\tilde{F}_{4}\left[\begin{array}{c}\{\frac{d}{2},\Delta,\Delta-\frac{d}{2}+\frac{1}{2},\Delta-\frac{d+2i\nu}{4},2\Delta-\frac{d}{2}\}\\\{\Delta+\frac{1}{2},\Delta-\frac{d}{2}+1,\Delta-\frac{d+2i\nu}{4}+1,2\Delta-d+1\}\end{array};1\right]\right.\\
 &\left.\quad+ \Gamma_{\Delta-\frac{d-2i\nu}{4}}\,{}_5\tilde{F}_{4}\left[\begin{array}{c}\{\frac{d}{2},\Delta,\Delta-\frac{d}{2}+\frac{1}{2},\Delta-\frac{d-2i\nu}{4},2\Delta-\frac{d}{2}\}\\\{\Delta+\frac{1}{2},\Delta-\frac{d}{2}+1,\Delta-\frac{d-2i\nu}{4}+1,2\Delta-d+1\}\end{array};1\right]\right)\,,
 \end{aligned}
 \end{align}
 where ${}_5\tilde{F}_{4}$ is the {\it regularized} generalized hypergeometric function. The result is valid in any dimensions, but the series that defines the hypergeometric function is divergent for $d+1\geq 4$. This is due to the fact that the bubble diagram has a UV divergence in this range of $d$\footnote{This is the same as what we have in flat space.}.
For $d=2$, i.e. on AdS$_3$, the expression simplifies to
 \begin{align}\label{eq:ads3Bnuexplicit}
 \tilde{B}(\nu)\quad \overset{d=2}{=} \quad \frac{i\left[\psi(\Delta-\frac{1+i\nu}{2})-\psi(\Delta-\frac{1-i\nu}{2})\right]}{8\pi \nu}\,.
 \end{align}
This precisely matches the result in \cite{Giombi:2017hpr}.
 
 \subsubsection{Generalization to $1/N$ corrections}\label{subsubsec:corrections}
The idea explained above can in principle be applied also to $1/N$ corrections. At the next order in the $1/N$ expansion, the four-point function in the singlet sector reads
\begin{align}
\begin{aligned}
&\frac{1}{N^2}\langle \phi^{i}(P_1)\phi^{i}(P_2)\phi^{k}(P_3)\phi^{k}(P_4)\rangle=\frac{1}{(P_{12})^{\Delta}(P_{34})^{\Delta}}\\
&+\frac{1}{N}\left[\frac{1}{(P_{13})^{\Delta}(P_{24})^{\Delta}}+\frac{1}{(P_{14})^{\Delta}(P_{23})^{\Delta}}+g_{12|34}\right] +\frac{1}{N^2}\left[g_{13|24}+g_{14|23}+g'_{12|34}\right]\,,
\end{aligned}
\end{align}
where $g'_{12|34}$ is the s-channel diagram for the $1/N^2$ correction to the four-point function. Since we already determined $g_{12|34}$, we also know $g_{13|24}$ and $g_{14|23}$. The OPE expansion of these two terms will again yield a collection of scalar double-trace operators without any shifts of the conformal dimension. However, for the same reason that we provided above, we do not expect such operators to exist in the full OPE expansion of the boundary conformal theory. This means that those double-trace contributions must be killed by the last term $g'_{12|34}$. This constrains the form of $g'_{12|34}$ and, with a little more assumption, it is likely that we can determine $g'_{12|34}$ without performing explicit diagrammatic computations. If successful, this would provide a recursive way to bootstrap $1/N$ corrections in this theory. It would be interesting to carry this out explicitly but we will leave it for future investigations.  

 It is worth pointing out that the bootstrap analysis performed in this subsection relies crucially on the fact that the correlators are meromorphic functions of the spectral parameter. By contrast, the scattering amplitude in flat space contains branch cuts, which make it difficult to perform analogous analysis. In this sense, the results in this subsection provide evidence that studying a theory on AdS rather than in flat space is not just extra complication but has real advantages.

    
\subsection{Analyzing the Correlators}
We now study the properties of the correlator that we derived above.

\begin{figure}[t]
\centering
\includegraphics[clip,height=7cm]{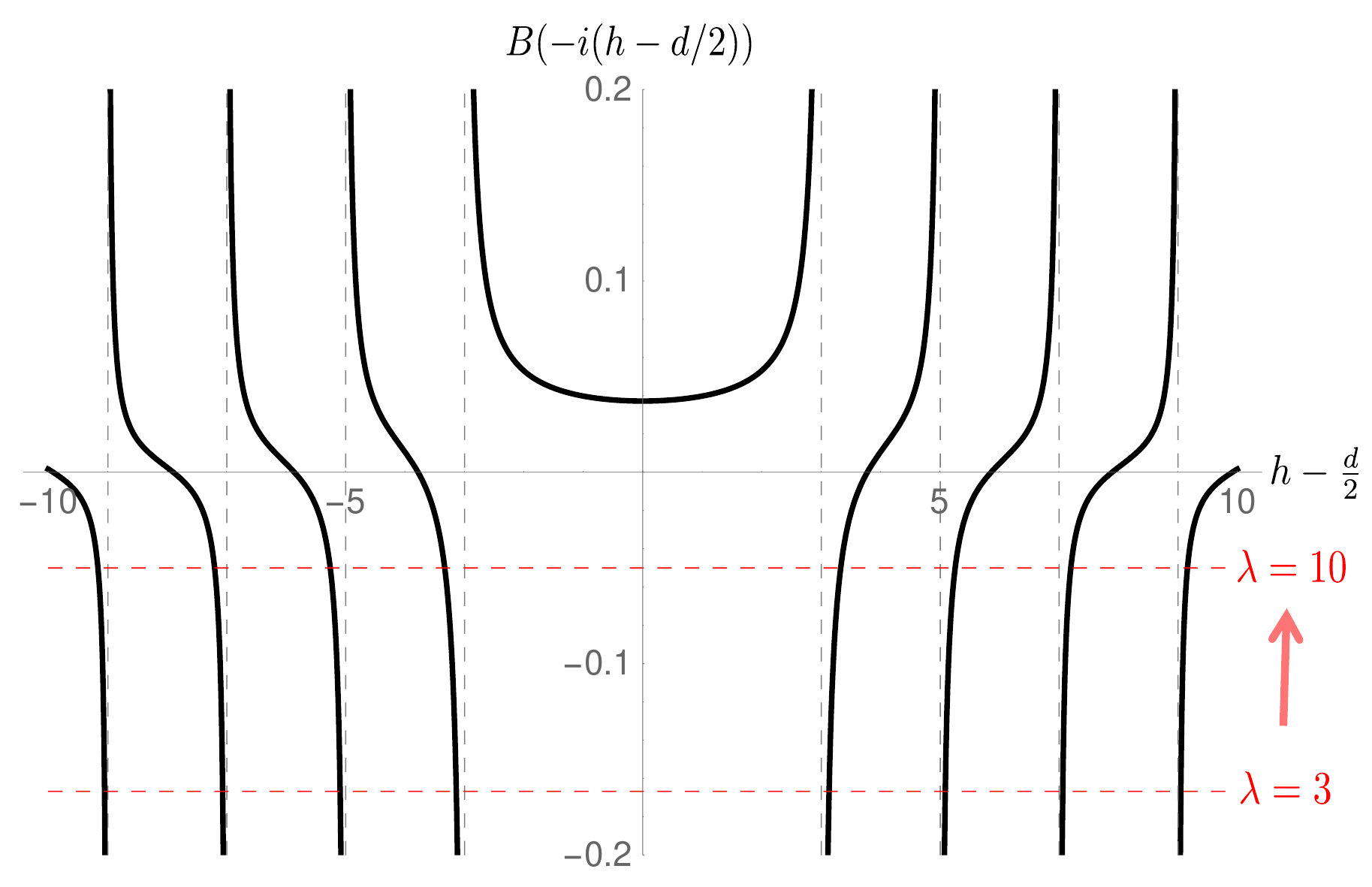}
\caption{The bubble function $\tilde{B}(\nu)$ for $d=2$ (AdS$_3$) and $\Delta=2$. The black curves denote the values of the bubble function $\tilde{B}(-i (h-d/2))$ and the horizontal axis is $h-d/2$ where $h$ is a conformal dimension. The spectrum of the operator can be read off from the intersection points of the black curve and the red dashed lines at $-\lambda^{-1}/2$. As shown in the figure, the red dashed line moves upward as we increase the coupling and eventually coincides with the horizontal axis.}\label{fig:readingoffdimensions}
\end{figure}

The simplest physical information that can be extracted from the correlator is the spectrum of the boundary conformal theory. As explained in the previous subsection, at $\mathcal{O}(1/N)$ the dimensions $h$ of the scalar double trace operators in the $O(N)$ singlet sector get finite shifts, which can be read off from the equation
\begin{align}
\lambda^{-1}+2 \tilde{B}\left(-i(h -\tfrac{d}{2})\right)=0\,.
\end{align}
As shown in figure \ref{fig:readingoffdimensions}, the dimensions determined by this equation start from the generalized-free-field spectrum and increase as we crank up the coupling $\lambda$, eventually receiving $\mathcal{O}(1)$ anomalous dimensions at strong coupling. Below we study several limits of this equation and discuss its physical consequences. For simplicity, we only present the results for AdS$_3$, but the qualitative features are the same\footnote{The only exception is the analysis on the critical point, which does not exist in AdS$_2$.} also for AdS$_2$.
\paragraph{Large Conformal Dimension}
Let us first analyze the large-$h$ behavior of the operator spectrum. In the limit $h\to \infty$, the bubble function $\tilde{B}(\nu)$ for AdS$_3$ can be expanded as
\begin{align}
\tilde{B}(-i (h-1))=\frac{\cot \left(\pi (\Delta-\tfrac{h}{2})\right)}{8h}\left[1+\mathcal{O}(1/h^2)\right]
\end{align} 
Therefore, the leading asymptotic operator spectrum is determined by the equation
\begin{align}
\lambda^{-1}+\frac{\cot \left(\pi (\Delta-\tfrac{h}{2})\right)}{4h}=0\,.
\end{align}
Incidentally the form of the equation resembles the one obtained for the SYK model with a quartic interaction ($q=4$) \cite{Maldacena:2016hyu, Polchinski:2016xgd}.

When the coupling is turned off, the solution to this equation is 
\begin{align}
h=2\Delta+2n \qquad n\in \mathbb{N}_{\geq 0}\qquad (\lambda=0)
\end{align}
and coincides with the scalar double-trace spectrum. On the other hand, in the infinite coupling limit, they are shifted by $1$, namely
\begin{align}
h=2\Delta+2n +1 \qquad n\in \mathbb{N}_{\geq 0}\qquad (\lambda=\infty)\,,
\end{align}
while in the intermediate range of the coupling, the operators receive anomalous dimensions $0\leq \delta h\leq 1$.
\paragraph{Conformal Limit}
As we will discuss in detail in section \ref{sec:Critical}, the bulk theory becomes conformal when $\Delta=1$ and $\lambda=\infty$. The operator spectrum in this limit is determined by the equation
\begin{align}
\tilde{B}(-i (h-1))=\frac{\cot \tfrac{\pi h}{2}}{8-8h}=0\,,
\end{align} 
whose solution is given by
\begin{align}
h=2\Delta+2n+1 =2n+3 \qquad n\in \mathbb{N}_{\geq 0}\,.\label{eq:sigmaconfspec}
\end{align}
This in particular contains the dimension $3$ operator which can be interpreted as the displacement operator.
Note also that this spectrum is identical\footnote{This fact seems to be related to the bulk conformal symmetry: From the bulk point of view, the conformal dimension (or more precisely the spectral parameter $\nu$) parametrizes the scale and the large dimension limit corresponds to the UV limit in the bulk. If the theory is at criticality, we expect that the observables do not qualitatively change under change of scales, which is what we found here. However, at the time of writing, we do not know how to make this argument more rigorous. For a related discussion on the consequence of the bulk conformality, see section \ref{sec:Critical}.} to the asymptotic spectrum that we derived above. In section \ref{sec:Critical}, we study the correlator at the critical point in more detail and compare with the known results.
\paragraph{Flat-Space Limit}
Let us next consider the flat-space limit. For this purpose, it is useful to reinstate the dependence on the AdS radius,\footnote{We use conventions where no factors of $L$ appear in the definition \eqref{eq:specrecall} of the spectral representation, hence $\tilde{F}$ carries the same dimension as $F$. This implies that when we reinstate $L$ the convolution identity becomes $\widetilde{F\star G} = L^{-d-1}\tilde{F}\tilde{G}$. In particular, the spectral transform of the delta function is $L^{-d-1}$.}
\begin{align}
\begin{aligned}\label{eq:totakeflat}
\langle \delta \sigma (x)\delta \sigma (y)\rangle&=\int^{\infty}_{\infty}d\nu\, \tilde{F}_{\delta\sigma\delta\sigma}(\nu)\,\Omega_{\nu}(x,y)\,,\\
\tilde{F}_{\delta\sigma\delta\sigma}(\nu)&=-\frac{1}{L^4}\frac{1}{(\lambda L)^{-1}+2 L^2\tilde{B}(\nu)}\,,
\end{aligned}
\end{align}
where $L^2 \tilde{B}(\nu)$ is given by \eqref{eq:ads3Bnuexplicit}. 
\begin{figure}[t]
\centering
\includegraphics[clip,height=6cm]{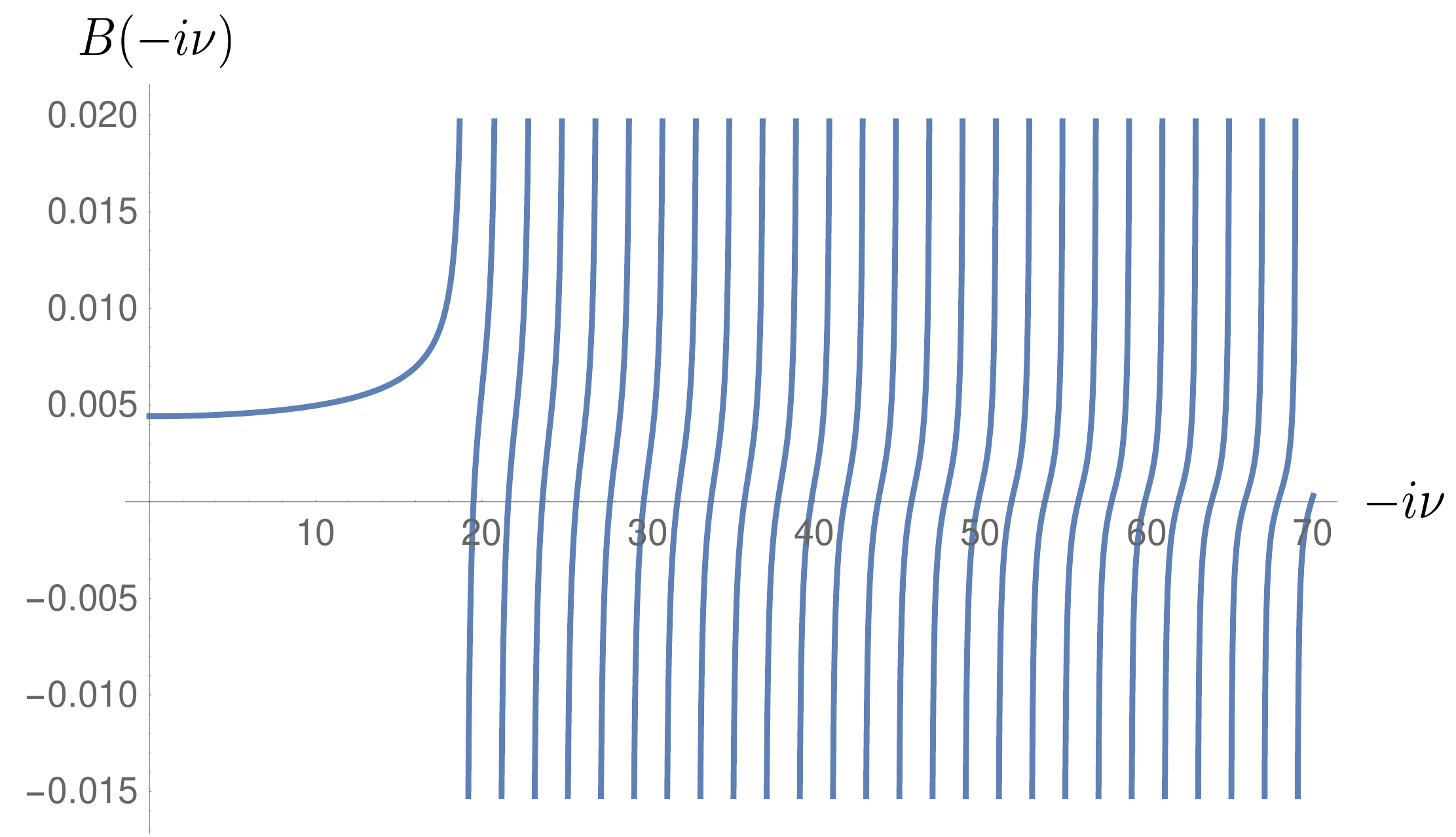}
\caption{The behavior of the bubble function $\tilde{B}(\nu)$ in AdS$_3$ for $\Delta=10$. As depicted in the figure, it has a sequence of poles (starting from $-i\nu= 2\Delta-1$) along the imaginary axis which condense into a two-particle branch cut upon taking the flat-space limit.\label{fig:sequenceofpoles}}
\end{figure}

To take the flat-space limit, we need to send the AdS radius $L$ to be infinite while keeping the parameters in the Lagrangian ($\lambda$ and $M^2$) fixed. 
We can then identify the spectral representation with the Fourier transformation in the radial direction as we show rigorously in Appendix \ref{sec:FslSpRep}:
 \begin{align}
 L^{d+1} \,\tilde{F}_{OO}(\nu = L |p|) \underset{L\to\infty}{\longrightarrow} \tilde{F}^{\rm flat}_{OO}(|p|)~.
 \end{align}
 Here $\tilde{F}_{OO}$ on the left-hand side is the spectral representation of the bulk two-point functions  $\langle OO\rangle$ and $\tilde{F}^{\rm flat}_{OO}$ is the radial Fourier transformation of the two-point function in flat space. If we perform the same rescaling of the spectral parameter to the function $\tilde{B}(\nu)$, we obtain
 \begin{align}
 L^3 \, \tilde{B}(L|p|)\underset{L\to\infty}{\longrightarrow} \tilde{B}^{\rm flat}(|p|)=\frac{{\rm arctan}(\tfrac{p}{2M})}{4\pi |p|}\,.\label{eq:fsscalar}
 \end{align}
 As expected, the quantity on the right hand side coincides with the bubble diagram in flat space $\mathbb{R}^3$:
 \begin{align}
 \tilde{B}^{\rm flat}(|p|)=\frac{{\rm arctan}(\tfrac{p}{2M})}{4\pi |p|}&=\int \frac{d^3q}{(2\pi)^3} \frac{1}{(q^2+M^2)((p+q)^2+M^2)}\,.
 \end{align}
 Thus the limit of the full two-point function correctly reproduces the result in flat space:
 \begin{align}
 \lim_{L\to \infty}L^{3} \,\tilde{F}_{\delta\sigma\delta\sigma}(\nu = L |p|)=-\frac{1}{\lambda^{-1}+2 \tilde{B}^{\rm flat}(|p|)}\,,
 \end{align}

Note that the flat-space limit of the bubble function $\tilde{B}^{\rm flat}(|p|)$ has a branch cut if we analytically continue the momentum to the imaginary value:
\begin{align}\label{eq:branchcutBflat}
\tilde{B}^{\rm flat}(-i p)=\frac{{\rm arctanh}(\tfrac{p}{2M})}{4\pi p}\,.
\end{align}
This branch cut, which starts from $p=2M$, is a familiar two-particle threshold in flat space. On the other hand, the bubble function $\tilde{B}(\nu)$ in AdS contains a collection of poles on the (negative) imaginary $\nu$ axis:
\begin{align}
L^2\tilde{B}(-i\nu)=\frac{\psi (\Delta-\tfrac{1-i\nu}{2})-\psi (\Delta-\tfrac{1+i\nu}{2})}{8\pi\nu} =\infty \quad \text{at }\nu=2\Delta+2n-1 \quad (n\in \mathbb{N}_{\geq 0})\,.
\end{align}
As shown in figure \ref{fig:sequenceofpoles}, these poles come close to each other and reproduce the branch cut in \eqref{eq:branchcutBflat} upon taking the flat-space limit.

\paragraph{Scale Dependence of the Correlator}
\begin{figure}
\centering
\begin{minipage}{0.49\hsize}
\centering
\includegraphics[clip,height=4.5cm]{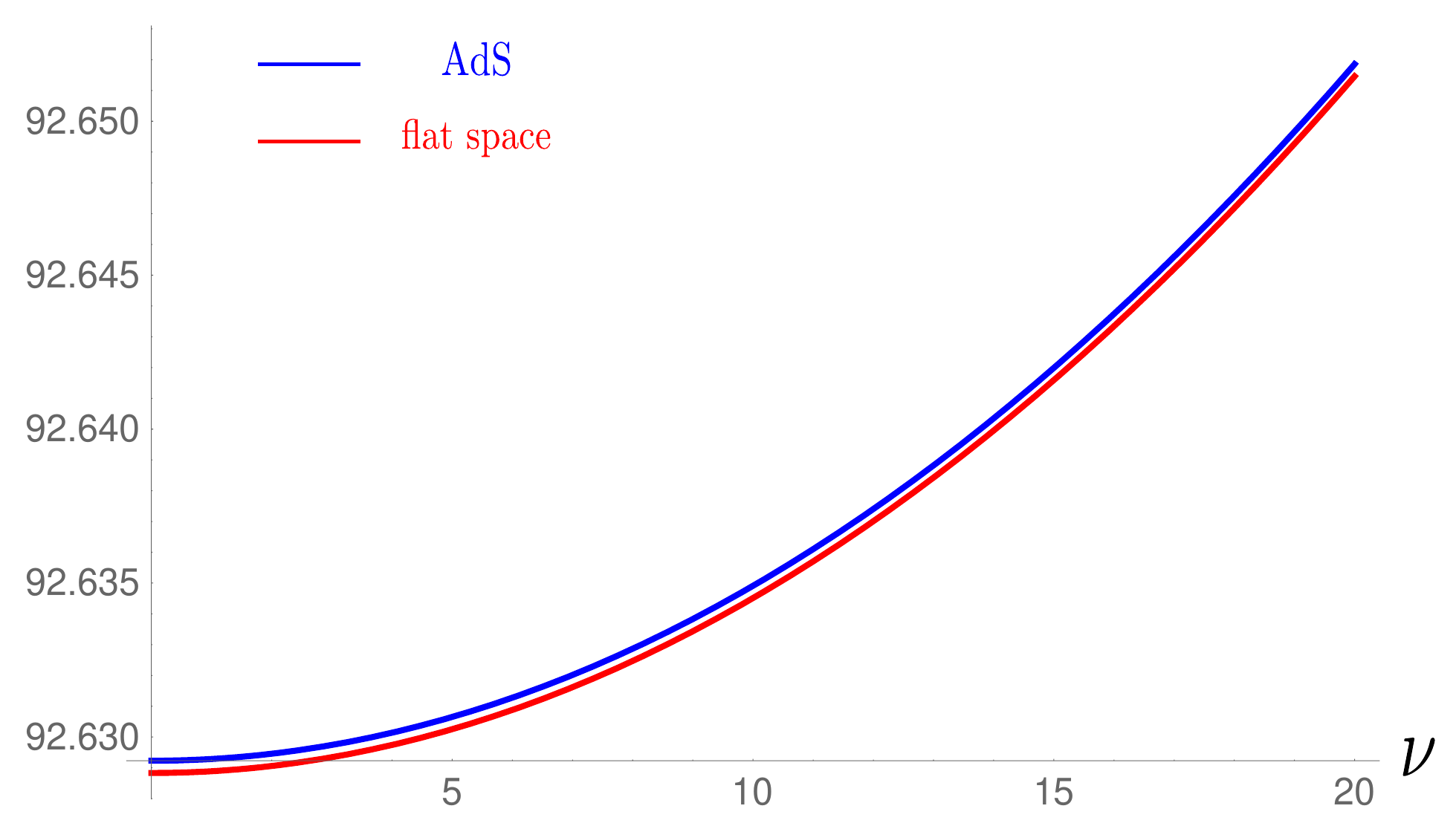}\\
(a) $\Lambda_{\rm AdS}=0.01$
\end{minipage}
\begin{minipage}{0.49\hsize}
\centering
\includegraphics[clip,height=4.5cm]{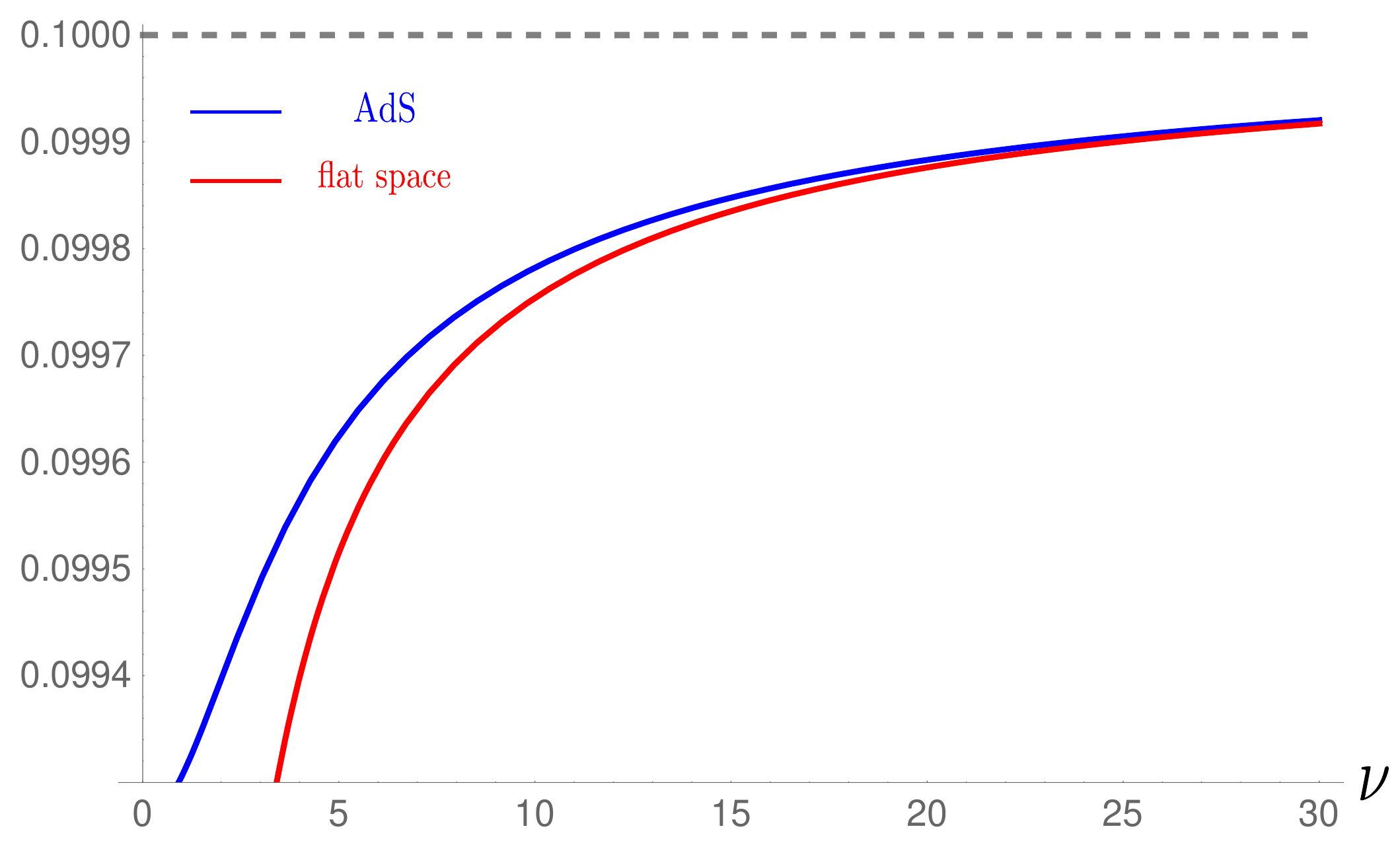}\\
(b) $\Lambda_{\rm AdS}=10$
\end{minipage}
\caption{Comparison of the two-point functions in AdS  and in flat space. In both figures, $\Lambda_{m}=1$ and the blue curve denotes the function in AdS ($\tilde{F}_{\delta\sigma\delta\sigma}(\nu)$) while the red curve denotes the function in flat space ($\tilde{F}^{\rm flat}_{\delta\sigma\delta\sigma}(\nu)$). (a) The plot for $\Lambda_{\rm AdS}=0.01$. Since $\Lambda_{\rm AdS}$ is small, the theory does not see the AdS curvature until it reaches the deep IR. Therefore the two functions stay close for a wide range of parameters. (b) The plot for $\Lambda_{\rm AdS}=10$. In this case, as soon as the functions deviate from the UV value (denoted by a gray dashed line), they start to differ significantly.\label{fig:comparewithflat}}
\end{figure}
Let us now analyze the behavior of the two-point function in a generic parameter regime. As discussed above, the spectral parameter plays the role of the radial momentum in flat space. Therefore, by analyzing the behavior of the correlator as a function of the spectral parameter, one can gain some information about the renormalization-group flow and the scale dependence of the theory. For this purpose, it is convenient to introduce the following dimensionless parameters:
\begin{align}
\Lambda_{m}= \frac{M}{\lambda}\,,\qquad \Lambda_{\rm AdS}=\frac{1}{\lambda L}\,.
\end{align}
Roughly speaking, $\Lambda_{m}$ parametrizes the mass scale of the theory while $\Lambda_{\rm AdS}$ is the scale which governs the ``finite-size correction'' coming from the AdS radius. In terms of these quantities, $\tilde{B}(\nu)$ and the spectral transform of the two-point function are given by
\begin{align}
\begin{aligned}
\tilde{B}(\nu)=&\frac{i\left[\psi \left(\tfrac{1-i\nu}{2}+\sqrt{1+\tfrac{\Lambda_{m}^2}{\Lambda_{\rm AdS}^2}}\right)-\psi \left(\tfrac{1+i\nu}{2}+\sqrt{1+\tfrac{\Lambda_{m}^2}{\Lambda_{\rm AdS}^2}}\right)\right]}{8\pi \nu}\,,\\
\tilde{F}_{\delta \sigma \delta\sigma}(\nu)&=\frac{1}{\Lambda_{\rm AdS}+2 \tilde{B}(\nu)}\,.
\end{aligned}
\end{align} 

Unless the theory is fine-tuned to be at the critical point, the UV behavior of the theory is governed by flat-space physics because the AdS curvature becomes negligible in the extreme UV. Therefore, the two-point function approaches the flat-space counterpart
\begin{align}\label{eq:flatspacecounterpart}
\tilde{F}_{\delta \sigma \delta\sigma}^{\rm flat}(\nu)=\left(\Lambda_{\rm AdS}+\frac{{\rm arctan}\left[\tfrac{\Lambda_{\rm AdS}\nu}{2\Lambda_{m}}\right]}{2\pi \nu}\right)^{-1}\,,
\end{align}
when $\nu$ is sufficiently large. Let us first consider the situation where $\Lambda_{\rm AdS}\ll 1$ and $\Lambda_{m}$ is finite. When $\Lambda_{\rm AdS}$ is small, we expect that the theory does not see the effect of the AdS curvature until we reach the deep IR regime and therefore the two-point function stays close to \eqref{eq:flatspacecounterpart} for a wide range of the spectral parameter. This is indeed the case as shown in figure \ref{fig:comparewithflat}-(a).
Only when the spectral parameter becomes of order $\Lambda_{\rm AdS}$ does the theory start seeing the effect of the AdS radius, and from there the two functions start to differ. 
On the other hand, if $\Lambda_{\rm AdS}$ is much larger than $1$, there is basically no regime in which the flat-space approximation is valid. Therefore, as soon as the two-point function deviates from the UV value, the two functions start differing significantly (see figure \ref{fig:comparewithflat}-(b)).

\begin{figure}[t]
\centering
\begin{minipage}{0.49\hsize}
\centering
\includegraphics[clip,height=4.5cm]{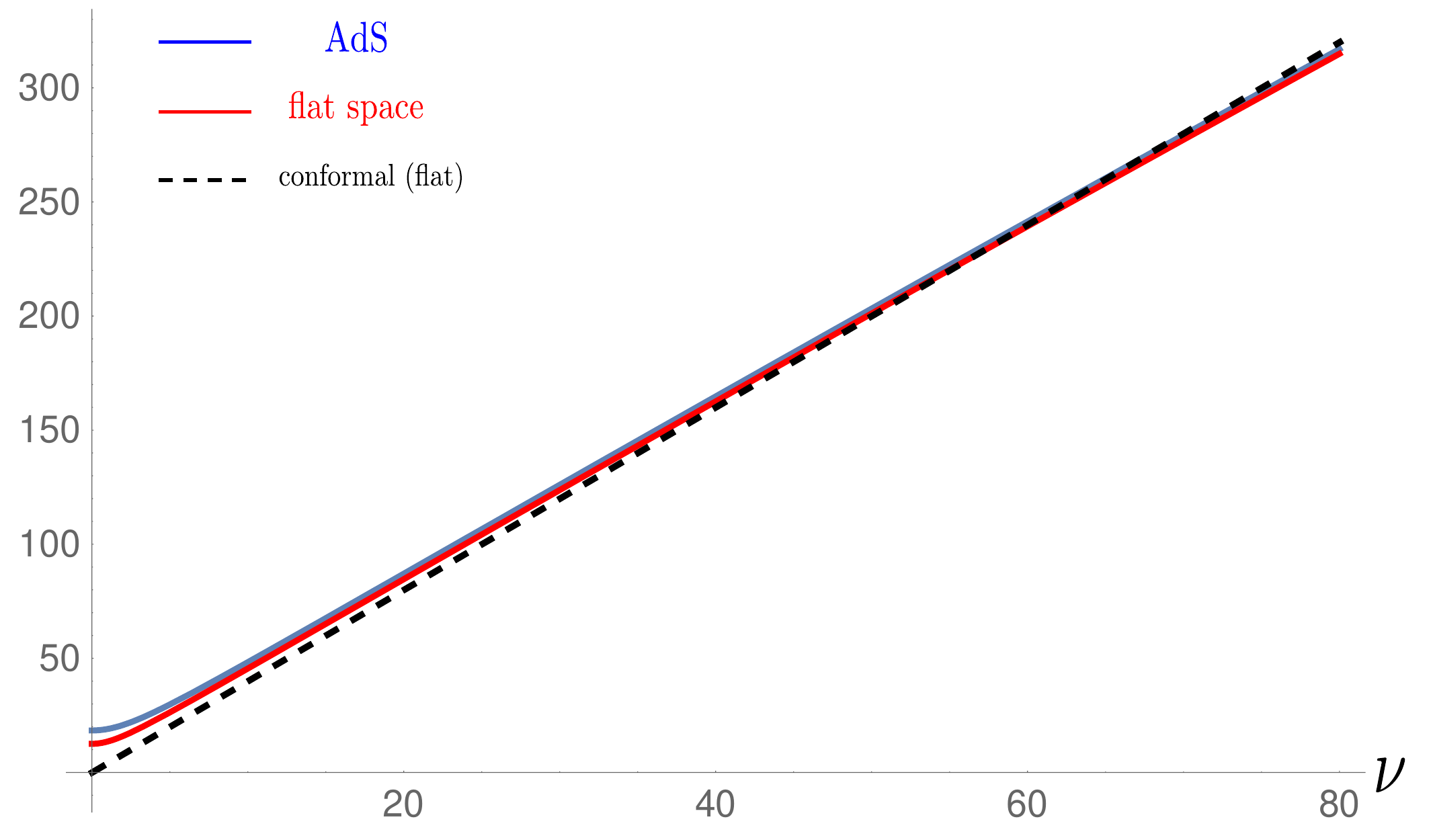}\\
(a) $\Lambda_m=\Lambda_{\rm AdS}=0.0001$
\end{minipage}
\begin{minipage}{0.49\hsize}
\centering
\includegraphics[clip,height=4.5cm]{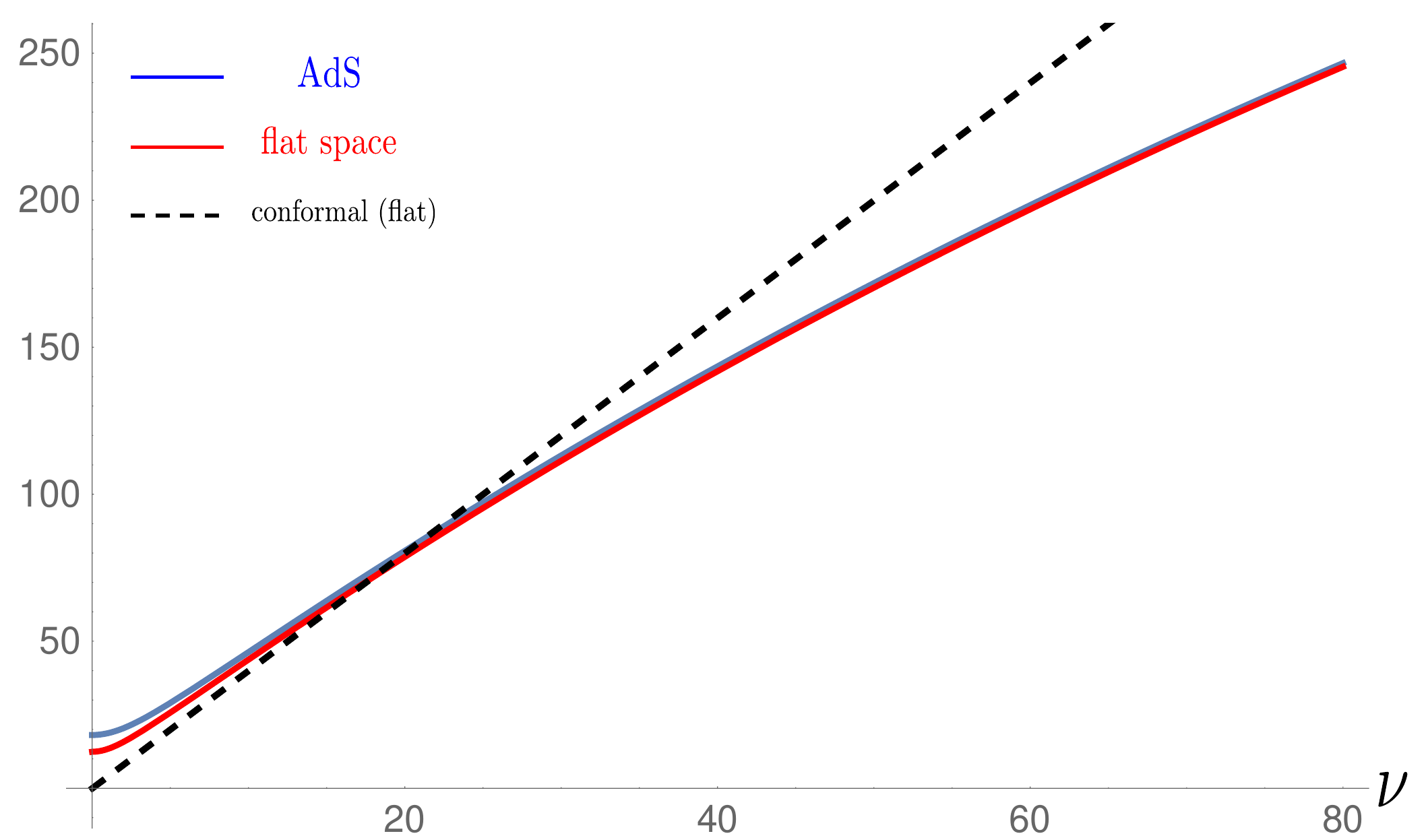}\\
(b) $\Lambda_m=\Lambda_{\rm AdS}=0.001$
\end{minipage}
\caption{Flow to the flat-space critical point. In both figures, the blue and red curves denote the two-point functions in AdS ($F_{\delta\sigma\delta\sigma}(\nu)$) and in flat space ($F^{\rm flat}_{\delta\sigma\delta\sigma}(\nu)$) respectively while the black dashed line denotes the two-point function at the critical point in flat space ($F^{\text{flat, conformal}}_{\delta\sigma\delta\sigma}(\nu)=4\nu$). When both $\Lambda_{m}$ and $\Lambda_{\rm AdS}$ are small, the theory does not see any scale until deep in the IR. Therefore, the two-point function exhibits conformal behavior in flat space in some range of the spectral parameter. (a) The plot for $\Lambda_m=\Lambda_{\rm AdS}=0.0001$. As shown in the figure, the three curves are close to each other for a wide range of the spectral parameter. (b) If we increase $\Lambda_m$ and $\Lambda_{\rm AdS}$, the curves start to deviate from each other (although there is still a small range of the spectral parameter in which they coincide.)\label{fig:comparewithconformal}}
\end{figure}

Finally, when both $\Lambda_{\rm AdS}$ and $\Lambda_{m}$ satisfy $\Lambda_{\rm AdS}, \Lambda_{m}\ll 1$, the theory does not see any scale until deep in the IR, and therefore flows close to the critical point in flat space. This can also be verified explicitly using our result as shown in figure \ref{fig:comparewithconformal}.

It would be interesting to study in more detail the scale dependence of the theory by formulating a renormalization-group equation for quantum field theories in AdS. In particular, it would be nice to understand the differences and the similarities between the usual finite-size scaling \cite{Fisher:1972zza,Brezin:1981gm} and the corrections induced by the AdS radius.


\subsection{Correlators in the Symmetry-Breaking Phase}\label{subsec:corrsymbreak}
We now consider the symmetry-breaking phase with $m^2<m_0^2$ and non-zero VEV\footnote{The quantities $m^2/\lambda$ and $m_0^2/\lambda$ are UV divergent and depend on the choice of regularization scheme, but the VEV $|\Phi|^2$ is physical and independent of such choices.}
\begin{align}
|\Phi|^2 = \frac{m_0^2-m^2}{2\lambda} ~,\\
\frac{m_0^2}{\lambda}\equiv\tr\left(-\frac{1}{\square}\right)~.
\end{align}  
We will see some interesting new phenomena in this phase. In particular, using the correlators computed above, we will provide an example of the AdS analogue of a resonance in a scattering amplitude. We will also comment on the implication of the existence of the bulk Goldstone bosons for the boundary conformal theory.

Without loss of generality we take the VEV along the $N$-th component $\Phi^i = \delta^{iN} |\Phi|$. We decompose the $O(N)$ vector in the radial mode $\rho$ and the Goldstone bosons $\pi^i$  
\begin{align} 
\rho & \equiv \delta{\phi}^N  = \phi^N - \sqrt{N} |\Phi|~,\\
 \pi^i & \equiv \phi^i~,~~i=1,\dots,N-1~.
\end{align} 
In this phase $M^2=0$, which implies a non-zero VEV $\Sigma = -\frac{m^2}{2} $ and 
\begin{equation}
\delta \sigma = \sigma + \sqrt{N}\frac{m^2}{2}~.
\end{equation}

The effective Lagrangian in these variables is, up to a constant
\begin{align}
\mathcal{L}_{\rm eff} & = \frac12 (\partial \rho)^2 +\frac12 (\partial \pi^i)^2   - \frac{1}{2\lambda} (\delta\sigma)^2 + 2 |\Phi| \, \delta\sigma \rho+ \frac{1}{\sqrt{N}}  \,\delta\sigma (\pi^i)^2 \nonumber   \\ & + \sqrt{N}\,\left(\frac{m^2}{2} + \lambda |\Phi|^2\right)\delta\sigma + \frac{N}{2} \, \tr \log\left(-\square +  \frac{2}{\sqrt{N}}\delta\sigma \right)~.\label{eq:expsymmact}
\end{align}
Note that the terms linear in $\delta\sigma$ cancel, as they should. 

\subsubsection{Resonance in AdS}
In eq. \eqref{eq:expsymmact} we see that the symmetry-breaking VEV induces a mixing between $\rho$ and $\delta\sigma$. As a consequence, even at infinite $N$ there is an $\mathcal{O}(1)$ interaction between $\rho$ and two pions. Before we start the analysis in AdS, let us remind the reader that in flat space this implies that the $\rho$ particle, which at tree-level has mass $m^2_\rho = 4\lambda |\Phi|^2$, becomes unstable at the quantum level. The associated pole gets an imaginary part, and it manifests as a resonance in the 2 to 2 amplitude of the pions. This is discussed in the original paper \cite{Coleman:1974jh}, and we will see it emerge from the flat-space limit of our result.

The quadratic terms in the action involving $\rho$ and $\sigma$ can be written in matrix notation as follows
\begin{align}
&\int_x \int_y \,\frac 12 \left( \delta\sigma(x) ~\rho(x) \right) K(x,y)
\begin{pmatrix} \delta\sigma(y) \\ \rho(y)\end{pmatrix}~,\\
& K(x,y) \equiv \begin{pmatrix}
-\frac{1}{\lambda}\delta^{d+1}(x,y)- 2B(x,y) & 2 |\Phi| \delta^{d+1}(x,y) \\
2 |\Phi|\delta^{d+1}(x,y)  &  -\square_y \delta^{d+1}(x,y) \\ 
\end{pmatrix}
\end{align}
where $\int_x \equiv \int_{AdS_{d+1}} d^{d+1} x \sqrt{g(x)}$ and similarly for $y$, and $\delta^{d+1}(x,y)$ is the delta function on AdS. The bubble function $B(x,y)$ here is evaluated at $M^2 =0$, i.e. with $\Delta =2$. Inverting the 2-by-2 kernel $K(x,y)$ above one obtains the corresponding matrix of two-point functions, at leading order at large $N$. This computation becomes algebraic if we employ the spectral representation.
Defining as usual
\begin{equation}
K(x,y) = \int_{-\infty}^\infty d\nu \, \tilde{K}(\nu) \Omega_\nu(x,y)~,
\end{equation}
we have
\begin{equation}
\tilde{K}(\nu) = \begin{pmatrix}
-\frac{1}{\lambda}- 2 \tilde{B}(\nu) & 2 |\Phi| \\
2 |\Phi| &  \nu^2+\frac{d^2}{4} \\ 
\end{pmatrix}~.
\end{equation}
Therefore, the matrix of two-point function can be simply expressed in terms of the function $\tilde{B}(\nu)$ as follows
\begin{align}
& \langle \begin{pmatrix} \delta\sigma \\ \rho \end{pmatrix} \left( \delta\sigma ~\rho\right) \rangle(\nu)  = (\tilde{K}(\nu))^{-1} = \frac{1}{\det \tilde{K}(\nu)}\begin{pmatrix}
\nu^2+\frac{d^2}{4}  & -2 |\Phi| \\
-2 |\Phi|  &  -\frac{1}{\lambda}- 2  \tilde{B}(\nu)\\ 
\end{pmatrix}  ~.\label{eq:matrixcorr}
\end{align}
Instead of diagonalizing the matrix, if we are interested in the spectrum of boundary operators contributing to either of the two correlators, we can simply look at the zeroes of the determinant
\begin{align}
\det \tilde{K}(\nu) &= \left(\nu^2+\frac{d^2}{4}\right)\left(-\frac{1}{\lambda}- 2 \tilde{B}(\nu)\right) - 4|\Phi|^2 = -\frac{1}{\lambda}\left[f(\nu)+ 2(m^2_0-m^2)\right] \label{eq:detres}\\ f(\nu) & \equiv \left(\nu^2+\frac{d^2}{4}\right)\left(1+ 2\lambda \tilde{B}(\nu)\right)~, \label{eq:deff}
\end{align}
along the imaginary $\nu$ axis. The scaling dimensions $h$ of these operators is related to the location of the zeroes by $-i \nu = \pm (h - \tfrac{d}{2})$. 

By attaching bulk-to-boundary propagators of four external pions to the bulk two-point function of $\delta \sigma$, similarly to what we did in the previous section in the massive phase, we obtain a boundary four-point correlator that is the AdS version of the 2 to 2 scattering amplitude of pions. The poles the two-point function of $\delta \sigma$ ---the upper diagonal entry of  \eqref{eq:matrixcorr}--- then correspond to the dimension of operators exchanged in this four-point function. By inspection of  \eqref{eq:matrixcorr} we see that these poles are again simply given by the zeroes of $\det \tilde{K}(\nu)$.

In the limit $\lambda \to 0$ we recover classical physics in AdS, and indeed the equation $\det \tilde{K}(\nu) = 0$ reduces to 
\begin{equation}
\nu^2+\frac{d^2}{4}  + 2(m^2_0-m^2) = 0~,\label{eq:freerho}
\end{equation}
that corresponds to a $\rho$ particle of mass $m^2_\rho = 4 \lambda |\Phi|^2 = 2(m^2_0-m^2)$. 

\begin{figure}
\hspace{2.1cm}
\begin{subfigure}{0.5\textwidth}
\centering
  \includegraphics[width=1.51\linewidth]{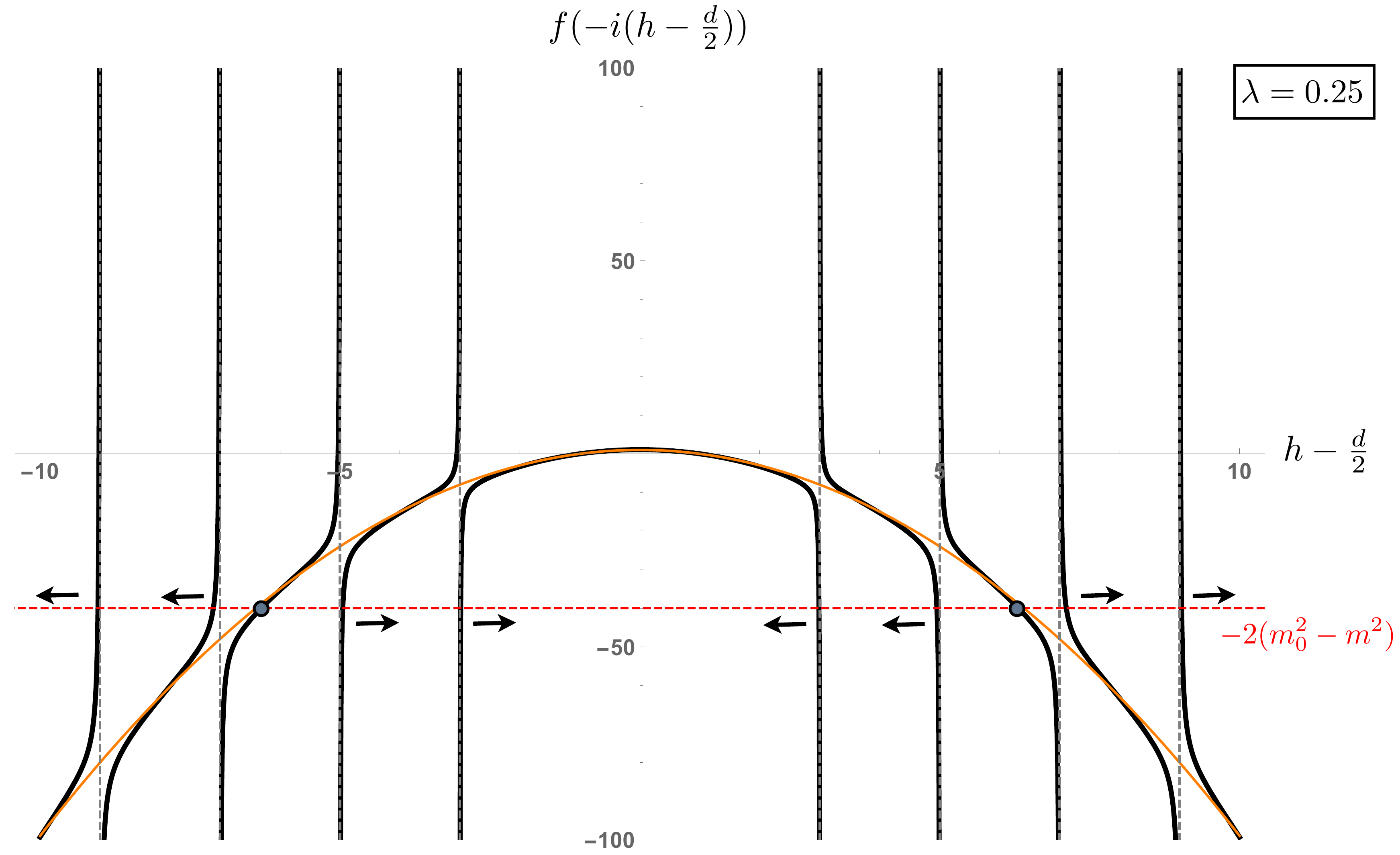}
  \label{fig:sub1}
\end{subfigure}
\newline
\phantom{1}\hspace{2cm}
\begin{subfigure}{0.5\textwidth}
\centering
  \includegraphics[width=1.51\linewidth]{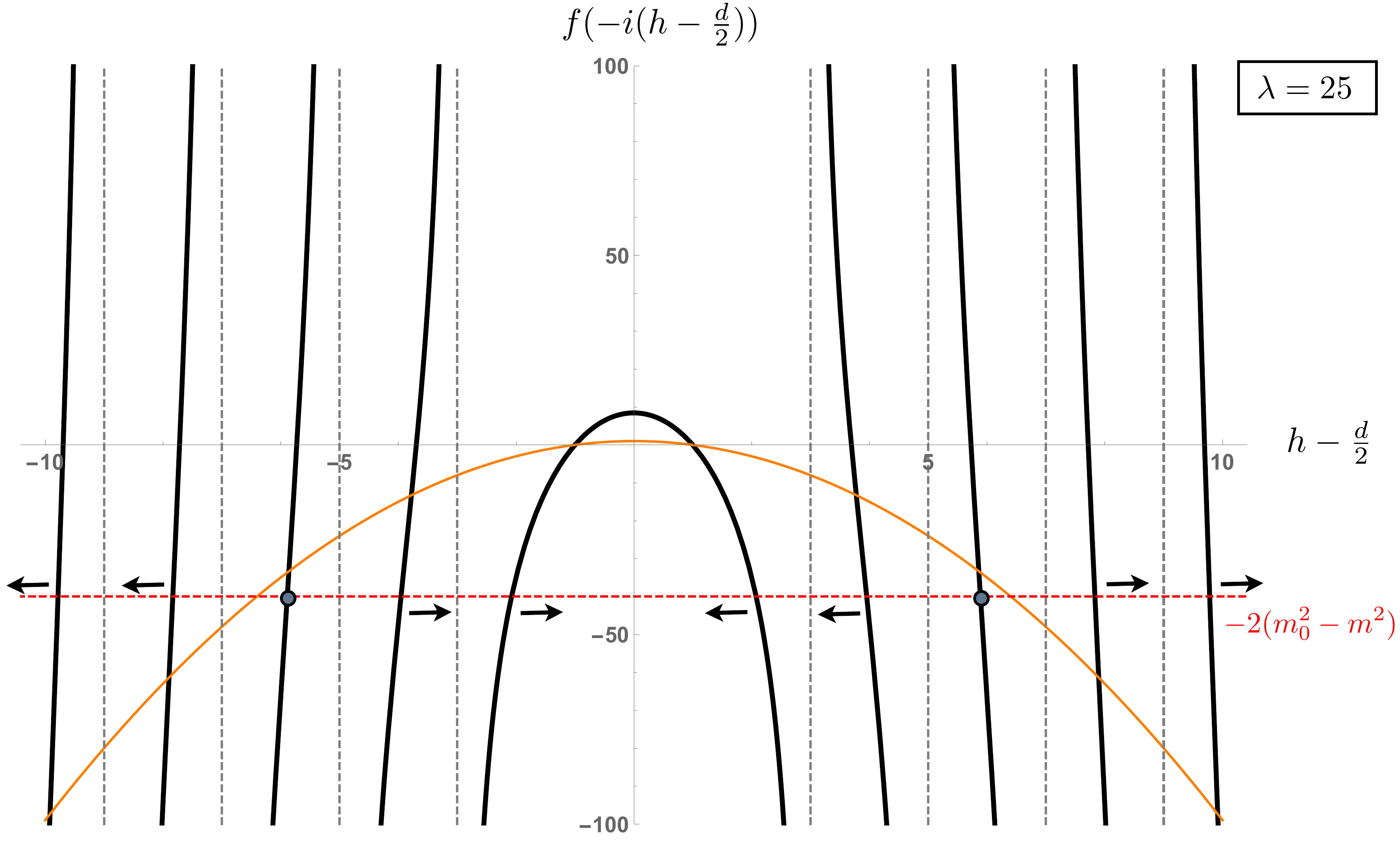}
  \label{fig:sub3}
\end{subfigure}
\caption{In black, the function $f(-i(h-\tfrac{d}{2}))$ defined in eq. \eqref{eq:deff} as a function of $h-\tfrac{d}{2}$, for $d=2$, i.e. AdS$_3$. The red dashed line is the constant $-2(m_0^2 - m^2)$ which we fixed to $-40$. Its intersections with the black curve determine the values $h$ of the scaling dimensions of boundary operators contributing to the AdS amplitude of pions. The gray vertical dashed lines are the dimensions of double-trace operators in the free theory. The orange curve is the limit $\lambda \to 0$ of the black curve, and its intersection with the red line corresponds to the $\rho$ particle at tree-level. The dotted intersection between the black and the red curve is the continuation of the $\rho$ particle to finite $\lambda$. The arrows denote the sign of the double-trace anomalous dimensions.}
\label{fig:Resonance}
\end{figure}

Turning on $\lambda$, we generate an additional infinite set of poles associated to the field $\delta\sigma$, that correspond to the finite-coupling version of the double-trace operators / two-pion states. These poles are analogous to the ones that we discussed in the massive phase, see fig. \ref{fig:readingoffdimensions}. For small $\lambda$, the spectrum is illustrated in the upper plot in fig. \ref{fig:Resonance}: there is a pole very close to the free-particle pole of eq. \eqref{eq:freerho}. Moreover, all the ``double-trace" poles are close to their values in the free theory, and there is a distinctive feature in the pattern of their anomalous dimensions, namely they flip sign when they cross the $\rho$-particle pole. Note that this is different from the situation in the massive phase, depicted in fig. \ref{fig:readingoffdimensions}, in which all the anomalous dimensions have the same sign. As we crank up $\lambda$, the distinction between the $\rho$-particle pole and the double-trace poles becomes obscured, but the qualitative features of the pattern of the anomalous dimensions persist as shown in the lower figure of fig. \ref{fig:Resonance}.
\begin{figure}[t]
\centering
\begin{minipage}{0.49\hsize}
\centering
\includegraphics[clip, height=5cm]{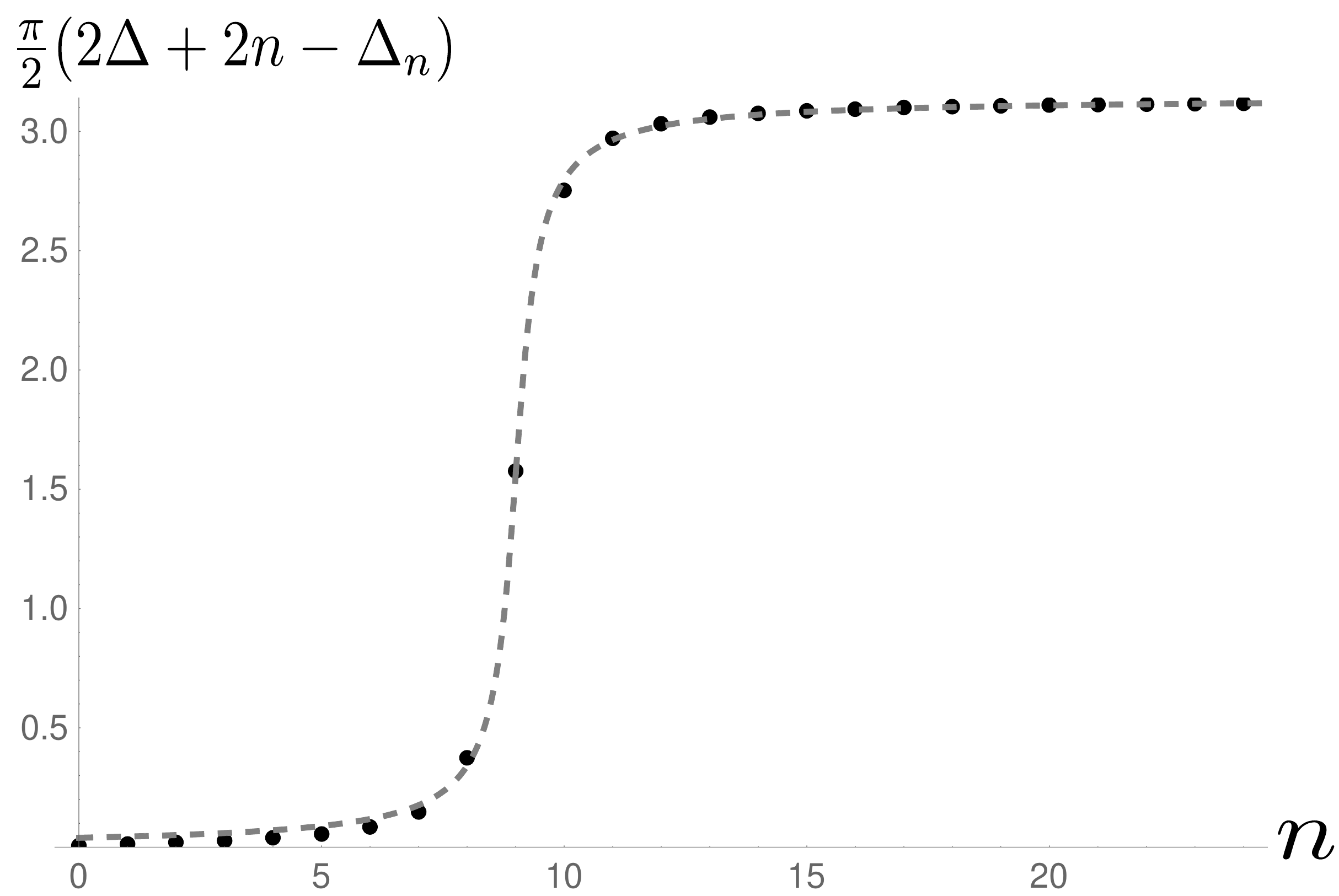}\\
(a) $\lambda=4$, $x=8.99$, $y=0.35$
\end{minipage}
\begin{minipage}{0.49\hsize}
\centering
\includegraphics[clip, height=5cm]{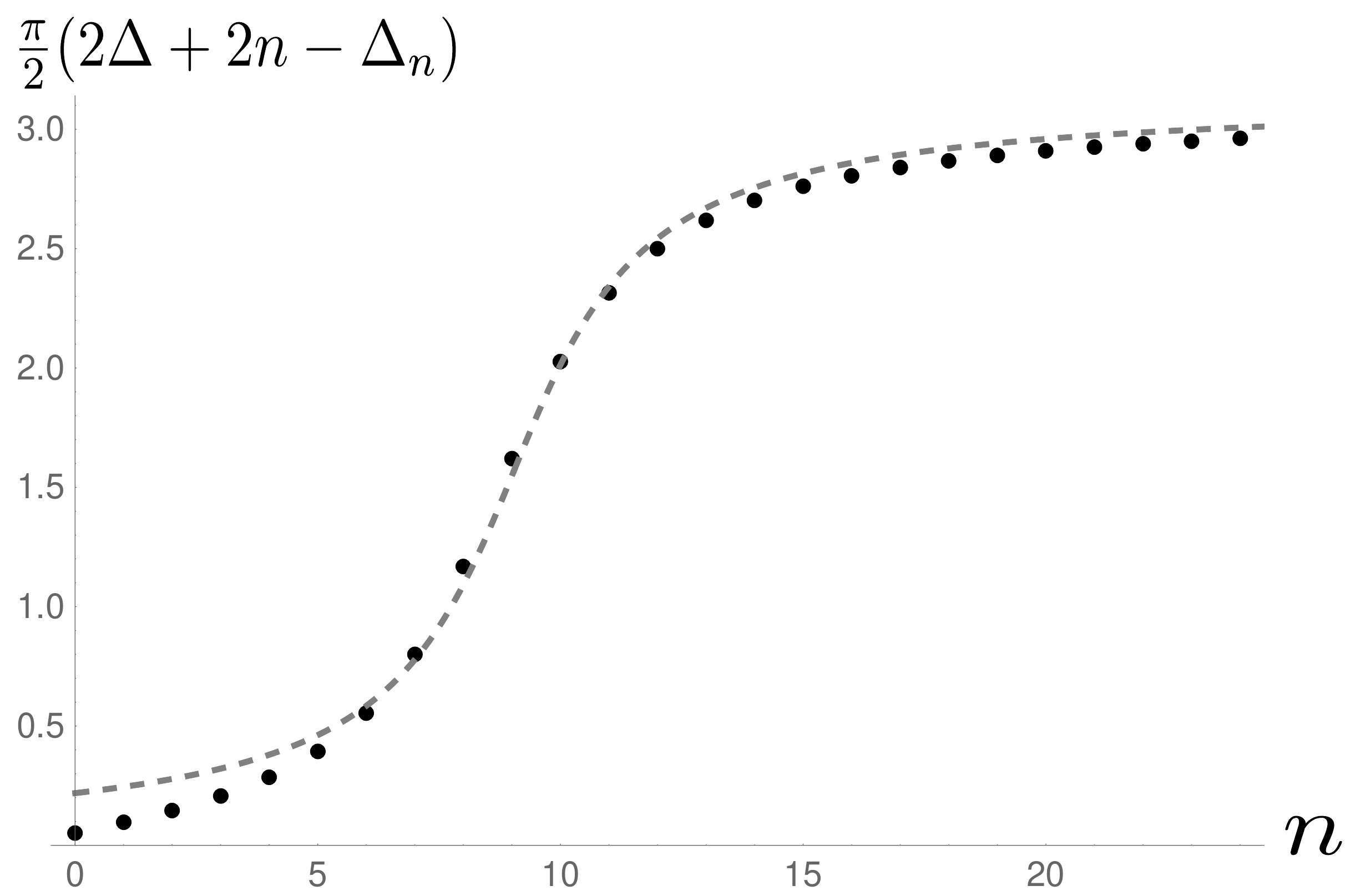}\\
(b) $\lambda=30$, $x=9.03$, $y=2.01$
\end{minipage}
\caption{The anomalous dimensions of the double-trace operators $\Delta_n-(2\Delta+2n)$. In both figures, we considered AdS$_3$ and set $\Delta=2$ (i.e. $M^2$ = 0) and $2(m_0^2 - m^2) = 400$. The black dots are the actual values of the anomalous dimensions while the dashed curve is the phase of the Breit-Wigner pole, $\arg \left(\dfrac{-1}{n-x+iy}\right)$. The values of $x$ and $y$ are determined by fitting the result for the anomalous dimensions. (a) When the coupling is small, the anomalous dimension undergoes a quick shift by $\pi$ around the $\rho$ pole. Correspondingly the imaginary part of the Breit-Wigner pole $y$ is small. (b) As we increase $\lambda$, the slope becomes less steep, which implies that the resonance becomes broad.}\label{fig:phaseshiftano}
\end{figure}

This sign-flip is reminiscent of one of the signatures of a resonance in a 2-to-2 scattering amplitude in flat space, namely that the phase shift gets shifted by $\pi$ across the resonance. The connection can be made more explicit using the relation\footnote{Roughly speaking, the exponential on the right hand side measures the AdS analogue of the phase shift, namely the relative phase shift over one period of the AdS time between the free propagation of the two particles in AdS and the actual particle/operator that appears in the OPE of four-point function. For a more precise definition and derivation, see the original paper \cite{Paulos:2016fap}.
} proposed in \cite{Paulos:2016fap},
\begin{align}
e^{2i\delta_{l}(s)}=\lim_{\Delta_{i}\to \infty} \langle e^{-i\pi (\Delta-\Delta_1-\Delta_2-l)}\rangle\,,
\end{align}
where $\Delta_i$ are the dimensions of the external operators while $\langle \ast\rangle$ denotes the average over the operators with dimension $\Delta\simeq\sqrt{s}$ weighted by the structure constants squared. If we neglect the weights coming from the structure constants and assume that the dimensions of the operators are close to those of the double-trace operators in the free theory, this simplifies to the following relation between the anomalous dimensions of the double-trace operators and the phase shift $\delta_l(s)$,
\begin{align}\label{eq:simplerelanomalousphase}
\delta_l (s)\sim \frac{\pi}{2}(\Delta_1+\Delta_2+2n+l-\Delta_n^{(l)})\qquad \sqrt{s}\sim \Delta_1+\Delta_2+2n+l\,,
\end{align}
where $\Delta_n^{(l)}$ is the $n$-th lightest operator with spin $l$. As we show in fig. \ref{fig:phaseshiftano}, the plot of the anomalous dimensions\footnote{To apply the formula \eqref{eq:simplerelanomalousphase}, we should regard the $\rho$-pole as one of the double-trace poles. This is physically reasonable since there is no clear distinction between $\rho$ particle and the double-trace states at finite $\lambda$.} of our result neatly reproduces the behavior of the phase shift coming from a simple pole of the Breit-Wigner type, $\frac{-1}{n-x+iy}$. In particular, the slope becomes less steep as we increase the coupling constant (see fig. \ref{fig:phaseshiftano}-(b)) being in line with the flat-space intuition that the particles are easier to decay when the coupling is strong. These results provide evidence that the pattern of the anomalous dimensions that we found is tied to the existence of the resonance in flat space.\footnote{The emergence of a resonance in the flat-space limit of AdS/CFT was discussed in \cite{Fitzpatrick:2011hu} using the Mellin amplitude. Our result provides an explicit realization of such a mechanism and also shows that the resonance-like behavior can be seen already at finite AdS radius.}

Alternatively, we can check explicitly that the resonance emerges upon taking the flat-space limit of eq. \eqref{eq:detres}. For simplicity we will restrict ourselves to $d=2$, i.e. AdS$_3$, in which the bubble function for $M^2 = 0$ simplifies to
\begin{equation}
\tilde{B}(\nu)\vert_{d=2, M^2=0} = -\frac{1}{2\pi(1+\nu^2)} + \frac{\tanh\left(\frac{\pi \nu}{2}\right)}{8\nu}~.
\end{equation}  
Plugging $\nu = |p|$ in \eqref{eq:detres}, and then taking $|p|$, $\lambda$ and $|\Phi|$ to $\infty$ with $\lambda/|p|$ and $|\Phi|^2/|p|$ fixed, we obtain 
\begin{equation}
-\lambda \det \tilde{K}(\nu)\longrightarrow p^2\left(1+\frac{\lambda}{4 |p|}\right) +4 \lambda |\Phi|^2~.\label{eq:fspres}
\end{equation}
This expression is equivalent to the flat-space result of \cite{Coleman:1974jh} so we can simply rely on their analysis. We have a square-root branch point at $p^2 = 0$ due to the dependence on $|p| = \sqrt{p^2}$, that is produced by the condensation of the ``double-trace" poles. This branch-cut is  interpreted as the two-pion cut in the 2-to-2 scattering amplitude of massless pions. The complex $|p|$ plane thus contains both the first sheet ($\mathrm{Re} |p| > 0$) and the second sheet ($\mathrm{Re} |p| < 0$) for the variable $p^2$, with the physical region $p^2<0$ corresponding to the negative imaginary axis for $|p|$. Setting \eqref{eq:fspres} to zero and solving for $|p|$ we find the following two poles
\begin{equation}
|p| = -\frac{\lambda}{8} \pm i \sqrt{4\lambda |\Phi|^2 - \frac{\lambda^2}{64} }~,
\end{equation}
that are both on the second sheet, and approach the classical result $m_\rho^2 = 4\lambda |\Phi|^2$ for small $\lambda$. We see that indeed, as anticipated from the AdS result, the width increases with $\lambda$, at least as long as $4\lambda |\Phi|^2 - \frac{\lambda^2}{64} > 0$. On the other hand some features of the flat-space result do not have an evident counterpart in the $\nu$ space analysis of the AdS correlator above. Namely, there are actually two distinct poles on the second sheet in flat-space, while we could only detect one ``resonance-like" feature in the anomalous dimensions. Moreover for large values of $\lambda$, such that $4\lambda |\Phi|^2 - \frac{\lambda^2}{64} < 0$, both poles lie on the real negative $|p|$ axis, and one of them gets closer and closer to the physical sheet as $\lambda \to \infty$. It would be interesting to better elucidate the relation between the (simple) analytic structure of the boundary four-point function in $\nu$ space, and the resulting (complicated) analytic structure in $p^2$ that emerges in the flat-space limit. The example presented here can provide a useful playground for further studies of this problem.

\subsubsection{Goldstone Bosons and Conformal Manifold} One of the most basic aspects of the physics of the symmetry-breaking phase is the presence of massless Goldstone bosons, namely the $N-1$ pions $\pi^i$ with mass-squared $M^2=0$. In AdS background, the boundary values of these fields define marginal operators in the boundary conformal theory. Therefore, at least for $N=\infty$ and any $\lambda > 0$, the set of boundary conformal theories defined by the $O(N)$ model at any of the symmetry-breaking vacua together form a conformal manifold, parametrized by exactly marginal couplings. Since the value of the exactly marginal couplings corresponds to the constant expectation value of the bulk fields, this conformal manifold has the same geometry as the space of vacua in the bulk, namely it is an $N-1$-dimensional sphere. By analogy with flat space in $d+1=3$, it seems reasonable to assume that the the symmetry-breaking vacua in AdS will continue to exist also at finite $N$, suggesting that the conformal manifold should remain unlifted by $1/N$ corrections (the situation in AdS$_2$ is somewhat special, as we discussed in section \ref{eq:AdS2}). 

More generally, the low-energy effective theory at a vacuum with symmetry breaking of a continuous symmetry is governed by the structure of a quotient between the broken and the unbroken group, $G/H$ \cite{Coleman:1969sm,Callan:1969sn}. When this occurs in AdS, we expect that the boundary conformal theory has a conformal manifold which coincides with the coset $G/H$, and the global symmetry group at each point in the conformal manifold is given by a residual symmetry group $H$. This quotient structure is reminiscent of the construction of conformal manifolds in superconformal field theories \cite{Green:2010da}. 

Perhaps it might look surprising that such a simple construction would give rise to a conformal manifold in a non-supersymmetric theory, that is famously hard to find examples of (see \cite{Bashmakov:2017rko, Behan:2017mwi, Hollands:2017chb, Sen:2017gfr} for recent discussions of the implications of conformal manifolds in general conformal field theories). However the key point is that here we are considering a  {\it conformal theory} rather than a {\it conformal field theory}, by which it is meant that the boundary theory is not local, i.e. it does not contain a stress-tensor. As recently pointed out in the context of the long-range Ising model \cite{Behan:2017emf}, relaxing the constraint of locality makes it easier to find examples of conformal manifolds.\footnote{The observation in \cite{Behan:2017emf} is weaker, namely it concerns continuous families of conformal theories, where the continuous parameter does not necessarily correspond to an exactly marginal operator. For instance, a massive scalar in AdS, or equivalently a generalized-free-field theory, comes within a family of theories parametrized by the bulk mass, or the boundary scaling dimension. However there is no local operator in the conformal theory that couples to this continuous parameter. Similarly, in the context of the $O(N)$ model changing $\lambda$ (in units of the AdS radius) gives rise to a continuous family of boundary theories, which in this case are not simple generalized free fields. On the other hand, the example that we are discussing here is stronger, because we do have exactly marginal operators in the conformal theory. Other examples of conformal manifolds in non-supersymmetric theories were recently found by relaxing the constraint of unitarity \cite{Caetano:2016ydc, Mamroud:2017uyz}.}   An even simpler example of a conformal manifold without locality is just given by a free theory of massless scalar fields in AdS background, which has a shift symmetry. The symmetry-breaking phase in the $O(N)$ model is a more appealing example because it is an interacting theory, and it provides a dynamical mechanism that enforces the presence of massless scalar fields. 

If, on the other hand, the boundary conformal theory is local, the bulk theory must contain dynamical gravity. It is generally believed that in quantum gravity all the symmetries are gauged \cite{Banks:2010zn}. In such situations, our construction of the conformal manifold would not work since the symmetry breaking in the bulk is accompanied by the Higgs phenomenon, and induces a mixing between the massless Goldstone bosons and the gauge fields that makes all the fields massive. On the boundary side, this phenomenon can be interpreted as the mixing between the current multiplet $J_{\mu}$, which is dual to the gauge field, and the marginal operators $O$. Together they recombine into a single long multiplet satisfying $\partial^{\mu}J_{\mu} \sim O$. This observation suggests that two seemingly different facts, the difficulty of finding a conformal manifold in non-supersymmetric conformal field theories and the absence of global symmetries in quantum gravity, might be related to each other.

The relation between symmetry breaking in the bulk and conformal manifolds also have interesting implications for the exact-marginality conditions: Suppose there exist at least two marginal operators in the theory, which we denote with $O$ and $O^{\prime}$. When one perturbs the theory by a marginal coupling $\int g \, O$, one can show using first-order conformal perturbation theory that the other marginal operator $O^{\prime}$  remains marginal only when the OPE coefficient $C_{OOO^{\prime}}$ vanishes. In our context, this constraint is trivially satisfied since in the low-energy effective action there is no cubic term in the pions. (More generally, any term with an odd number of pions is not allowed, with the exception of Wess-Zumino terms when the bulk is even dimensional \cite{Witten:1983tw}). 

Beyond leading order, one also finds constraints on higher-point functions of the marginal operators whose exploration have started only recently. In particular the next-to-leading constraints were studied in the previously mentioned references \cite{Bashmakov:2017rko, Behan:2017mwi, Hollands:2017chb}. It would be interesting to verify beyond leading order that the dynamics of the low-energy effective action in the bulk automatically leads to boundary correlators that satisfy the marginality constraints.\footnote{A similar problem was studied in \cite{Bashmakov:2017rko}. In that paper bulk theories of massless scalar fields were considered, and it was observed that the constraint of marginality leads to a theory with derivative interactions.} Another interesting direction for the future would be to reverse the logic, and try to derive the higher order marginality constraints on correlation functions using the effective Lagrangian of the Goldstone bosons in the bulk. 


\section{Critical Point}\label{sec:Critical}

In flat-space $\mathbb{R}^3$ the $O(N)$ model undergoes a second-order phase transition at a certain (scheme-dependent) value of the mass-squared parameter $m^2$, which separates the symmetry-breaking and the symmetry-preserving phases. The IR physics for that tuned value of $m^2$ is described by an interacting CFT with $O(N)$ global symmetry. This fixed point can be ---at least formally--- dimensionally continued to $\mathbb{R}^{d+1}$ with $2 < d+1 < 4$, and can be studied perturbatively in $4-\epsilon$ \cite{Wilson:1973jj} and $2+\epsilon$ \cite{Brezin:1975sq} expansion. A continuation to $4 < d+1 < 6$ has been proposed in \cite{Fei:2014yja}.

As we discussed in section \ref{sec:Phases}, on AdS background for intermediate values of the mass-squared parameter we have both symmetry-breaking vacua and a symmetry-preserving one, hence the physics is more similar to that of a first-order phase transition. Nevertheless, we will provide evidence that the in the symmetry-preserving vacuum, for a particular value of $m^2$, the theory enjoys conformal symmetry in the bulk. We will refer to this case as the ``critical point". In order to argue for its existence, we will first discuss more generally how to diagnose the presence of conformal symmetry on AdS background. 

As a preliminary observation, note that, contrarily to flat-space, we cannot define the critical point in terms of an enhancement of the spacetime symmetry. In fact, a generic quantum field theory on (Euclidean) AdS$_{d+1}$ background enjoys a symmetry under the isometry group $SO(d+1,1)$. If it is a conformal field theory, via a Weyl transformation it can be mapped to $\mathbb{R}_+ \times \mathbb{R}^d$
\begin{equation}
\frac{dz^2 + (d\vec{x})^2}{z^2} \to dz^2 + (d\vec{x})^2~,\label{eq:Weyl}
\end{equation} 
where $z > 0$ and $\vec{x} \in \mathbb{R}^d$. Hence it is equivalent to a boundary conformal field theory (BCFT) with $d+1$ dimensional bulk, which also has spacetime symmetry $SO(d+1,1)$. 

On the other hand, the conformality of the bulk theory implies that among the boundary operators there must exist a scalar operator of protected dimension $d+1$, the so-called displacement operator. Moreover, even though on $\mathbb{R}_+ \times \mathbb{R}^d$ the larger bulk conformal symmetry $SO(d+2,1)$ is broken by the presence of the boundary, it still organizes the local bulk operators in conformal multiplets, and it constrains the bulk OPE. Upon Weyl rescaling, these properties are also imported to the AdS background. These considerations suggests that to detect bulk conformality, besides checking the existence of a scalar boundary operator of dimension $d+1$, we need to look at the properties of correlators when the insertion points are far away from the boundary, or equivalently very close to each other in the bulk. We will make this idea concrete in the following subsection, and then we will apply it to find the AdS critical point of the $O(N)$ model. For the reason explained above, this is equivalent to finding conformal boundary conditions for the $O(N)$ model in flat space, and in fact it will allow us to extract data of the associated BCFT.

\subsection{Conformal Symmetry in AdS: Bulk Two-point Functions}  

Consider the two-point function of a scalar operator $O$ on AdS$_{d+1}$. Due to AdS isometries it can only depend on the insertion points through their distance, e.g. the chordal distance, whose square we denote with $\zeta$
\begin{equation}
\langle O(x_1) O(x_2) \rangle  = F_{OO}(\zeta)~,~~\zeta  \equiv  \frac{(\vec{x}_1 - \vec{x}_2)^2 + (z_1-z_2)^2}{z_1 z_2}~.
\end{equation}
Alternatively, we can use the spectral representation (see appendix \ref{app:SpRep}) to view the two-point function as a function of $\nu$ 
\begin{equation}
\langle O(x_1) O(x_2) \rangle = \int_{-\infty}^{+\infty}d\nu \, \tilde{F}_{OO}(\nu) \Omega_\nu (x_1, x_2)~.\label{eq:nuspace}
\end{equation}
If the theory has bulk conformal symmetry, we can perform the Weyl rescaling \eqref{eq:Weyl} to $\mathbb{R}_+ \times \mathbb{R}^d$. Using the transformation law for primary operators operators under Weyl rescalings, and assuming the correlator is not affected by a Weyl anomaly, we obtain that
\begin{equation}
\langle O(x_1) O(x_2) \rangle_{\mathbb{R}_+ \times \mathbb{R}^d} = \frac{1}{(z_1 z_2)^\Delta}F_{OO}(\zeta)~,\label{eq:Wrescal}
\end{equation}
where $\Delta$ is the scaling dimension of $O$. The standard parametrization of bulk two-point function in BCFT is
\begin{equation}
\langle O(x_1) O(x_2) \rangle_{\mathbb{R}_+ \times \mathbb{R}^d} = \frac{1}{(4z_1z_2)^\Delta} f_{OO}(\xi)~,\label{eq:btwop}
\end{equation} 
where $\xi$ is the $SO(d+1,1)$-invariant cross-ratio
\begin{equation}
\xi = \frac{(\vec{x}_1 - \vec{x}_2)^2 + (z_1-z_2)^2}{4 z_1 z_2} = \frac{\zeta}{4}~.
\end{equation}
Comparing \eqref{eq:Wrescal} and \eqref{eq:btwop} we find the relation between the two parametrizations of the two-point function
\begin{equation}
f_{OO}(\xi) = 4^\Delta F_{OO}(4 \xi)~.\label{eq:WAdS}
\end{equation}

In BCFT, the function $f_{OO}$ admits expansions in two distinct OPE channels, a boundary channel and a bulk channel. The boundary-channel expansion is around $\xi \to \infty$, which means that the insertion points approach the boundary, and it is obtained by replacing both bulk operators with their bulk-to-boundary OPE and summing the resulting boundary two-point functions. When the bulk operators are scalars only boundary scalar operators contribute. The contributions of the conformal family of a certain boundary primary $\hat{O}$ of dimension $\hat{\Delta}$ can be resummed in the boundary block \cite{McAvity:1995zd}
\begin{equation}
f_{OO}^{\rm bdy}(\hat{\Delta},\xi) = \xi^{-\hat{\Delta}} {}_2F_1\left(\hat{\Delta},\hat{\Delta} - \tfrac{d}{2}+\tfrac 12,2\hat{\Delta}-d+1,-\tfrac{1}{\xi} \right)~.\label{eq:bdycb}
\end{equation}
On the other hand, the bulk-channel expansion is around $\xi \to 0$, which means that the insertion points are approaching each other in the bulk, or equivalently they are far away from the boundary. In this case the expansion is obtained by replacing the two operators with their bulk OPE and summing the resulting bulk one-point functions. Only scalar operators can have a non-zero one-point function. Summing over the conformal family of  a given bulk primary $O'$ of scaling dimension $\Delta'$, one finds the bulk block \cite{McAvity:1995zd}
\begin{equation}
f_{OO}^{\rm bulk}(\Delta',\xi) = \xi^{-\Delta + \frac{\Delta'}{2}} {}_2F_1\left(\tfrac{\Delta'}{2},\tfrac{\Delta'}{2}, \Delta' - \tfrac{d}{2}+\tfrac 12,-\xi \right)~.\label{eq:bulkcb}
\end{equation}
Given the identification \eqref{eq:WAdS}, a necessary condition for bulk conformality is that the function $F_{OO}$ should similarly admit a sensible expansion in both channels, i.e. both in boundary blocks and in bulk blocks. 

However, as it turns out, the expansion in boundary blocks is not useful to diagnose bulk conformality, because an analogous expansion exists for a general quantum field theory on AdS background, even if massive.
One way to derive this boundary-channel expansion is to use that $\tilde{F}_{OO}(\nu)$ has a sequence of poles at $\nu_n = \pm i(\frac d2-\hat{\Delta}_n )$ on the imaginary $\nu$ axis, labeled by a discrete parameter $n$. Therefore, we can close the contour in eq. \eqref{eq:nuspace} and use Cauchy's theorem to rewrite the two-point correlator as a sum over $n$ of $\Omega_{\nu_n} (x_1, x_2)$. The functions $\Omega_{\nu_n} (x_1, x_2)$ coincide with the boundary conformal blocks \eqref{eq:bdycb} for $\hat{\Delta}=\hat{\Delta}_n$. The equivalence between the two set of functions is due to the fact that they are solution to the same eigenvalue problem: as shown in \cite{Liendo:2012hy}, the boundary block is an eigenfunction of the quadratic Casimir of the boundary conformal group $SO(d+1,1)$. Regarding $SO(d+1,1)$ as the isometries of AdS$_{d+1}$, this Casimir operator is naturally mapped to the AdS Laplacian, whose eigenfunctions are the functions $\Omega_{\nu} (x_1, x_2)$.\footnote{As an example of the boundary-channel expansion, for a free scalar $\phi$ in AdS with $m^2 = \Delta_\phi(\Delta_\phi -d)$, either conformally-coupled or not, the two-point function is the bulk-to-bulk propagator \eqref{eq:bulktobulk}, which can be seen as a single boundary block of a boundary scalar primary operator $\hat{\phi}$ of dimension $\Delta_\phi$ (which is either $\leq$ or $> \tfrac d 2$ depending on the boundary condition). In fact, in the free theory there is only one operator in the bulk-to-boundary OPE of $\phi$, the operator that in the language of holography is ``dual" to the bulk scalar field. 

Two-point functions of composites of $\phi$ also admit a boundary-channel decomposition for any value of $m^2$. In this case there is an infinite sum over boundary blocks, because the bulk-to-boundary OPE contains infinitely many multi-trace operators built out of the generalized-free-field $\hat{\phi}$. This is a way to understand the evaluation of the bubble diagram ---that can be thought of as the two point function of $\phi^2$ in a free scalar theory--- as an infinite sum of double-trace contributions, in agreement with \cite{Fitzpatrick:2010zm, Fitzpatrick:2011hu} and with our calculation of the diagram in section \ref{sec:Corr}. 

The two-point function of the field $\delta \sigma$ derived in section \ref{sec:Corr} is a further example, this time an interacting one. Also in this case we see that for any value of the parameters, hence irrespectively of bulk conformality, the spectral representation has a family of isolated poles on the imaginary axis of $\nu$, and therefore the two-point function can be written as a sum of boundary blocks, with the dimension of the boundary operators determined by the position of the poles as we described. 

One can wonder whether it is possible that in some cases such poles would accumulate, or that the spectral representation would have branch-cuts on the imaginary axis. Given the interpretation of a singularity in terms of the exchange of the conformal family of a boundary primary operator, we can rule out these possibilities if we assume that the boundary conformal theory has a discrete spectrum of primary operators, without accumulation points. It might be possible to prove this assumption, but we leave this for future work.}

On the other hand, the existence of the bulk-channel expansion does require bulk conformality. Indeed, the bulk blocks are eigenvalues of the Casimir of $SO(d+2,1)$, the bulk conformal symmetry \cite{Liendo:2012hy}. Equivalently, the form of the blocks is fixed by using the conformal OPE in the bulk, and the latter is not valid when the bulk theory is massive. This leads to the idea that in order to test bulk conformality we should check whether two-point functions admit an expansion in the bulk blocks of eq. \eqref{eq:bulkcb}. 

A practical way to check this, in the case of two identical operators, is to focus on the leading contributions in the limit $\xi \to 0$. If the bulk is conformal, a scalar primary $O'$ of dimension $\Delta'$ appearing in the $O\times O$ OPE contributes a power-law $\xi^{-\Delta + \frac{\Delta'}{2}}$. Its scalar descendants $\square^k O'$ contribute powers shifted by an integer $\xi^{-\Delta + \frac{\Delta'}{2} + k}$, with the sum over $k\in \mathbb{N}$ encoded in the bulk block. If the operators are identical, the leading power comes from the identity operator with $\Delta'=0$, and in this case the full block is just a single power-law $\xi^{-\Delta}$. In particular, since the identity does not have descendants, there should not be integer-shifted powers $\xi^{-\Delta + k}$. We see in examples (but did not attempt to prove in general) that in fact the AdS two-point functions admit an expansion in powers of $\zeta$ as $\zeta \to 0$, both in conformal and massive theories. However, in massive theories, denoting the leading power with the same symbol $\zeta^{-\Delta}$ (even though $\Delta$ loses its interpretation of bulk scaling dimension), it comes accompanied by a tower of ``pseudo-descendant" terms $\zeta^{-\Delta+k}$, $k\in \mathbb{N}$. The coefficient of these terms are functions of the parameters of the theory, and the bulk-conformality constraint on the parameters therefore comes from setting these coefficients to zero. This condition is not sufficient to ensure bulk-conformality,  but since the number of parameters/couplings is finite, the fact that a solution exists for all $k$'s is a strong evidence that the theory indeed is conformal for those values of the parameters. We hasten to clarify that strictly speaking the condition is also not necessary, because there could be a primary scalar operator with integer scaling dimension $m$ in the $O\times O$ OPE, whose conformal family produces the integer-shifted powers with $k\geq m$. In generic theories we do not expect such an operator to exist, but one should keep this possible exception in mind.

We can also derive a similar ---but less rigorous--- condition on the two-point function in $\nu$ space. This is important for our purposes, because we computed the two-point function of the field $\delta \sigma$ in $\nu$ space. To this end, note that the derivation of the flat-space limit in section \ref{sec:FslSpRep} in the appendix can be also understood as the statement that the bulk OPE limit $\zeta \to 0$ maps to $\mathrm{Re}(\nu)\to \infty$. More concretely, we can look at the spectral representation of a power law in AdS$_{d+1}$, which is derived in section \ref{app:power} in the appendix and reproduced here
\begin{equation}
F_\Delta(\zeta) = \left(\frac{\zeta}{4}\right)^{-\Delta} \longleftrightarrow \tilde{F}_\Delta(\nu) = c_{\Delta}\frac{\Gamma(-\tfrac d2+\Delta\pm i \nu )}{\Gamma(\tfrac 12 \pm i \nu)}~, c_{\Delta} \equiv (4\pi)^{\tfrac{d+1}{2}}\frac{\Gamma(\tfrac{d+1}{2}-\Delta)}{\Gamma(\Delta)}~.\label{eq:nupower}
\end{equation}
As explained in more detail in the appendix, the definition of the transform in this case requires an analytic continuation in the parameter $\Delta$. The transform of the power has the following asymptotics at large and real $\nu$
\begin{equation}
\tilde{F}_\Delta(\nu) \underset{\nu\to \infty}{\sim} c_{\Delta} \,\nu^{2\Delta -d-1}\left(1 +\sum_{n>1} \frac{c_n}{\nu^{2n}}  \right)~,\label{eq:nupowerexpand}
\end{equation}
where $c_n$ are $\Delta$ and $d$ dependent coefficients. From this result we confirm that an expansion in powers of $\zeta$ in the OPE limit $\zeta \to 0$ maps to an expansion at large $\nu$, i.e. larger positive powers of $\zeta$ map to larger negative powers of $\nu$. Therefore, the condition that in position space the leading contribution must come from the block of the identity is mapped to the following condition in $\nu$ space: the relative coefficients of the leading powers at large $\nu$ should agree with the asymptotic expansion of the spectral transform of a power law $\tilde{F}_\Delta(\nu)$, for a certain value of $\Delta$. What makes this condition less rigorous than the position-space version is that it requires us to commute two operations on the two-point function: $(i)$ taking the spectral transform, and going to the large-$\nu$ limit, and $(ii)$ the bulk OPE expansion. It is not clear that commuting these operations is always allowed.\footnote{In flat space without a boundary, it was observed in \cite{Dymarsky:2014zja} that naively commuting the Fourier transform and the OPE sometimes leads to wrong results.} One might be especially wary of commuting with the OPE expansion if the spectral representation is strictly-speaking not well-defined for the terms that are retained, as is the case for the power-law contribution of the identity, whose spectral representation can only be defined via an analytic continuation.\footnote{We thank the anonymous referee for their comments on this point.} One step towards a more rigorous version of this argument would require to compute the spectral transform of a generic bulk block,\footnote{Note that the crossing kernel in $\alpha$-space for a $d$-dimensional BCFT was computed in \cite{Hogervorst:2017kbj}. This result gives the the spectral transform of a certain combination of a bulk block and its shadow, and this seems like a promising starting point for the calculation of the transform of the block. Moreover, in presence of a small expansion parameter, like $1/N$ in our setup,  the scaling dimensions will be expanded in the small parameter. Therefore to apply this method one also needs to understand the behavior in $\nu$ space of the derivatives of the blocks w.r.t. the dimensions. The large $\nu$ behavior of the OPE data for the four-point function in conformal field theories was analyzed rigorously in \cite{Mukhametzhanov:2018zja} by using complex tauberian theorem.} and study its large-$\nu$ behavior. We leave this as an interesting direction for the future. In the following we will proceed as if the leading contribution of operators other than the identity could indeed be estimated by commuting the OPE power series with the large-$\nu$ limit of the spectral transform, and we will see that this allows us to make progress in the specific examples we will consider.

\paragraph{Check: Conformally-coupled Free Scalar} Consider a free scalar $\phi$ in AdS$_{d+1}$ background. The only free parameter is the mass-squared coupling $m^2 = \Delta_\phi(\Delta_\phi-d)$, and we can ask for what value of $m^2$, or equivalently $\Delta_\phi$, the theory has conformal symmetry in the bulk. In this case we already know the answer, i.e. the conformally-coupled scalar has $m^2 = - \frac{d^2-1}{4}$,\footnote{Recall that the conformal coupling on a $d+1$ dimensional background is $\frac{1}{4}\frac{d-1}{d} R\, \phi^2$, where $R$ is the Ricci scalar. On AdS$_{d+1}$ with radius $1$, we have $R = - d(d+1)$.} and we can check that the criterion proposed above reproduces this answer.

For generic $m^2$, and either choice of boundary condition, the leading $\zeta \to 0$ behavior of the two-point function \eqref{eq:bulktobulk} is 
\begin{equation}
\langle\phi(x_1)\phi(x_2) \rangle \underset{\zeta\to 0}{\sim} \frac{\Gamma(\tfrac{d-1}{2})}{4\pi^\frac{d+1}{2}} \,\zeta^{-\frac{d-1}{2}}~.\label{eq:leadfreesc}
\end{equation}

In this case, the full two-point function in position space is a power times an hypergeometric function of argument $- \frac{4}{\zeta}$, which we can rewrite as a sum of hypergeometric functions with inverted argument, so it is easily checked that for generic $\Delta_\phi$ the ``pseudo-descendant" powers $\zeta^{-\frac{d-1}{2}+k}$ resum to give 
\begin{equation}
\frac{\Gamma(\tfrac{d-1}{2})}{4\pi^\frac{d+1}{2}} \,\zeta^{-\frac{d-1}{2}}\,{}_2F_1\left(\Delta_\phi-\tfrac{d-1}{2},\tfrac{d+1}{2}-\Delta_\phi,\tfrac{3-d}{2},-\tfrac{\zeta}{4}\right)~.
\end{equation}
We see that we can set to zero the coefficient of all the $k>0$ powers at once, by setting $\Delta_\phi = \frac{d\pm1}{2}$. These two values indeed correspond to a conformally-coupled scalar, with the sign fixed by the boundary condition: The case $\Delta_\phi = \frac{d-1}{2}$ corresponds to Neumann boundary conditions, with the boundary operator being just the restriction of $\phi$ to the boundary, while the case $\Delta_\phi =\frac{d+1}{2}$ corresponds to Dirichlet boundary condition, with the boundary operator being the the restriction of the normal derivative of $\phi$ to the boundary.

More interestingly, the same answer can also be obtained just from the knowledge of the spectral representation of the two-point function, i.e.
\begin{equation}
\tilde{F}_{\phi\phi}(\nu) = \frac{1}{\nu^2 + (\Delta_\phi -\tfrac d2)^2}~.\label{eq:freesp}
\end{equation}
In the limit of large $\nu$ large this goes like $\nu^{-2}$, which by eq. \eqref{eq:nupowerexpand} indeed corresponds to the power-law \eqref{eq:leadfreesc} in position space. Moreover, by the argument explained above, not only the leading term, but also all the subleading powers $\nu^{-(2+2 n)}$, $n\in \mathbb{N}$ should agree with the transform of power law \eqref{eq:leadfreesc}. Using eq. \eqref{eq:nupower}, we find that the transform of the power-law \eqref{eq:leadfreesc} is 
\begin{equation}
\tilde{F}_{\Delta = \frac{d-1}{2}}(\nu)  = \frac{(4\pi)^{\frac{d+1}{2}}}{\Gamma(\frac{d-1}{2})} \frac{1}{\nu^2 + \frac 14}~.\label{eq:freespleadpow}
\end{equation}
Comparing \eqref{eq:freesp} and \eqref{eq:freespleadpow}, we see that in order to match also all the subleading powers $\nu^{-(2+2 n)}$ at once, we need to set $\Delta_\phi$ to the conformal value $\Delta_\phi = \frac{d\pm1}{2}$. 

It is worth noticing that in this simple example the spectral representation and the bulk OPE expansion commute. To see this, note that the full spectral representation of the two-point function is proportional to the spectral representation of the identity bulk block, i.e. the leading power-law. This implies that the contribution of the bulk block of the operator $\phi^2$, even though it is non-vanishing in position space, gives no contribution in $\nu$ space. Indeed we can confirm this by noting that the bulk block of $\phi^2$ is an infinite series of positive integer powers of $\zeta$, whose spectral representation vanishes according to \eqref{eq:nupower}. Note that more generally this means that in mean-field theory, if the spectral representation and the bulk OPE commute, the bulk blocks of double-trace operators will drop from the spectral representation of the two-point function.

\subsection{Critical Point of the $O(N)$ Model on AdS$_3$}\label{subsec:critON}

We will now apply the method to detect bulk conformality described above to the two-point function of the field $\delta \sigma$ in the $O(N)$ model, computed in sec. \ref{sec:Corr}. Differently from the rest of the paper, in this subsection we will denote the scaling dimension of the $O(N)$-vector boundary operator with $\hat{\Delta}$, in compliance with the common usage in the BCFT literature, and reserve the symbol $\Delta$ without a hat for bulk scaling dimensions. We reproduce here the result of sec.  \ref{sec:Corr} for the spectral representation of the two-point function of $\delta\sigma$
\begin{align}
\tilde{F}_{\delta\sigma\delta\sigma}(\nu) & = -\frac{1}{\lambda ^{-1} + 2\tilde{B}(\nu)}~, \label{} \\
\tilde{B}(\nu)&  = \tfrac{i}{8\pi\nu}(\psi(\hat{\Delta} -\tfrac 12 - i \tfrac{\nu}{2})-\psi(\hat{\Delta} -\tfrac 12 + i \tfrac{\nu}{2}))~,
\end{align}
where $\hat{\Delta} = \hat{\Delta}_+ = 1 + \sqrt{M^2 + 1}$. We want to determine for what values of the parameters $\lambda$ and $M^2$, if any, the two-point function is compatible with bulk conformal symmetry.

The leading behavior of the bubble diagram for large real $\nu$ is 
\begin{equation}
\tilde{B}(\nu)\underset{\nu\to \infty}{\sim} \frac{1}{8 \nu} +\mathcal{O}(\nu^{-2})~.
\end{equation}
As a consequence, for $\lambda < \infty$, the two-point function behaves like $\mathrm{const.} + \nu^{-1}$, and by eq. \eqref{eq:nupowerexpand} the only compatible assignment of scaling dimension is $\Delta_\sigma = 1$. This is the scaling dimension of the scalar bilinear in the free UV theory, hence for $\lambda < \infty$ we cannot find any interesting critical point. We can still ask for what values of the parameters do the series of powers $\nu^{-1-2n}$, $n \in \mathbb{N}$ have coefficients which are compatible with bulk conformality. Not surprisingly, we find that the only solution is the trivial one $\lambda = 0$, which sets the two-point function to zero altogether, as we would expect in the free UV theory. There is no constraint on $M^2$, because in this limit $\delta \sigma$ is decoupled from $\delta \phi$.

In order to find a non-trivial scaling we need to set $\lambda = \infty$. In this case the two-point function behaves like $\nu$ at large $\nu$, and the compatible assignment of scaling dimension is $\Delta_\sigma = 2$. This is indeed the scaling dimension of $\sigma$ in the critical $O(N)$ model. However the parameter $M^2$, or equivalently $\hat{\Delta}$, is still free, and for generic values we do not have bulk conformal symmetry. 

To determine the conformal value of $\hat{\Delta}$, we look at the large real $\nu$ expansion of the two-point function for $\lambda = \infty$
\begin{equation}
\tilde{F}_{\delta\sigma\delta\sigma}(\nu) \underset{\lambda = \infty}{=} \!-\frac{1}{2\tilde{B}(\nu)} \!\underset{\nu \to \infty}{\sim} \!\!  - 4\left(\nu + \frac{4(\hat{\Delta}-1)}{\pi} + \frac{16(\hat{\Delta}-1)^2}{\pi^2\nu} + \mathcal{O}(\nu^{-2}) \right)~.\label{eq:explinf}
\end{equation}
Before going ahead and matching this expansion with the expansion of the block of the identity, we need to open a brief parenthesis to explain what are the other possible contributions from other scalar operators in the bulk OPE of two $\sigma$'s. 

At leading order at large $N$ $\sigma$ is just a generalized-free-field, and besides the identity the OPE contains a tower of double-trace operators, the scalars ones being schematically $\sigma \square^n \sigma$, of dimension $4 + 2n$. Moreover, recall that $\sigma$ has a non-zero 1-point function $\sqrt{N}\Sigma$ (in the normalization in which the connected two-point function starts at $\mathcal{O}(1)$), leading to a disconnected contribution to the two-point function of order $N$, larger than the identity contribution. In the OPE expansion of the two-point function, this disconnected constant piece must be reproduced by the leading-order contribution of the scalar double-trace operators, which also get a 1-point function of order $N$. Considering the two-point function in $\nu$ space, under the assumption that we can commute it with the bulk OPE, the disconnected constant term and the leading contribution of the scalar double-traces trivially match, because they both vanish. This is due to the fact ---that was already observed in the example of the conformally-coupled free scalar--- that bulk blocks of double-trace operators only contain positive integer powers of the cross-ratio, whose spectral representation vanishes according to \eqref{eq:nupower}, and similarly for the constant disconnected term.

The connected part of the two-point function, i.e. the two-point function of $\delta \sigma$, starts at order $\mathcal{O}(1)$ at large $N$, and it potentially receives contributions in the bulk OPE from: $(i)$ the identity operator, $(ii)$ the field $\sigma$ itself, due to a compensation between the $\mathcal{O}(1/\sqrt{N})$ OPE coefficient $C_{\sigma\sigma\sigma}$ and the $\mathcal{O}(\sqrt{N})$ 1-point function, and $(iii)$ the next-to-leading double-trace contribution, namely the correction to their scaling dimension and to their (OPE coefficient $\times$ one-point function). Let us analyze what this would imply for the spectral representation of the two-point function, if we allow ourselves to naively commute the bulk OPE with the spectral transform. In $\nu$ space, the contribution $(iii)$ can only come from the anomalous dimensions of the double-trace operators, because the contribution from the correction to the (OPE coefficient $\times$ one-point function) has the same functional form in position space as the leading-order contribution, which we already argued to vanish in the spectral representation. Moreover, when $d+1 = 3$ there is a dramatic simplification in this OPE, because both the OPE coefficient $C_{\sigma\sigma\sigma}$ \cite{Petkou:1994ad} and the full set of $1/N$ anomalous dimensions of the double-trace operators \cite{Lang:1993ct} happen to vanish precisely at this value of the spacetime dimension. Hence, specifically for $d+1=3$, we are led to the conclusion that only the bulk block of the identity contributes to the two-point function in $\nu$ space at this order.  

Using eq. \eqref{eq:nupower} we find that  the spectral representation of the block of the identity in this case is
\begin{equation}
\tilde{F}_{\Delta = 2}(\nu)  =  -16 \pi^2 \nu \coth(\pi \nu)\underset{\nu \to \infty}{\sim} -16 \pi^2 \nu +\mathcal{O}(e^{-2\pi\nu})~.\label{eq:idblocksigma}
\end{equation}
Comparing with \eqref{eq:explinf}, we see that in order to make the first two subleading powers agree with the block of the identity, we necessarily have to set $\hat{\Delta} = 1$. Plugging $\hat{\Delta} = 1$ back in the two-point function, it simplifies to
\begin{equation}
\tilde{F}_{\delta\sigma\delta\sigma}(\nu) \underset{\lambda = \infty, \hat{\Delta} =1}{=} - 4 \,\nu \coth(\tfrac{\pi\nu}{2})\underset{\nu \to \infty}{\sim} -4\,\nu +\mathcal{O}(e^{-\pi\nu})~,\label{eq:conftwopsigma}
\end{equation}
and we see that actually all the subleading powers match, not just the first few shown in equation \eqref{eq:explinf}. Moreover, the spectrum of boundary operators that contribute to this two-point function was worked out in eq. \eqref{eq:sigmaconfspec}, that we reproduce here for convenience
\begin{align}
\hat{\Delta}_n =2n+3 \qquad n\in \mathbb{N}_{\geq 0}\,,\label{eq:sigmaconfspec2}
\end{align}
and we see that in particular there is a scalar operator of dimension 3, that can be the displacement operator.
These observations strongly hint that indeed for $\lambda= \infty$, $\hat{\Delta} = 1$ we have bulk conformal symmetry.  On the other hand, we note that there are exponentially small contributions at large $\nu$ that do not agree with the transform of the block of the identity. At finite $\nu$, the two-point function for $\hat{\Delta} =1$ simply does not agree with the spectral representation of the block of the identity. This could indicate that actually for no values of $(\lambda, \hat{\Delta})$ the theory is conformal. More likely, this mismatch is due to the subtlety that we mentioned in the commutation between the bulk OPE and the spectral transform. Even though we fall short of rigorously proving that for $\lambda =\infty$ and $\hat{\Delta} = 1$ the theory has conformal symmetry in the bulk, in what follows we will provide additional evidence that this is the case, by successfully matching boundary conformal data extracted from the AdS correlators with the known results about the conformal boundary conditions of the $O(N)$ model computed in flat space. 

\subsection{$O(N)$ BCFT Data from the AdS Correlators}\label{sec:ONBCFT}
The conformal boundary conditions for the large-$N$ critical $O(N)$ model were studied in flat space in \cite{Bray:1977tk, Ohno:1984iy, McAvity:1995zd}. These references considered generic $d+1$ between 2 and 4, and studied two distinct conformal boundary conditions, dubbed {\it ordinary transition} and {\it special transition}, that can be understood as the IR fixed point of the RG started respectively with Dirichlet or Neumann boundary conditions for the $O(N)$ vector fields. In our analysis in AdS$_3$ we focused on the $+$ boundary condition, that is the AdS version of the Dirichlet boundary condition, so we expect to match with the ordinary transition. 

The first BCFT datum that we can match is the scaling dimension of the boundary $O(N)$-vector operator: Above we found $\hat{\Delta} = 1$ for $d+1=3$, and indeed the flat-space analysis gives that this scaling dimension is $d-1$ (recall that $d$ here is the boundary dimension) at the ordinary transition.

Secondly, we can look at the dimension of the boundary operators appearing in the bulk-to-boundary OPE of the field $\sigma$. The analysis in flat space at the ordinary transition finds infinitely many such operators $\hat{\sigma}_n$ labeled by a non-negative integer $n$, and their scaling dimensions are
\begin{equation}
\hat{\Delta}_n = d +1+ 2n~.
\end{equation}
 Note that in particular there always exists a scalar operator with dimension $d+1$, the displacement operator. Again, we find a match with the spectrum determined from the AdS$_3$ correlator, that we showed in eq. \eqref{eq:sigmaconfspec2}. From the residue of the spectral representation of the two-point function \eqref{eq:conftwopsigma} at the poles, we can easily read-off the bulk-to-boundary OPE coefficients to be
\begin{equation}
b_{\sigma\hat{\sigma}_n}^2 = 32(n+1)~,
\end{equation}
at leading order at large $N$. 

 Taking as an input from the flat-space studies that for generic $d$ the conformal point is $\lambda = \infty$ and $\hat{\Delta} = d-1$, we verified more generally that for these values of $\lambda$ and $\hat{\Delta}$ the poles of the AdS$_{d+1}$ correlator are located precisely at $\nu_n =-i(\hat{\Delta}_n-\tfrac{d}{2})$. This is a non-trivial check of our general-$d$ formula \eqref{eq:finalBnu} for the bubble diagram.

With no additional work, we can also compute the connected four-point function of the $O(N)$-vector boundary operators at order $1/N$, simply by plugging $\lambda = \infty$ and $\hat{\Delta} = 1$ in the AdS result of eq.s \eqref{eq:scalarfourpfun}-\eqref{eq:g1234explicit}. Taking the residue of the $\nu$ integrand at the $\hat{\sigma}_n$ poles, we compute the leading-order boundary OPE coefficients between two boundary $O(N)$-vector operators and a $\hat{\sigma}_n$, obtaining
\begin{equation}
C_{\hat{\phi}\hat{\phi}\hat{\sigma}_n}^2 = \frac{1}{N}\frac{\pi \Gamma(\frac32+2n)^2}{2^{4n+1}(n!)^2}~.
\end{equation}

These are just some examples of BCFT observables that we could easily extract from the AdS correlators. It might be useful more generally to approach perturbative calculations in BCFT from the point of view of AdS, especially with the aid of the spectral representation. A concrete motivation to explore more this direction is that the computation of $1/N$ corrections in the $O(N)$ BCFT via Feynman diagrams on $\mathbb{R}^2\times\mathbb{R}_+$ is often a daunting task, and the ``bootstrap" approach of subsection \ref{sec:Corr41} could potentially turn out to be more effective. An alternative way would be to analyze them by using the standard conformal bootstrap; namely by studying the crossing equation for the BCFT \cite{Liendo:2012hy}. See \cite{Bissi:2018mcq} for a recent work in that direction. 


\section{Correlators  of the Gross-Neveu Model in AdS}\label{sec:fermion}

In this section we study correlation functions of the Gross-Neveu (GN) model in AdS$_{d+1}$. The GN model describes $N$ spin $1/2$ fermions, with a quartic interaction preserving $U(N)$ symmetry. It was introduced in \cite{Gross:1974jv} as a solvable two-dimensional model displaying asymptotic freedom and chiral symmetry breaking, and later generalized to higher dimension (see \cite{ZinnJustin:1991yn, Rosenstein:1990nm} and references therein). In $d +1 > 2$ the model is non-renormalizable, so it should be regarded as a low-energy effective theory. It is still sensible to study the scattering of low energy excitations. Moreover, at large $N$ it is possible to show the existence of a UV fixed point, whose observables can be computed in $1/N$ perturbation theory.
\subsection{Computation of the Correlators\label{subsec:GNcompcorr}}
The Lagrangian is
\bea
\mm{L} = \bar{\Psi}^i  \, \gamma\cdot \nabla \Psi^i + m \bar{\Psi}^i \Psi^i +\frac{g}{2N} (\bar{\Psi}^i \Psi^i)^2~,
\eea 
where $\Psi^i$ are Dirac fermions, and $i= 1, \ldots , N$, and $\gamma$ are the gamma matrices in $d+1$ dimensions. In Poincar\'e coordinates $(z, \vec{x})$, we have gamma matrices with flat indices $\gamma^a = (\gamma^0,\vec{\gamma})$ and to contract with bulk vectors we use the vielbein $e^a_\mu = \frac{1}{z}\delta^a_\mu$. Introducing a HS field $\s$, we can rewrite the Lagrangian as
\bea
\mm{L} = \bar{\Psi}^i  \, \gamma\cdot \nabla \Psi^i  + m \bar{\Psi}^i \Psi^i -\frac{1}{2 g}\s^2 + \frac{1}{\sqrt{N}} (\bar{\Psi}^i \Psi^i)\s~.
\eea
The equation of motion of $\s$ gives
\bea
\s=  \frac{g}{\sqrt{N}} \bar{\Psi}^i \Psi^i~.
\eea
The effective Lagrangian that defines the $1/N$ perturbation theory is\begin{equation}\label{eq:Leffferm}
\mm{L}_{\rm eff} = \bar{\Psi}^i  \, \gamma\cdot \nabla \Psi^i  + m \bar{\Psi}^i \Psi^i -\frac{1}{2 g}\s^2 + \frac{1}{\sqrt{N}} (\bar{\Psi}^i \Psi^i)\s- N \, \tr \log\left(\gamma\cdot \nabla + m + \frac{1}{\sqrt{N}}\sigma \right)~.
\end{equation}
The resulting effective potential can induce a VEV for the field $\sigma$, causing a shift of the mass of the fermions. We will denote the physical mass as
\begin{equation}
M = m + \Sigma~,~~\Sigma \equiv \frac{\langle \sigma\rangle}{\sqrt{N}}~.
\end{equation}
It is actually always consistent to start with $m=0$, because this choice is protected by symmetry (a discrete chiral symmetry left unbroken by the quartic interaction if $d+1$ is even, or parity --by which we mean a reflection w.r.t. to one of the boundary coordinates-- if $d+1$ is odd). The VEV for $\sigma$ is then a spontaneous breaking of the symmetry that protects $m=0$. We will consider $m=0$ in the following, so that $M = \Sigma$. For future reference we note that the vacuum equation for $\Sigma$, obtained by taking a derivative of the effective potential in \eqref{eq:Leffferm}, can be written as follows
\begin{equation}
\frac{1}{g}\Sigma = -\tr \left[\frac{1}{ \gamma\cdot \nabla + \Sigma}\right] ~.\label{eq:gap}
\end{equation}
This is the familiar gap equation, straightforwardly generalized to AdS background by using the appropriate Dirac operator.

Near the boundary, a solution of the Dirac equation behaves as
\begin{equation}
\Psi^i(z, \vec{x}) \underset{z\to 0}{\longrightarrow}  z^{\Delta_+}(\psi_+^i(\vec{x})+\mathcal{O}(z) ) +  z^{\Delta_-}(\psi_-^i(\vec{x})+\mathcal{O}(z) )~,
\end{equation}
where $\Delta_{\pm} = \frac{d}{2}\pm M$ and the modes $\psi_\pm^i(\vec{x})$ satisfy $\gamma_0 \psi_\pm^i = \pm \psi_\pm^i$. The two possible $U(N)$-symmetric boundary conditions consist in setting to zero $\psi^i_-$ or $\psi^i_+$, and we call them $+$ and $-$ boundary condition, respectively. Choosing the $\pm$ condition, the boundary theory contains $U(N)$-vector fermionic operators of scaling dimension $\Delta_\pm$. These operators always have half of the components of their parent bulk field, and they are Weyl (Dirac) fermions on the boundary, if the bulk is odd (even, respectively) dimensional. When $d+1$ is even, the real parameter $M$ that enters the mass/dimension formula is obtained by absorbing the possible phase in a chiral rotation, and the leftover discrete chiral transformation flips the sign of $M$. When $d+1$ is odd, the two boundary conditions are exchanged by parity, that flips at the same time the sign of the bulk mass and the chirality of the boundary condition.

For the computation of the AdS correlators we will adopt the same strategy as for the scalar case of section~\ref{sec:Corr}. We will first parametrize the two-point function of the auxiliary field in terms of an unknown bubble function, and then from that we will obtain the boundary four-point function of the $U(N)$-vector operators. Finally, the consistency of the OPE of the boundary theory will allows us to compute the bubble function. To our knowledge, the fermionic bubble diagram in AdS has not been computed before.

\paragraph{``Bootstrapping" the Fermionic Bubble} The bulk two-point function of $\d \sigma \equiv \sigma - \langle\sigma\rangle$ coming from the effective Lagrangian \eqref{eq:Leffferm} is
\begin{align}\label{eq:delsigxyfermion}
\langle \delta\sigma(x) \delta\sigma(y)\rangle = -\left[\frac{1}{g}\mathds{1} -  B_F \right]^{-1}(x, y)~,
\end{align}
where $B_F(x,y)$ is a product of 2 bulk-to-bulk propagators
\begin{align}\label{eq:productofbtobfermion}
B_F(x,y) =  \Tr \left[\left(\frac{1}{ \gamma\cdot \nabla + M}\right)^2(x,y)\right]~,
\end{align}
and $\Tr$ denotes the trace over the fermion indices (to be distinguished from the functional trace $\tr$). We denote by $\tilde{B}_F(\nu )$ the spectral representation of the fermionic bubble
 \begin{align}
 B_F(x,y)  =\int_{-\infty}^{\infty} d\nu \,\tilde{B}_F(\nu )\Omega_{\nu}(x,y)~.
 \end{align}
Hence, we can write eq.~\eqref{eq:delsigxyfermion} as
 \begin{align}\label{eq:spectralforsigmafermion}
\langle\delta \sigma (x)\delta\sigma (y)\rangle=-\int_{-\infty}^{\infty} d\nu \,\frac{1}{g^{-1}- \tilde{B}_F (\nu)}\Omega_{\nu}(x,y)\,.
\end{align}
We can then proceed to compute the boundary 4-point function $$\langle  \bar{\Psi}^{i}(P_1,S_1) \Psi^{j}(P_2,S_2)   \bar{\Psi}^{k}(P_3,S_3) \Psi^{l}(P_4,S_4)  \rangle~. $$ We are using the embedding-space formalism, and $S$ denotes the spinor-polarization variable for the boundary operators, that keeps track of the spinor indices, i.e. $\Psi(P,S)\equiv \bar{S}\Psi(P)$. See ref.s \cite{Weinberg:2010fx, Iliesiu:2015qra, Iliesiu:2015akf, Isono:2017grm} for more details about the embedding formalism for spinors in CFT, and  \cite{Nishida:2018opl} for its extension to the bulk fermions. We will use the conventions of \cite{Nishida:2018opl}. The spinor-polarization variable for the bulk fermions will be denoted by $S_b$.
 
The bulk-to-boundary propagators of the fermion fields with boundary scaling dimension $\Delta$, choosing the + boundary condition for definiteness, are \cite{Nishida:2018opl}
\begin{align}
K^F_{\D} (X,\bar S_b,P,S_\pa) & = \sqrt{ \mm{C}^{F}_{\D} }\frac{  \bar S_b \Pi_- S_\pa  }{(-2X \cdot P)^{\D+\frac{1}{2}}}~,
\\
\bar{K}^F_{\D} (X,  S_b,P, \bar S_\pa) & = \sqrt{ \mm{C}^{F}_{\D} } \frac{   \bar S_\pa   \Pi_-   S_b }{(-2X \cdot P)^{\D+\frac{1}{2}}}~,
\\
\text{with}~~~\mm{C}^{F}_{\D}  & = \frac{1}{\pi^{d/2}}  \frac{\G(\D+\frac{1}{2})}{\Gamma(\D+\frac{1-d}{2})}~,
\end{align}
where $\Pi_{\pm}$ are chiral projectors in embedding space. Let us define
\begin{equation}
S_{12|34} =  ( \bar S_{2\pa} \Pi_- S_{1\pa} ) ( \bar S_{4\pa} \Pi_- S_{3\pa})~,~~ S_{14|32} = (\bar{S}_{4\pa} \Pi_- S_{1\pa} )(  \bar S_{2\pa} \Pi_- S_{3\pa})~.
\end{equation}
At leading order at large $N$ the four-point function is just given by the mean-field theory answer
\begin{align}
&\langle  \bar{\Psi}^{i}(P_1,S_1) \Psi^{j}(P_2,S_2)   \bar{\Psi}^{k}(P_3,S_3) \Psi^{l}(P_4,S_4)  \rangle\vert_{\mathcal{O}(1)} \nonumber\\ &~~~~~~~~~~~~~~~~~~~= \frac{\delta^{ij}\delta^{kl}S_{12|34}}{(P_{12})^{\Delta+\frac12}(P_{34})^{\Delta+\frac12}}-\frac{\delta^{il}\delta^{jk}S_{14|32}}{(P_{14})^{\Delta+\frac12}(P_{23})^{\Delta+\frac12}}~.
\end{align}
Note that the only allowed Wick contraction is between $\Psi$ and $\bar{\Psi}$, so differently from the scalar case here we only have the s-channel and u-channel contraction, but no t-channel.

At order $1/N$ the four-point function receives contributions from the exchanges of the field $\sigma$ in the bulk, both in the s-channel and the u-channel. Since this diagram is a scalar exchange, the spinor polarization structure is the same as that of the disconnected part. Therefore, we can write the $1/N$ four-point function as follows
\begin{align}
 &\langle  \bar{\Psi}^{i}(P_1,S_1) \Psi^{j}(P_2,S_2)   \bar{\Psi}^{k}(P_3,S_3) \Psi^{l}(P_4,S_4)  \rangle\vert_{\mathcal{O}(1/N)} \nonumber \\ &~~~~~~~~~~~~~~~~~~~= \frac{\delta^{ij}\delta^{kl}S_{12|34}\, g_{12|34} - \delta^{il}\delta^{jk}S_{14|32}\, g_{14|32}}{N}~,
\end{align}
where
\begin{align}
S_{12|34}\, g_{12|34} & = \int d X_1 dX_2  \langle\delta\sigma (X_1)\delta\sigma(X_2)\rangle 
  (\pa_{S_{1b}} \pa_{\bar{S}_{2b}}) \bar{K}^F_{\Delta}(X_1,  S_{1b},P_1, \bar S_{1\pa}) K^F_{\Delta}(X_1,\bar S_{2b},P_2,S_{2\pa})
\nn\\
& \times(\pa_{S_{3b}} \pa_{\bar{S}_{4b}})  \bar{K}^F_{\Delta}(X_2,  S_{3b},P_3, \bar S_{3\pa})K^F_{\Delta}(X_2,\bar S_{4b},P_4,S_{4\pa})~,\label{eq:1ovNfer}
\end{align}
and similarly for $g_{14|32}$. The derivatives w.r.t. the bulk polarizations implement the contraction of the indices at the interaction vertex. We can evaluate the expression in \eqref{eq:1ovNfer} without much effort by relating it to the calculation that we already did for the scalar. This is possible because, after the bulk spinor indices are appropriately contracted, the product of two fermionic bulk-to-boundary operators becomes proportional to the product of two scalar ones with shifted dimension, namely\footnote{The relation between an AdS scalar exchange with external fermions and a scalar exchange with external scalars, upon shifting the dimensions by $\frac12$, was noted already in \cite{Kawano:1999au}, see eq. (2.18) therein, and more recently in \cite{Faller:2017hyt}. The fact that here we are exchanging a ``composite" scalar, rather than an elementary one, does not affect the argument.}
\begin{align}
 & (\pa_{S_{1b}} \pa_{\bar{S}_{2b}})  \bar{K}^F_{\Delta}(X_1,  S_{1b},P_1, \bar S_{1\pa}) K^F_{\Delta}(X_1,\bar S_{2b},P_2,S_{2\pa}) \nonumber \\ & =(2\Delta+1-d) (\bar S_{1\pa} \Pi_- S_{2\pa}) K_{\Delta+\frac 12}(X_1,P_1) K_{\Delta+\frac 12}(X_1,P_2)~.
\end{align}
The $\Delta$-dependent prefactor comes from the different normalization of the scalar and fermionic propagators. Using this identity and the scalar result in eq. \eqref{eq:g1234explicit} we obtain\footnote{The scalar result is also multiplied by an additional combinatorial factor of $4$.}
\begin{align}
g_{12|34} & = -\frac{1}{(P_{12})^{\Delta+\frac12}(P_{34})^{\Delta+\frac12}} \nonumber\\ & \times\int\frac{d \nu}{2\pi}\frac{1}{g^{-1}- \tilde{B}_F (\nu)}\frac{\Gamma_{\Delta+\frac{1}{2}-\frac{d+2i\nu}{4}}^2\Gamma_{\Delta+\frac{1}{2}-\frac{d-2i\nu}{4}}^2\Gamma_{\frac{d+2i\nu}{4}}^4}{4\pi^{\frac d2} \Gamma^2_{\Delta +\frac 12}\Gamma^2_{\frac 12 -\frac{d}{2} +\Delta}\Gamma_{i \nu}\Gamma_{\frac d2 + i \nu}}\mm{K}_{\frac{d}{2}+i\nu}(z,\bar{z})~,\label{eq:fermg1234}
\end{align}
in terms of the cross-ratios $z,\bar{z}$, where $\mathcal{K}$ as above denotes the scalar conformal block.

Next, we project the four-point function to the $U(N)$ singlet sector in the s-channel (namely $12\to 34$ channel). This is achieved by contracting the correlator with the tensor $\frac{\delta_{ij}\delta_{kl} }{N^2} $. The result reads
\begin{align}
&\frac{1}{N^2}\langle  \bar{\Psi}^{i}(P_1,S_1) \Psi^{j}(P_2,S_2)   \bar{\Psi}^{k}(P_3,S_3) \Psi^{l}(P_4,S_4)  \rangle =
\nn\\
& \frac{ S_{12|34}}{(P_{12})^{\Delta+\frac{1}{2}}(P_{34})^{\Delta+\frac{1}{2}}}+\frac{1}{N} \Big[- \frac{S_{14|32}}{ (P_{14})^{\Delta+\frac{1}{2}}(P_{23})^{\Delta+\frac{1}{2}}}+ S_{12|34}\,g_{12|34}  \Big]   +\mathcal{O}(\frac{1}{N^2})\,.\label{eq:fourpointferm}
\end{align}

Just like we saw in the scalar calculation, the projection mixes different orders in the large-$N$ expansion. At $\mathcal{O}(1/N)$ after the projection we get contribution from the u-channel of mean-field theory, and the s-channel of the $\sigma$-exchange diagram.  Note that, besides the poles coming from the propagator of $\sigma$, the exchange diagram $g_{12|34}$ has spurious poles in the lower-half plane at 
    \begin{equation}
    \frac{d}{2}+ i\nu = 2\Delta+2n+1~,~~n\in\mathbb{N}_{\geq0}
    \end{equation}
    with residue proportional to the structure $S_{12|34}$. By analogy with the scalar case, we expect these poles to cancel with scalar double-trace operators (products of two fermionic operators), of scaling dimension $2\Delta+2n+1$, arising from the u-channel of mean field theory
\begin{align}
\label{eq:GFFOPEfermion}
\frac{S_{14|32}}{(P_{14})^{\Delta+\frac{1}{2}}(P_{23})^{\Delta+\frac{1}{2}}}=   \frac{S_{12|34} }{(P_{12})^{\Delta+\frac{1}{2}}(P_{34})^{\Delta+\frac{1}{2}}}    \sum_{n}c_n^2 \,\mathcal{K}_{2\Delta+2n+1,l=0}(z, \bar z)+\dots~,
\end{align} 
where $\dots$ denote other double-trace contributions that we will not need to consider. In the appendix \ref{app:fermgff} we compute the coefficients $c_n^2$ in $d=2$ and $d=1$.

Requiring the cancelation between the spurious poles in the lower-half plane and the double-trace operators we obtain the relation
 \begin{align}
\frac{1}{g^{-1}-\tilde{B}_F(\nu)}    &  \frac{  \Gamma_{\Delta+\frac{1}{2}-\frac{d+2i\nu}{4}}^2\Gamma_{\Delta+\frac{1}{2}-\frac{d-2i\nu}{4}}^2\Gamma_{\frac{d+2i\nu}{4}}^4}{4 \pi^{\frac{d}{2}}\Gamma_{\Delta+\frac{1}{2}}^2\Gamma_{\frac{1}{2}-\frac{d}{2}+\Delta}^2\Gamma_{i\nu}\Gamma_{\frac{d}{2}+i\nu}}   \nonumber\\& ~~~~~\overset{\frac{d}{2}+i\nu \sim 2\Delta+2n+1}{\sim} ~-\frac{c_n^2}{\frac{d}{2}+i\nu -(2\Delta+2n+1)}\,.
 \end{align}
   To compensate for the double-pole in the numerator of the l.h.s., the function $\tilde{B}_F(\nu)$ must have simple poles with appropriate residues at these points. Up to a possible constant shift, we can then identify the function $ \tilde{B}_F(\nu)$ with the resulting sum over poles. In the case of the scalar bubble diagram in $d+1<4$, we were then able to fix the constant term, by imposing that the diagram should vanish as a power-law in the large-$\nu$ limit, as prescribed by the flat-space limit. For the fermionic bubble in $d+1\geq 2$, on the other hand, we have two important differences, related to each other: first, the same diagram in flat space {\it grows} with momentum at large momentum, hence we cannot use this constraint to fix the constant shift; secondly, the resulting sum over poles is divergent (we explicitly checked this for the cases in which we computed the coefficients $c_n^2$, namely $d=1,2$). The latter is an expected UV divergence, that of course is also present for the flat-space fermionic bubble.  Therefore, we can write
 \begin{align}
 \label{eq:polfermion}
 &(\tilde{B}_F(\nu))_{\rm reg}  =  \frac{1}{\pi^{\frac{d}{2}}\Gamma_{\Delta+\frac{1}{2}}^2\Gamma_{\Delta-\frac{d-1}{2}}^2 }  \\
 &\times \left[\sum_{n=0}^\infty  \frac{ \Gamma_{2\Delta+n+1-\frac{d}{2}}^2\Gamma_{\Delta+n+\frac12}^4}{ (n!)^2\Gamma_{2\Delta+2n+1-\frac{d}{2}}\Gamma_{2\Delta+2n+1}} \frac{1}{c_n^2} \frac{2(\frac{d}{2}-2\Delta-2n-1)}{(\frac{d}{2} - 2\Delta-2n-1))^2+\n^2}\right]_{\rm reg}+ C(\Delta,d)_{\rm reg} ~.\nonumber
 \end{align}
 Here the square bracket denotes that the sum can only be defined with some regulator, and we introduced a constant shift $C(\Delta,d)_{\rm reg}$, that depends on the choice of the regulator and cannot be fixed by the knowledge of the poles. Note that one would find an analogous ambiguity for the scalar bubble in $d+1\geq4$. 
 
 While for the application to the GN model we will actually need to fix the constant shift, if one is only interested in the $\nu$ dependence of the diagram this ambiguity is not relevant, as we will now show. We will consider the specific cases $d=2$ and $=1$ (i.e. AdS$_3$ and AdS$_2$) for which we computed the OPE coefficients $c_n^2$ in appendix \ref{app:fermgff}. 
 
 In $d=2$, plugging eq.~\eqref{eq:ope2} in eq.~\eqref{eq:polfermion} we obtain the following sum, e.g. with a hard cutoff regulator
\begin{align}
 (\tilde{B}_F(\nu))_{d=2,\text{cutoff}}  & =  -\frac{2}{\pi   } \sum_{n=0}^{n_{\text{max}}}   \frac{(n+1)(2\Delta+n-1)}{( 4(n+\Delta)^2+ \n^2 ) } \nonumber\\
& = -\frac{i( 4(\Delta-1)^2+\n^2 )}{8\pi \n  }  \Big( \psi(\Delta+\frac{i\n}{2})- \psi(\Delta-\frac{i\n}{2}) \Big) - \frac{n_{\text{max}}}{2\pi} \label{eq:fermBcut} \\
&~~~~~~~~~~~~~~~~~~~~~~~~~~~~~~~~~~~~~~~ + C(\Delta,d=2)_{\rm cutoff}+\mathcal{O}(1/n_{\text{max}})~.\nonumber
\end{align}
We can also regularize it with a ``naive'' Pauli-Villars regulator, i.e. subtracting the same diagram as if we added to the theory a bosonic Dirac field with large mass $M_{PV} >0$ 
\begin{align}
(\tilde{B}_F(\nu))_{d=2,\text{PV}} & =  -\frac{2}{\pi  }  \sum_{n=0}^\infty \Big[ \frac{(n+1)(2\Delta+n-1)}{( 4(n+\Delta)^2+ \n^2 ) } -  \frac{(n+1)(2(1 + M_{PV})+n-1)}{( 4(n+1 + M_{PV})^2+ \n^2 ) } \Big] \nonumber\\
& = -\frac{i( 4(\Delta-1)^2+\n^2 )}{8\pi \n  }  \Big( \psi(\Delta+\frac{i\n}{2})- \psi(\Delta-\frac{i\n}{2}) \Big) - \frac{M_{PV}}{2\pi} \label{eq:fermBPVregu} \\
&~~~~~~~~~~~~~~~~~~~~~~~~~~~~~~~~~~~~~~~ + C(\Delta,d=2)_{\rm PV}+\mathcal{O}(1/M_{PV})~.\nonumber
\end{align}
A third possible regularization consists in subtracting the same summand at $\n=0$
  \bea
(\tilde{B}_F(\nu))_{d=2,\text{sub}}=  -\frac{2}{\pi  } \sum_{n=0}^\infty \Big[ \frac{(n+1)(2\Delta+n-1)}{( 4(n+\Delta)^2+ \n^2 ) } -  \frac{(n+1)(2\Delta+n-1)}{( 4(n+\Delta)^2 ) } \Big]
\nn\\
= -\frac{i ( 4(\Delta-1)^2+\n^2 )}{8\pi \n }  \Big( \psi(\Delta+\frac{i\n}{2})- \psi(\Delta-\frac{i\n}{2}) \Big)+C(\Delta,d=2)_{\rm sub}~.\label{eq:fermBsub}
\eea
 As we would expect on physical grounds, comparing the different regularizations we see that the result contains a universal $\nu$-dependent part, plus a constant scheme-dependent shift. 
 
Similarly, we can compute the fermionic bubble diagram in $d=1$. In this case we were only able to perform the sum using a subtraction at $\nu=0$ as UV regulator. Plugging eq.~\eqref{eq:ope1} in eq.~\eqref{eq:polfermion}, and subtracting the $\n=0$ term, we obtain a sum that can be written in terms of a hypergeometric function as follows
 \begin{align}\label{eq:finalBnufermion}
 &\left(\tilde{B}_F(\nu)\right)_{\rm sub}= \mathcal{B}_{\text{AdS}_2} (\Delta,\nu)+C(\Delta, d=1)_{\rm sub}~,
 \end{align}
 with
 \begin{align}
& \mathcal{B}_{\text{AdS}_2} (\Delta,\nu)=\frac{i\nu\Gamma_{\frac{1}{4}+\Delta}\Gamma_{\frac{1}{2}+\Delta}^2\Gamma_{\frac{1}{2}+2\Delta}}{8\sqrt{\pi}}\times\\
 &\left(\Gamma_{\Delta+\frac{1}{4}+\frac{i\nu}{2}}\,{}_6\tilde{F}_{5}\left[\begin{array}{c}\{\frac{3}{2},\frac{1}{4}+\Delta,\frac{1}{2}+\Delta,\frac{1}{2}+\Delta,\frac{1}{2}+2\Delta,\frac{1}{4}+\Delta+\frac{i\nu}{2}\}\\\{2\Delta,1+\Delta,1+\Delta,\frac{5}{4}+\Delta,\frac{5}{4}+\Delta+\frac{i\nu}{2}\}\end{array};1\right]-(\nu \leftrightarrow -\nu)\right)~,\nn
 \end{align}
 where ${}_6\tilde{F}_{5}$ is the regularized generalized hypergeometric function.
  
 Let us emphasize again that all these results are subject to a constant shift $C(\Delta,d)$ since we only determined the bubble function by specifying the locations and the residues of the poles. To unambiguously determine the shift, one needs to resort to a different physical input, which we will provide in the next subsection.
 
\subsection{Parity-Preserving Pauli-Villars Regularization}
Using the results in the previous subsection, one can in principle compute the two-point function of $\sigma$ as
\begin{equation}
\tilde{F}_{\delta\sigma\delta\sigma}(\nu) = -\frac{1}{g^{-1} - \tilde{B}_F(\nu)}~,
\end{equation}
where $g^{-1}$ is determined by the gap equation \eqref{eq:gap} as\footnote{Recall that $\Sigma=M=\Delta-\frac d2$.}
\begin{align}\label{eq:gapusedtodetermineg}
Mg^{-1}=-{\rm tr}\left[\frac{1}{\gamma\cdot \nabla + M}\right]\,,
\end{align}
while $\tilde{B}_F(\nu)$ is the bubble function given in \eqref{eq:polfermion}. However, both of these quantities are separately divergent and have to be regulated. Furthermore, the bubble function is subject to a constant shift $C(\Delta,d)$ as we discussed in the previous subsection. Therefore, to proceed, we need to
\begin{enumerate}
\item Choose a regularization scheme which regularizes both the gap equation and $\tilde{B}_F$ at the same time.
\item Use some additional input to determine the constant shift $C(\Delta,d)$.
\end{enumerate}
After carrying out these procedures, the difference of $g^{-1}$ and $\tilde{B}_F$ is expected to become finite and independent of the regularization scheme although each of them is separately scheme-dependent.

\paragraph{Regularization} Let us first discuss the regularization scheme. As shown in the previous subsection, there are multiple ways to regulate the bubble function $\tilde{B}_F$. However, most of the regularizations cannot be applied to the computation of the gap equation, with the exception being the ``naive'' Pauli-Villars regularization performed in \eqref{eq:fermBPVregu}. Applying the Pauli-Villars regularization, the gap equation \eqref{eq:gapusedtodetermineg} changes to
\begin{align}\label{eq:regulateddirac}
M(g^{-1})_{\rm reg}=-{\rm tr}\left[\frac{1}{\gamma\cdot \nabla + M}\right]+{\rm tr}\left[\frac{1}{\gamma\cdot \nabla + M_{\rm PV}}\right]\,.
\end{align}
The right hand side can be computed using the spectral representation of ${\rm tr}\left[\frac{1}{\gamma\cdot \nabla +M}\right]$ given in Appendix \ref{apsubsec:femionloop},
\begin{align}\label{eq:diracloop}
\begin{aligned}
&{\rm tr}\left[\frac{1}{\gamma\cdot \nabla + M}\right]=\\
&c_{d+1}\int_{-\infty}^{\infty} d\nu \left[\frac{1}{\nu^2+(M-\tfrac{1}{2})^2}-\frac{1}{\nu^2+(M+\tfrac{1}{2})^2}\right]\frac{\Gamma(\tfrac{d}{2})\Gamma(1+\tfrac{d}{2}\pm i\nu)}{4 \pi^{\frac{d}{2}+1}(d+1)\Gamma(d)\Gamma(\pm i\nu)}\,,
\end{aligned}
\end{align}
where $c_{d+1}$ is the number of components of Dirac spinor in $d+1$ dimensions (e.g.~$c_3=2$). Now, evaluating the right hand side of \eqref{eq:regulateddirac} using \eqref{eq:diracloop}, we find that the divergence is {\it not} regulated, contrary to our naive expectation. The same problem shows up also in the analysis in flat space, and it is essentially because the naive Pauli-Villars regularization does not preserve the discrete symmetry of the original Lagrangian which flips the sign of the mass term. The resolution to this problem in flat space is explained in Appendix \ref{app:PV}: The idea is to add $N/2$ bosonic Dirac fields with mass $M_{\rm PV}$ and $N/2$ bosonic Dirac fields with mass $-M_{\rm PV}$ instead of adding $N$ bosonic Dirac fields with mass $M_{\rm PV}$. When the $\sigma $ field takes a non-zero expectation value, these masses are shifted to $M_{+}=M+M_{\rm PV}$ and $M_{-}=M-M_{\rm PV}$. With this new parity-preserving Pauli-Villars regularization, the gap equation reads
\begin{align}
\begin{aligned}
&M(g^{-1})_{\rm reg}=-{\rm tr}\left[\frac{1}{\gamma\cdot \nabla + M}\right]+\frac{1}{2}\left({\rm tr}\left[\frac{1}{\gamma\cdot \nabla + M_{+}}\right]+{\rm tr}\left[\frac{1}{\gamma\cdot \nabla + M_{-}}\right]\right)\,.
\end{aligned}
\end{align}
Using the spectral representation \eqref{eq:diracloop}, we can check that the right hand side is now convergent. The result for $d=2$ (AdS$_3$) reads
\begin{align}\label{eq:Mginverse1}
M(g^{-1})_{{\rm PV}, \,d=2}= \frac{2M^2-M_{+}^2+M_{-}^{2} -\tfrac{1}{2}}{4\pi}=\frac{M^2}{2\pi}-\frac{M M_{\rm PV}}{\pi}-\frac{1}{8\pi}\,,
\end{align}
while the result for $d=1$ (AdS$_2$) reads
\begin{align}\label{eq:Mginverse2}
\begin{aligned}
M(g^{-1})_{{\rm PV}, \,d=1}&=\frac{2M\psi(M)-M_{+}\psi (M_{+})-M_{-}\psi(-M_{-})+1}{2\pi}\\
&=\frac{2M \psi (M)-2M\log M_{\rm PV}-2M+1}{2\pi }+\mathcal{O}(1/M_{\rm PV})\,.
\end{aligned}
\end{align}
For the bubble function $\tilde{B}_{F}$, this regularization gives the same result as the naive Pauli-Villars regularization \eqref{eq:fermBPVregu} and therefore removes the divergence as expected. 
\paragraph{Determination of the Constant Shift} Having identified the correct regularization scheme, the next task is to determine the constant shift of the bubble function. For this purpose, let us consider a free fermion\footnote{To emphasize that we are not considering the Gross-Neveu model, here we use a different notation for the fermions.} $\psi$ with mass $M$ and analyze the connected two-point function of a composite operator $\bar{\psi}\psi$. Since this two-point function is precisely given by the bubble diagram, it can be expressed in terms of the bubble function as\footnote{The minus sign comes from our convention of the bubble function.}
\begin{align}
\langle\bar{\psi}\psi (x)\,\,\bar{\psi}\psi (y) \rangle_{\rm connected}=-B_{F}(x,y)\,.
\end{align}
On the other hand, the one-point function of the same operator is given by a trace of the propagator. Therefore we have the relation
\begin{align}
\langle \bar{\psi}\psi (x)\rangle=-{\rm tr}\left[\frac{1}{\gamma\cdot \nabla +M}\right]\,.
\end{align}
Now comes the crucial observation: In free-fermion theory, these two quantities are related by differentiation with respect to mass $M$; more precisely we have the relation\footnote{This can be seen explicitly in the path integral formalism in which the one-point function is given by
\begin{align}
\langle \bar{\psi}\psi(x)\rangle =\frac{1}{Z}\int \mathcal{D}\psi\,\,\bar{\psi}\psi(x)\,\,e^{-\int d^{d+1}y \sqrt{g}\left(\bar{\psi}\gamma\cdot \nabla\psi +M\bar{\psi}\psi\right)}\,,
\end{align}
 The differentiation with respect to the mass brings down an extra $-\bar{\psi}\psi$ from the action. In addition, it acts on the partition function $Z$ producing a disconnected correlator. In total, we obtain the integrated connected correlator.}:
\begin{align}
\begin{aligned}
\frac{\partial}{\partial M}\langle \bar{\psi}\psi (x)\rangle =&-\int d^{d+1}y \sqrt{g(y)}\langle\bar{\psi}\psi (x)\,\,\bar{\psi}\psi (y) \rangle_{\rm connected}\,.
\end{aligned}
\end{align}
Translating this relation to the bubble function and using the spectral representation of the integrated correlator \eqref{eq:niceFOO}, we obtain
\begin{align}
-\frac{\partial}{\partial M}{\rm tr}\left[\frac{1}{\gamma\cdot \nabla +M}\right] =\int d^{d+1} \sqrt{g (y)}B_F(x,y)=\tilde{B}_F(\tfrac{id}{2})\,.
\end{align}
Of course, both sides of the relation are divergent and must be regulated using the regularization that we just explained. Once regulated, the relation allows us to determine the value of $\tilde{B}_F (\nu)$ evaluated at $\nu=id/2$, hence the constant shift, from ${\rm tr}\left[\frac{1}{\gamma\cdot \nabla +M}\right]$ which we already computed in \eqref{eq:Mginverse1} and \eqref{eq:Mginverse2}.

Carrying out these procedures, we get for $d=2$ (AdS$_3$),
\begin{align}\label{eq:BFPVregd2}
\begin{aligned}
(\tilde{B}_F(\nu))_{{\rm PV},\,d=2}=&\frac{-i(4M^2+\nu^2)}{8\pi  \nu}\left(\psi (M+1+\tfrac{i\nu}{2})-\psi (M+1-\tfrac{i\nu}{2})\right)\\
&+\frac{1+2M-4M_{\rm PV}}{4\pi}\,,
\end{aligned}
\end{align}
while for $d=1$ (AdS$_2$), we get
\begin{align}\label{eq:BFPVregd1}
\begin{aligned}
(\tilde{B}_F(\nu))_{{\rm PV},\,d=1}=&\mathcal{B}_{\text{AdS}_2} (M+\tfrac{1}{2},\nu)-\mathcal{B}_{\text{AdS}_2} (M+\tfrac{1}{2},\tfrac{i}{2})\\
&+\frac{M\psi^{(1)}(M)+\psi (M)-1-\log M_{\rm PV}}{\pi}\,,
\end{aligned}
\end{align}
where $\psi^{(1)}(x)\equiv d\psi (x)/dx$.
\paragraph{Two-point function of $\sigma$} We are now in the position to compute the two-point function of $\sigma$,
\begin{align}
(F_{\delta\sigma\delta\sigma}(\nu))^{-1}=-(g^{-1})_{\rm PV}+(\tilde{B}_F(\nu))_{\rm PV}\,.
\end{align}
 For $d=2$ (AdS$_3$), the difference of \eqref{eq:Mginverse1} and \eqref{eq:BFPVregd2} gives
 \begin{align}
 (F_{\delta\sigma\delta\sigma}(\nu))^{-1}=\frac{-i(4(\Delta-1)^2+\nu^2)}{8\pi  \nu}\left(\psi(\Delta+\tfrac{i\nu}{2})-\psi(\Delta-\tfrac{i\nu}{2})\right)+\frac{2\Delta-1}{8\pi (\Delta-1)}\,.
 \end{align}
 For $d=1$ (AdS$_2$), the difference of \eqref{eq:Mginverse2} and \eqref{eq:BFPVregd1} gives
 \begin{align}
 (F_{\delta\sigma\delta\sigma}(\nu))^{-1}=\mathcal{B}_{\text{AdS}_2} (\Delta,\nu)-\mathcal{B}_{\text{AdS}_2} (\Delta,\tfrac{i}{2})+\frac{(\Delta-\tfrac{1}{2})\psi^{(1)}(\Delta-\tfrac{1}{2})}{\pi}-\frac{1}{\pi (2\Delta-1)}\,.
 \end{align}
 
\subsection{Bound State in AdS}

We will now study the physical properties of the correlators obtained above. We will restrict ourselves to $d=2$, i.e. AdS$_3$.

\begin{figure}[t]
\centering
\includegraphics[clip,height=7cm]{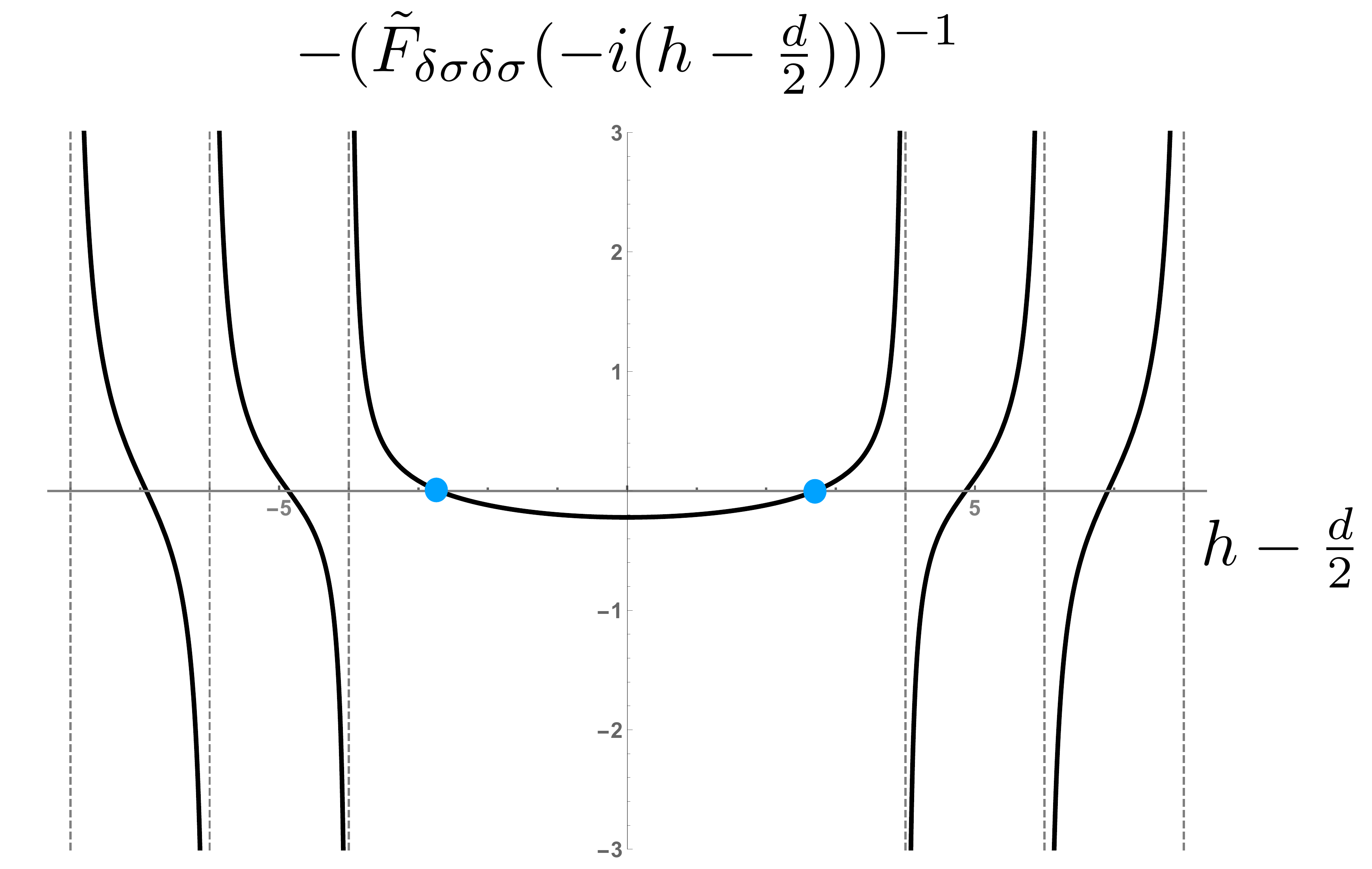}
\caption{The (-)inverse of the connected two-point function of $\sigma$ in the massive GN model in $d=2$ (i.e. AdS$_3$), for $\Delta = 2$. Since we used the gap equation to determine $g^{-1}$, there is no additional tunable parameter. Besides the zeroes associated to double-trace operators (whose tree-level dimension is denoted with vertical dashed gray lines), there is an additional zero highlighted by a dot. The associated operator corresponds to a bound state of two fermions in AdS that does not exist in the free theory. \label{fig:boundstate}}
\end{figure}

The result of the calculation in the previous subsections is that the connected two-point function of $\sigma$ in the massive phase is free of any divergence or ambiguity and takes the following form
\begin{equation}
\tilde{F}_{\delta\sigma\delta\sigma}(\nu) = \frac{1}{-\frac{i( 4(\Delta-1)^2+\n^2 )}{8\pi\n  }  \Big( \psi(\Delta+\frac{i\n}{2})- \psi(\Delta-\frac{i\n}{2}) \Big) +\frac{2\Delta-1}{8\pi(\Delta -1)}} ~.\label{eq:twopointGN}\\
\end{equation}
The resulting spectrum of boundary operators is illustrated in \ref{fig:boundstate}. For the ease of comparison with fig. \ref{fig:readingoffdimensions} that describes the $O(N)$ model, we plotted the $(-)$inverse of the function $\tilde{F}_{\delta\sigma\delta\sigma}(\nu)$.

We see the by-now-familiar sequence of poles originating from the double-trace operators of the free-theory, whose associated anomalous dimensions are positive and of $\mathcal{O}(1)$ for $\Delta\sim\mathcal{O}(1)$ (as in fig. \ref{fig:boundstate} which has $\Delta =2$), and become small at large $\Delta$. Notably, besides the sequence of double-trace poles, there is an additional pole with associated dimension $\Delta_{b} < 2\Delta + 1$, whose origin can be traced to the factor $( 4(\Delta-1)^2+\n^2 )$ in the denominator of \eqref{eq:twopointGN}. Comparing with the analogous plot for the massive $O(N)$ model in fig. \ref{fig:readingoffdimensions}, we see that the main difference is that in the GN case the central, u-shaped part of the curve intersects the axis, giving rise to this additional pole. We interpret the additional state as a bound state of two fermions in AdS, that only exists at finite coupling. This state manifests itself in the boundary conformal theory as a scalar operator in the spectrum.
 
We will now verify that in the flat-space limit the additional state indeed approaches the bound-state that is known to exist in the GN model in $\mathbb{R}^3$ \cite{ZinnJustin:1991yn, Rosenstein:1990nm}.

\paragraph{Flat-Space Limit}
Let us start by considering the fermionic bubble diagram with parity-preserving PV regulator, written in eq. \eqref{eq:BFPVregd2}. Taking the limit $L\to\infty$ with $\nu=|p|L$, $\Delta=ML$ and the rescaling $M_{PV}\to L M_{PV}$, we find perfect agreement with the flat-space result \eqref{eq:flatPVbubble} in the appendix. 

Next, we can compare the fermionic trace that appears in the gap equation \eqref{eq:gap}. Again, starting from the AdS answer in eq. \eqref{eq:gap} and taking the limit as above, we find perfect agreement with the flat space result in eq. \eqref{eq:1pointMPV}.

It is nice to observe the even when considering correlation functions that depend on a UV regulator, as long as the regulator can be defined also for the theory on AdS, the dependence on the regulator in flat space can be precisely recovered from the AdS result. This is in agreement with the intuition that short distance effects are not affected by the curvature of the background.

The two checks above immediately imply that also the correlator of $\delta\sigma$ also has the correct behavior in flat-space limit
 \begin{equation}
 \lim_{L\to \infty}L^{3} \,\tilde{F}_{\delta\sigma\delta\sigma}(\nu = L |p|)=-\frac{1}{(g^{-1})_{PV} - (\tilde{B}_F^{\rm flat}(|p|))_{PV}} =\frac{1}{(p^2 + 4 M^2)\tilde{B}(p^2,M^2) }~,
\end{equation}
which has a bound-state pole at $-p^2 = M^2_b = 4 M^2$, i.e. the onset of the two-fermion threshold. It can be also easily verified that the scaling dimension $\Delta_b$ of the AdS bound state grows asymptotically like $2\Delta$ at large $\Delta$, matching with the mass of the bound state $M_b = 2 M$.

\subsection{Critical Point of the GN Model on AdS$_3$}\label{subsec:critGN}

In flat space $\mathbb{R}^3$, the RG flow of the massless (i.e. $\Sigma = M =0$) GN model can be followed towards the UV at large $N$. In the limit of $g\to\infty$ the correlation functions display scaling behavior, with dimensions that can be systematically computed in $1/N$ expansion, thus providing evidence for the existence of a UV fixed point. The latter fixed point is believed to exist also at finite $N$, and a UV completion has been proposed in \cite{ZinnJustin:1991yn} in terms of the so-called Gross-Neveu-Yukawa model (see \cite{Fei:2016sgs} for a recent perturbative study of this fixed point). 

Using the approach of section \ref{sec:Critical} and the correlators computed in the previous subsection, we can now look for this UV conformal fixed point on AdS$_3$ background. Similarly to the $O(N)$ case, the question is for what values of the dimensionless parameters $g$ and $M$, measured in units of the AdS radius, there exists conformal symmetry in the bulk. The manifestations of this symmetry that we will look for are: $(i)$ the existence of an expansion in bulk conformal blocks for the two-point function of $\sigma$, and $(ii)$ the existence of a displacement operator in the spectrum of boundary operators. As we discussed in section \ref{sec:Critical}, the point $(i)$ has the shortcoming that it requires us to commute the spectral representation with the bulk OPE expansion, which is not a rigorous operation, but we saw that nevertheless it allowed us to detect the critical point in the $O(N)$ example. 

In this subsection we will denote the boundary dimension of the $U(N)$-vector fermionic operators with $\hat{\Delta}$, and use $\Delta$ without hats for the scaling dimensions of bulk operators. The two-point function of $\sigma$ in eq. \eqref{eq:twopointGN} is only a function of $\hat{\Delta}$, because it was obtained using the gap equation to determine $g^{-1}$. Naively, inspired by flat space, one might think that the critical point in AdS$_3$ must be at the massless point $\Sigma = M = 0$, which corresponds to $\hat{\Delta} = 1$\footnote{Note that the connected two-point function of $\sigma$ in eq. \eqref{eq:twopointGN} actually vanishes at the massless point. However recall that it was derived under the assumption that $\hat{\Delta} >1$, with a strict inequality.}. On the other-hand, it is possible to have conformal boundary conditions for the GN model that actually break the symmetry under parity, meaning that we can have a BCFT with non-zero bulk one-point function of $\sigma$. This is analogous to the so-called {\it extraordinary transition} in the context of the 3d Ising model, in which the $\mathbb{Z}_2$ symmetry is broken by the boundary condition and the spin operator has a non-zero one-point function. We can then proceed using the two-point function of eq. \eqref{eq:twopointGN}, keeping in mind that we are looking for parity-breaking conformal boundary conditions.

Let us analyze what are the possible contribution to the bulk OPE expansion of the two-point function of $\sigma$ at the putative bulk conformal point. Similarly to the $O(N)$ model, at leading order at large $N$ $\sigma$ is a generalized free field of dimension $\Delta_\sigma = 1$. Hence in the leading-order bulk OPE there are double-trace operators, and the scalar ones ---that schematically are $\sigma\square^n \sigma$ and have scaling dimension $2 + 2n$--- can get a non-zero one-point function and therefore contribute to the two-point function. Indeed the two-point function of $\sigma$ at leading order $\sim \mathcal{O}(N)$ is the disconnected piece, due to the non zero one-point function $\langle \sigma\rangle = \sqrt{N}\Sigma$. This disconnected piece is reproduced by the leading order contribution of the double-trace operators in the OPE (which is of the same order because the OPE coefficients are of order $\sim\mathcal{O}(1)$ and the one-point functions are of order $\sim\mathcal{O}(N)$). Again in analogy with the $O(N)$ model, these leading-order terms drop in the spectral representation, because they are a series of positive integers powers of the cross-ratio.

Next, in the connected two-point function, i.e. the two-point function of $\delta\sigma$, that starts at $\sim\mathcal{O}(1)$, we have contributions in the bulk OPE from the identity operator, and from the $1/N$ corrections of the double-trace contributions. Note that, differently from the $O(N)$ case, in the GN model there can never be a contribution from $\sigma$ itself, because of the discrete symmetry that flips the sign of $\sigma$, enforcing the bulk OPE coefficient to vanish $C_{\sigma\sigma\sigma} = 0$ (for the $O(N)$ model this is only an accident that happens to be true for $d+1=3$, but there is no symmetry visible along the RG that enforces it). Another difference from the $O(N)$ model is that in this case we expect the double-trace operators $\sigma \square^n \sigma$ to receive an anomalous dimension at order $1/N$. Recall that in the spectral representation the contribution from the order $\sim\mathcal{O}(1/N)$ correction to the (OPE coefficient $\times$ one-point function) drops, because it has the same dependence on the cross-ratio as the leading-order contribution. 

Summarizing the last two paragraphs, in the bulk OPE expansion of $F_{\delta\sigma\delta\sigma}$ at the conformal point we expect to find the power-law contribution from the bulk block of the identity operator, and a contribution from the $1/N$ expansion of the blocks of the double-trace operators. Under the assumptions that this OPE can be mapped to the large-$\nu$ limit of $\tilde{F}_{\delta\sigma\delta\sigma}$, in this limit we expect to have a contribution from the transform of the power-law block of the identity, and additional contributions from the spectral transform of derivatives of power-laws, associated to the anomalous dimensions of double-trace operators.
 
The first few terms in the expansion for large real $\nu$ of the two-point function are
\begin{align}
&\tilde{F}_{\delta\sigma\delta\sigma}(\nu) \underset{\nu \to \infty}{\sim} \frac{8}{\nu}\left[1 + \frac{4(\hat{\Delta} - \frac 12)(\hat{\Delta} - \frac 32)}{\pi(\hat{\Delta}-1)\nu}  + \frac{1}{\nu^2}\left(\left(\frac{4(\hat{\Delta} - \frac 12)(\hat{\Delta} - \frac 32)}{\pi(\hat{\Delta}-1)}\right)^2 - 4 (\hat{\Delta} -1)^2\right)\right. \nonumber\\ &\left.+ \frac{1}{\nu^3}\left(\left(\frac{4(\hat{\Delta} - \frac 12)(\hat{\Delta} - \frac 32)}{\pi(\hat{\Delta}-1)}\right)^3 - \frac{64 (\hat{\Delta} -1)(\hat{\Delta} - \frac 12)(\hat{\Delta} - \frac 32)}{3\pi}\right) + \mathcal{O}\left(\frac{1}{\nu^4}\right)\right]~.\label{eq:sigmatwopointGN}
\end{align}
Comparing the leading power $\nu^{-1}$ to eq. \eqref{eq:nupowerexpand}, we find that the only compatible assignment of scaling dimension to $\sigma$ is $\Delta_\sigma = 1$, which is precisely the expected value at the UV fixed point of the GN model. Using this value of the dimension in eq. \eqref{eq:nupower}, we find that the spectral representation of the bulk block of the identity operator is
\begin{equation}
\tilde{F}_{\Delta = 1}(\nu) = 8\pi^2 \frac{\coth(\pi \nu)}{\nu} \underset{\nu \to \infty}{\sim} 8\pi^2\frac{1}{\nu}(1+\mathcal{O}(e^{-2\pi \nu}))~.
\end{equation}
The bulk blocks of the double-trace operators $\sigma\square^n\sigma$ in position space are a series of powers $\zeta^{n + k}$, with $k$ a non-negative integer labeling descendants. Hence in $\nu$ space the contributions from their anomalous dimensions are proportional to a derivative w.r.t. $\Delta$ of the transform of the power law $\zeta^{-\Delta}$, evaluated at $\Delta = -n-k$. Among all these contributions, the leading one at large $\nu$ comes from $n=k=0$ and is
\begin{equation}
\left(\frac{d}{d\Delta}\tilde{F}_{\Delta}(\nu)\right)\vert_{\Delta = 0} = -\frac{4\pi^2 \coth(\pi \nu)}{\nu+\nu^3}\underset{\nu \to \infty}{\sim} -4\pi^2\frac{1}{\nu^3}\left(1+\mathcal{O}\left(\frac{1}{\nu^2}\right)\right)~.
\end{equation}
Hence, in order for the two-point function to have a sensible bulk OPE, it should agree with the bulk block of the identity at large $\nu$ at least up to order $\nu^{-3}$. By inspection of eq. \eqref{eq:sigmatwopointGN} we see that this requires $\hat{\Delta} = \frac 32$ (the solution $\hat{\Delta} = \frac 12$ is outside the regime of the $+$ boundary condition, i.e. $\hat{\Delta} \geq 1$; it is tempting to speculate that $\hat{\Delta} = \frac 12$ is the solution in the case of $-$ boundary condition, however a more careful analysis is needed to verify this). Specifying to this value, the two-point function simplifies to
\begin{equation}
\tilde{F}_{\delta\sigma\delta\sigma}(\nu)\vert_{\hat{\Delta}=\frac 32} =  \frac{8}{\nu} \coth(\tfrac{\pi\nu}{2})\frac{1}{1+\frac{1}{\nu^2}}~.\label{eq:sigmatwopointGNconf}
\end{equation}
The resulting two-point function deviates at large $\nu$ from the block of the identity due to the additional powers coming from the factor $\frac{1}{1+\frac{1}{\nu^2}}$. These power-law deviations precisely match the possible powers from the anomalous dimensions of double-trace operators. Moreover, the spectrum of boundary operators that contribute to this two-point function (i.e. the poles of $\tilde{F}_{\delta\sigma\delta\sigma}(-i(h-1))$ as a function of $h$) is
\begin{equation}
\hat{\Delta}_n =2n+3 \qquad n\in \mathbb{N}_{\geq 0}\,.
\end{equation}
In particular, we find a boundary scalar operator of dimension 3, that can be the displacement operator. We view these facts as a strong hint that the theory at $\hat{\Delta} = \frac 32$ and $g$ fixed by the gap equation has bulk conformal symmetry.

It is interesting to observe that for the $O(N)$ model we found the conformal value of the boundary scaling dimension to be $\hat{\Delta}=1$, which is the boundary scaling dimension for a free-fermion theory with either choice of boundary condition (the corresponding boundary operator is just the restriction of some components of the bulk fermionic field to the boundary), while for the GN model we found $\hat{\Delta} = \frac 32$, which is the boundary scaling dimension for a free-scalar theory with Dirichlet boundary condition (the corresponding boundary operator is the normal derivative of the bulk scalar field, restricted to the boundary; the second solution $\hat{\Delta} = \frac 12$ would correspond to a free scalar with Neumann boundary condition). This observation is likely to have some interpretation in the context of the bosonization duality that these theories enjoy in presence of additional interactions with bulk Chern-Simons gauge fields \cite{Aharony:2012nh}. 

\subsection{GN BCFT Data from the AdS Correlators}
We can now straightforwardly repeat the analysis that we performed for the $O(N)$ case in subsection \ref{sec:ONBCFT} also for the GN model, and derive some of the data of the BCFT associated to the AdS critical point found in the previous subsection. To our knowledge, the conformal boundary conditions in the GN model was never studied before and our results provide the first predictions for its BCFT data.

We can directly read from the previous subsection that this BCFT contains $U(N)$-vector boundary fermionic operators with scaling dimension $\hat{\Delta} = \frac 32$, and that the bulk-to-boundary OPE of the operator $\sigma$ contains a family of scalar operators $\hat{\sigma}_n$ with scaling dimensions $3 + 2n$, with $n$ a non-negative integer, which includes the displacement operator.\footnote{Note that even though $\sigma$ is odd under parity, the boundary operators $\hat{\sigma}_n$ do not need to be, because the boundary condition that we are discussing breaks parity.} While the operators $\hat{\sigma}_n$ with $n>1$ are a continuation of the double-trace operators of the free-fermion theory, the displacement operator $\hat{\sigma}_0$ corresponds to the bound state in AdS. Through the AdS analysis we thus discover that the existence of this operator in the spectrum of the BCFT is connected to the existence of a bound state in the S-matrix of the gapped phase.
 
Looking at the residues of the two-point function \eqref{eq:sigmatwopointGNconf}, we obtain that the leading order bulk-to-boundary OPE coefficients of $\sigma$ with these operators are
\begin{equation}
b_{\sigma\hat{\sigma}_n}^2 = \frac{64(n+1)}{4n(n+2)+3}~.
\end{equation}
Plugging the two-point function of $\sigma$ in the four-point function of the $U(N)$-vector fermionic fields in \eqref{eq:fermg1234}-\eqref{eq:fourpointferm}, and looking at the residues of the integrand in $\nu$, we can also easily compute the boundary OPE coefficients between two $U(N)$-vector fields and a $\hat{\sigma}_n$ at first non-trivial order, i.e. $\mathcal{O}(1/N)$, finding
\begin{equation}
C_{\hat{\Psi}\hat{\Psi}\hat{\sigma}_n}^2 = \frac{1}{N}\frac{\pi (2n+1)(2n+3)\Gamma(\frac32+2n)^2}{2^{4n+4}(n!)^2}~.
\end{equation}

As we already remarked while analyzing the $O(N)$ BCFT, we stress that these data are not straightforward to obtain via a Feynman diagram calculation in flat-space. Hence it would be interesting to explore further whether the techniques used here can lead to an efficient calculation of additional data of the GN BCFT, perhaps including also subleading orders in $1/N$.

 \section{Conclusion}\label{sec:conclusion}
 
In this paper we studied the dynamics of large-$N$ vector models on AdS background. We demonstrated that the large-$N$ techniques familiar from flat space can be efficiently imported to AdS, with the spectral representation being the main technical tool required. The solvability at large $N$ allowed us to explore the finite-coupling regime of field theories on AdS, revealing phenomena that are not visible in the usual perturbative regime. Moreover, we explicitly saw in these examples that one can obtain both the flat-space S-matrix and the correlators in BCFT from the AdS correlation functions. 

In the context of the models considered in this paper, a clear direction for the future is to explore whether the simpler analytic structure of the correlators in AdS, as a function of $\nu$, allows to efficiently compute $1/N$ corrections. We saw a hint of how this could work in \ref{subsubsec:corrections}. Moreover, it would be desirable to clarify the validity of the large-$\nu$ analysis that we use to detect bulk conformality in sections \ref{subsec:critON} and \ref{subsec:critGN}, and to try to make this approach more rigorous.

More generally, it would be interesting to explore if techniques similar to the ones used for the vector models can be applied to other theories. The first possible generalization that naturally presents itself would amount to introducing Chern-Simons (CS) gauge-fields in AdS$_3$ coupled to the $O(N)$ (or $U(N)$) symmetry. These CS-matter theories are solvable at large $N$ in flat space, they enjoy a bosonization duality \cite{Aharony:2012nh}, and their S-matrix in the massive phase displays a non-standard crossing-symmetry property \cite{Jain:2014nza}. If the relevant diagrams can be resummed on AdS$_3$, it should be possible to see these properties of the S-matrix emerge from the boundary correlators,\footnote{The boundary correlators in AdS that map to the 2 to 2 amplitude of $O(N)$ vector particles are correlators of line operators in the bulk that end on charged operator insertions at the boundary.} and also study the conformal boundary conditions for the critical points of these theories, hopefully obtaining more checks of the duality on the way. This would require us to consider more general diagrams than the bubble diagrams encountered in this paper, e.g. ladder diagrams.

Another possibility is to consider theories on AdS$_n\times$S$^m$. When the bulk is conformal, this would describe a  defect conformal field theory \cite{Billo:2016cpy} since AdS$_n\times$S$^m$ can be mapped to $\mathbb{R}^{n+m}$ with a codimension-$(m+1)$ defect via a Weyl transformation. It would in particular be interesting to consider an analogue of the twist defect \cite{Gaiotto:2013nva,Yamaguchi:2016pbj,Soderberg:2017oaa} and understand how the mass deformation in AdS allows us to interpolate such defect CFT correlators with the flat-space S-matrix.
 
 We hope that the results of this paper convinced the reader that quantum field theory in AdS is a rich and interesting subject which connects various different physics. As the closing remark\footnote{This last remark is largely motivated by an inspiring talk given by Sasha Zamolodchikov in a bootstrap workshop in the Azores \cite{Zamolodchikov:2018}.}, let us suggest yet another potentially interesting direction which we did not explore in this paper. In conformal field theories, the operator product expansion is extremely powerful since we have a good understanding of its analytic property such as the radius of convergence \cite{Pappadopulo:2012jk,Hogervorst:2013sma}. By contrast, the general properties of the OPE in massive quantum field theories are poorly understood. One might naively think that the operator product expansion in such theories is at best asymptotic and has a zero radius convergence. However, in cases where one has analytical control of a theory, be it integrability or large $N$, one often finds much a better behavior \cite{Zamolodchikov:2018}. Addressing this question directly in flat space would be a hard problem, but one might be able to make some progress for theories in AdS since any correlation functions in AdS admit an alternative expansion, which is the {\it boundary} operator product expansion as we explained in section \ref{sec:Critical}, and their analytic properties are likely to be under better control than those of their flat-space counterparts. It would be interesting if the idea of placing QFTs in AdS will help solving such a foundational question.
 
\section*{Acknowledgements}
We thank Shira Chapman, Davide Gaiotto, Andrea Guerrieri, Daniel Kapec, Zohar Komargodski, Juan Maldacena, Marco Meineri, Matthijs Hogervorst, Marco Serone, Joao Penedones, Vladimir Rosenhaus and Pedro Vieira for useful discussions. DC and SK would like to thank Perimeter Institute for Theoretical Physics where part of this work was done. Research at Perimeter Institute is supported by the Government of Canada through Industry Canada and by the Province of Ontario through the Ministry of Research \& Innovation.The work of SK is supported by DOE grant number DE-SC0009988.

%

\appendix

\section{Scalar Bulk-to-Bulk Propagator}\label{app:BtoB}

The bulk-to-bulk propagator for a scalar field in AdS$_{d+1}$ is \cite{Burges:1985qq}
\begin{align}
&G_\Delta(x_1,x_2) \nonumber \\ & = \frac{\Gamma(\Delta)}{2\pi^\frac{d}{2} \Gamma(\Delta -\frac{d}{2}+1)}\frac{L^{2\Delta -d +1}}{\zeta(x_1,x_2)^\Delta} {}_2F_1\left(\Delta,\Delta - \tfrac{d}{2}+\tfrac 12,2\Delta-d+1,-\tfrac{4 L^2}{\zeta(x_1,x_2)} \right)~.\label{eq:bulktobulk}
\end{align}
$\zeta(x_1,x_2)$ is the square of the chordal distance between the points. In Poincar\'e coordinates
\begin{equation}\label{eq:chordaldistance}
\zeta(x_1,x_2) = L^2 \frac{(z_1-z_2)^2 + (\vec{x}_1-\vec{x}_2)^2}{z_1z_2}~.
\end{equation} 
The propagator satisfies
\begin{equation}
\left[-\square_{x_1} + \frac{1}{L^2}\Delta(\Delta-d)\right] G_\Delta(x_1,x_2) = \delta^{d+1}(x_1,x_2)~.\label{eq:Green}
\end{equation}
Taking either of the two points close to the boundary, the propagator in eq. \eqref{eq:bulktobulk} behaves like $z_{1,2}^{\Delta}$. Hence it is valid for either of the two possible boundary conditions, upon the identification $\Delta= \Delta_\pm$.

Using a transformation formula for the hypergeometric function, we can rewrite the propagator in the form presented in \cite{Burgess:1984ti, Inami:1985wu}
\begin{align}
&G_\Delta(x_1,x_2) \nonumber \\ & =
\frac{\Gamma(\Delta)}{2\pi^\frac{d}{2} \Gamma(\Delta -\frac{d}{2}+1)}\frac{L^{-d +1}}{\left(2 \cosh\left(\sigma(x_1,x_2)\right)\right)^\Delta} {}_2F_1\left(\tfrac{\Delta}{2},\tfrac{\Delta+1}{2},\Delta-\tfrac{d}{2}+1,\tfrac{1}{\cosh\left(\sigma(x_1,x_2)\right)^2} \right)~,\label{eq:btbJac}
\end{align}
where $\sigma$ is the geodesic distance between the two points, in units of the AdS radius
\begin{equation}
\sigma(x_1,x_2) = 2 \, {\rm arcsinh}\left(\frac{\sqrt{\zeta(x_1,x_2)}}{2L}\right)~.
\end{equation}

\section{Spectral Representation of Two-Point Functions}\label{app:SpRep}

In this appendix we review the spectral representation for two-point functions in AdS. 

Consider a two-point function of a scalar operator $O$ on AdS background. Thanks to the isometries of the background, this is a function of only one non-negative real variable, the distance between the two points
\begin{equation}
\langle O(x_1) O(x_2) \rangle = F_{OO}(\sigma(x_1,x_2))~.
\end{equation}
It is useful to introduce a transform operation on two-point correlators, which we refer to as {\it spectral representation}, because it can be seen as an expansion in a (continuous) basis of eigenfunctions of the Laplacian on AdS \cite{Cornalba:2007fs, Cornalba:2008qf, Penedones:2007ns, Penedones:2010ue}. This transform has many nice properties that makes it analogous to (radial) Fourier transform.

To start with, define the eigenfunctions of the Laplacian in terms of the bulk-to-bulk propagator evaluated at complex values of the scaling dimension
\begin{equation}
\Omega_\nu(x_1,x_2) \equiv L^{d-1}\frac{i \nu}{2 \pi}\left(G_{\frac{d}{2}+i\nu}(x_1,x_2)-G_{\frac{d}{2}-i\nu}(x_1,x_2)\right)~.\label{eq:eigen}
\end{equation}
We will use interchangeably the notation $\Omega_\nu(x_1,x_2)$ and $\Omega_\nu(\sigma)$, the latter when we want to view it as a function on $\mathbb{R}_+$. Using \eqref{eq:Green} we find the eigenvalue equation
\begin{equation}
-\square_{x_1} \,\Omega_\nu(x_1,x_2) = \frac{1}{L^2}\left(\frac{d^2}{4} +  \nu^2\right) \Omega_\nu(x_1,x_2)~.\label{eq:boxeigen}
\end{equation}
The spectral representation is then the following integral expression for the correlator
\begin{equation}
\langle O(x_1) O(x_2) \rangle = \int_{-\infty}^{+\infty}d\nu \, \tilde{F}_{OO}(\nu) \Omega_\nu (x_1, x_2)~,\label{eq:specrap}
\end{equation} 
in terms of the function $\tilde{F}_{OO}$ of the variable $\nu$. The inverse of this representation is
\begin{align}
\tilde{F}_{OO}(\nu) 
& =  \frac{{\rm Vol}(S^d)}{\Omega_\nu(0)} \int_0^\infty d\sigma \,(\sinh \sigma)^d \,F_{OO}(\sigma) \,\Omega_\nu(\sigma) \nonumber \\
& = \frac{1}{\Omega_\nu(0)} \int_{{\rm AdS}_{d+1}} d^{d+1} x_1 \sqrt{g(x_1)}\,\langle O(x_1) O(x_2)\rangle \,\Omega_\nu(x_1,x_2)~,\label{eq:inversespec}
\end{align}
where 
\begin{equation}
\Omega_\nu(0)=\Omega_\nu(x,x) =\frac{ \Gamma\left(\frac{d}{2}\right)}{4 \pi^{\frac{d}{2}+1}\Gamma(d)}\frac{\Gamma\left(\frac{d}{2} \pm i \nu\right)}{\Gamma(\pm i \nu)} 
~.\label{eq:coincpt}
\end{equation}
where $\Gamma(z\pm a)\equiv\Gamma(z+a)\Gamma(z-a)$. It is also useful to rewrite the kernel just in terms of a single hypergeometric function, symmetric under exchange $\nu\to-\nu$. This can be done in two ways, thanks to an identity of the hypergeometric function
\begin{align}
\frac{\Omega_\nu(\sigma)}{\Omega_\nu(0)} & = {}_2F_1\left(\tfrac{\tfrac d2 + i \nu}{2}, \tfrac{\tfrac{d}{2}-i \nu}{2}, \tfrac{d+1}{2},- \sinh(\sigma)^2\right) \label{eq:kern1}\\
& = {}_2F_1\left(\tfrac d2 + i \nu , \tfrac{d}{2}-i \nu , \tfrac{d+1}{2},- \sinh\left(\tfrac{\sigma}{2}\right)^2\equiv -\tfrac{\zeta}{4 L^2}\right)~.\label{eq:kern2}
\end{align}

The existence of this transform and its inverse can be derived by viewing it as a special case of the so-called Jacobi transform \cite{koornwinder1984jacobi}.\footnote{We thank M. Hogervorst for drawing our attention to this reference.} The latter is an integral transform that acts on functions on $\mathbb{R}_+$ and it is labeled by two complex parameters $\alpha$ and $\beta$. We refrain from reproducing here the definition, that can be found in section 2 of \cite{koornwinder1984jacobi}.\footnote{The Jacobi transform also appeared in the recent works on the ``alpha-space" representation of four-point function conformal blocks in $d=1$ CFTs \cite{Hogervorst:2017sfd} and two-point function conformal blocks in boundary and crosscap CFTs in arbitrary $d$ \cite{Hogervorst:2017kbj}. These papers also contain a nice review of the Jacobi transform, however using a different set of conventions compared to \cite{koornwinder1984jacobi}, to which instead we conform. In particular the spectral representation of two-point functions in AdS$_{d+1}$ is identical to the ``alpha-space" representation of two-point functions in boundary CFT$_{d+1}$, in the boundary OPE channel.} Comparing the kernel in \cite{koornwinder1984jacobi} with eq. \eqref{eq:kern1}, we see that the spectral representation $\tilde{F}_{OO}(\nu)$ is the Jacobi transform with parameters $\alpha = \frac{d-1}{2}$ and $\beta=-\frac 12$ of the function $F_{OO}(\sigma)$ of the variable $\sigma \in \mathbb{R}_+$. Alternatively, comparing with eq. \eqref{eq:kern2}, we can also identify the spectral representation $\tilde{F}_{OO}(\nu)$, regarded as a function of the variable $2\nu$, with the Jacobi transform with parameters $\alpha = \beta =\frac{d-1}{2}$ of $F_{OO}(\sigma)$, regarded as a function of the variable $\tfrac{\sigma}{2} \in \mathbb{R}_+$. 

From \cite{koornwinder1984jacobi} we see that the transform can be defined on the space of functions of $\sigma$ that are square-integrable in the measure $d \sigma (\sinh \sigma)^d$, and it maps isometrically to the space of square-integrable functions in the measure $d \nu \frac{1}{\Omega_\nu(0)}$. This condition restricts the behavior of the two-point correlator when either point is taken close to the boundary. In Poincar\'e coordinates, assuming a power-law behavior near the boundary at $z=0$
\begin{equation}
\langle O(x_1) O(x_2) \rangle \underset{z_{1,2} \to 0}{\propto} z_{1,2}^\Delta~,
\end{equation}
for some real number $\Delta$, the condition is $\Delta > \frac d2$.

A useful property of the spectral representation is that it maps convolutions to products. Given two functions $F$ and $G$ that depend on a couple of points $(x_1,x_2)$ in AdS only through the distance $\sigma$, the convolution
\begin{equation}
(F\star G)(x_1,x_2) = \int_{{\rm AdS}_{d+1}} d^{d+1} x' \sqrt{g(x')}\, F(x_1,x')\, G(x',x_2)~,
\end{equation}
also defines a function that only depends on the distance. Then we have
\begin{equation}
\widetilde{F\star G}(\nu) = L^{-d-1} \tilde{F}(\nu) \tilde{G}(\nu)~.
\end{equation}
This can be derived either as a special case of the similar property of the Jacobi transform, discussed in section 7 of \cite{koornwinder1984jacobi}, upon viewing AdS$_{d+1}$ as the quotient $SO(d+1,1)_+/SO(d+1)$, or alternatively following \cite{Penedones:2007ns}.

Another useful property of the spectral representation can be derived by observing that the kernel in \eqref{eq:kern1}-\eqref{eq:kern2} simplifies to 1 for $\nu = \pm i\frac d2$. This implies that the integrated two-point function can be obtained from a simple evaluation of the spectral representation
\begin{equation}\label{eq:niceFOO}
\tilde{F}_{OO}( \pm i\tfrac d2) = \int_{{\rm AdS}_{d+1}} d^{d+1} x_1 \sqrt{g(x_1)}\,\langle O(x_1) O(x_2)\rangle ~.
\end{equation}

\subsection{Flat-Space Limit of the Spectral Representation}\label{sec:FslSpRep}
Here we show that in the flat-space limit, the spectral representation reduces to ordinary Fourier transform.  As a starting point, we assume that in the limit $L\to\infty$ the two-point correlator in position space approaches the flat-space correlator in position space, which is a function of the distance $r = |x_1-x_2|$ in $\mathbb{R}^{d+1}$, namely
\begin{equation}
F_{OO}(\sigma = r/L ) \underset{L\to\infty}{\longrightarrow} F^{\rm flat}_{OO}(r)~,\label{eq:fslimpos}
\end{equation}
where
\begin{equation}
\langle O(x) O(0) \rangle_{\mathbb{R}^{d+1}} = F^{\rm flat}_{OO}(|x|)~.
\end{equation}
It is implicit in the assumption \eqref{eq:fslimpos} that we take the limit $L\to \infty$ by scaling the parameters of the theory in an appropriate way to obtain a sensible flat-space limit, e.g. keeping masses and bulk couplings fixed in the limit.

We will now prove that \eqref{eq:fslimpos} implies that
\begin{equation}
L^{d+1} \,\tilde{F}_{OO}(\nu = L |p| ) \underset{L\to\infty}{\longrightarrow} \tilde{F}^{\rm flat}_{OO}(|p|)~.\label{eq:specfsl}
\end{equation}
The function $\tilde{F}^{\rm flat}_{OO}$ is the Fourier transform of the two-point correlator on $\mathbb{R}^{d+1}$
\begin{align}
\tilde{F}^{\rm flat}_{OO}(|p|) & \equiv \langle \tilde{O}(p) \tilde{O}(-p) \rangle^{\rm flat} \nonumber \\
& = \int d^{d+1} x \,e^{i p \cdot x} \,\langle O(x) O(0) \rangle^{\rm flat} \nonumber \\
& = |p|^{-\frac{d-1}{2}}\,(2\pi)^{\frac{d+1}{2}}\int_0^{\infty}d r\, r^{\frac{d+1}{2}} \,  J_{\frac{d-1}{2}}(|p|r) \, F^{\rm flat}_{OO}(r)~,\label{eq:rFT}
\end{align}
where $J$ denotes the Bessel function of the first kind. To prove \eqref{eq:specfsl} we use the following integral representation of the hypergeometric function, valid for $\mathrm{Re}(c)>\mathrm{Re}(b)>0$ 
\begin{equation}
{}_2F_1(a,b,c,z) = \frac{\Gamma(c)}{\Gamma(b)\Gamma(c-b)}\int_0^1 dt\, t^{b-1} (1-t)^{c-b-1} (1-z t)^{-a}~.
\end{equation} 
Plugging this formula in the expression for the bulk-to-bulk propagator in
\eqref{eq:bulktobulk} with $\Delta = \frac{d}{2}+ i |p| L$, $\zeta=r^2$ and changing variable to $y = \frac{2 L t}{r}$, after some simplification we are left with the following integral 
\begin{equation}
G_{\frac{d}{2}+i |p| L }(\zeta = r^2)\propto \frac{r}{2 L}\int_0^{\frac{2 L}{r}} dy \, y^{-\frac{d+1}{2}} \left(1+ \frac{r}{2 y L}\right)^{-\frac{d}{2}-i |p| L} \left(1- \frac{r y}{2  L}\right)^{-\frac{1}{2} + i |p| L}~.
\end{equation}
For $L$ large we can approximate 
\begin{equation}
\left(1+ \frac{r}{2 y L}\right)^{-\frac{d}{2}-i |p| L} \left(1- \frac{r y}{2  L}\right)^{-\frac{1}{2} + i |p| L} \sim e^{-i |p| r \frac{1}{2}\left(y+\frac{1}{y}\right)}~.
\end{equation}
Restoring all the prefactors, and using Stirling formula to approximate the gamma functions, we find
\begin{align}
& \frac{d r}{L}\frac{{\rm Vol}(S^d) \,\Omega_{|p| L}(r/L) \,\sinh(r/L)^d}{\Omega_{|p|L}(0)} \nonumber \\ & \underset{L\to\infty}{\longrightarrow}  dr\,L^{-d-1}|p|^{-\frac{d-1}{2}}  (2\pi)^{\frac{d-1}{2}} r^{\frac{d+1}{2}} \int_0^\infty d y \left( i^{\frac{d+1}{2}}e^{-i |p| r \frac{1}{2}\left(y+\frac{1}{y}\right)} + c.c. \right) \nonumber \\ & = d r\, L^{-d-1}|p|^{-\frac{d-1}{2}} (2\pi)^{\frac{d +1}{2}} r^{\frac{d+1}{2}} \,J_{\frac{d-1}{2}}(|p|r)~.
\end{align}
This proves that the kernel of the transform in the first line of\eqref{eq:inversespec} reduces to the kernel of the radial Fourier transform in \eqref{eq:rFT} up to the factor $L^{-d-1}$, hence establishing \eqref{eq:specfsl}. Note that similar estimates of a hypergeometric function in terms of a Bessel function have been used in the CFT literature to obtain the asymptotic behavior of conformal blocks \cite{Komargodski:2012ek, Fitzpatrick:2012yx, Qiao:2017xif}.

\subsection{Spectral Representation of the Scalar Propagator}\label{sec:SpRepProp}
As a first example of a spectral representation, consider the bulk-to-bulk propagator itself, first in the case $\Delta = \Delta_+ > \frac{d}{2}$. With this boundary condition, the propagator belongs to the class of functions for which the spectral representation is well-defined. The following integral representation holds
\begin{equation}
G_{\Delta = \Delta^+}(x_1,x_2) = \frac{1}{L^{d-1}}\int_{-\infty}^{+\infty} d\nu \,\frac{1}{\nu^2 + \left(\Delta -\frac{d}{2}\right)^2} \, \Omega_\nu(x_1,x_2)~.\label{eq:srprop}
\end{equation}
To prove this formula, instead of computing explicitly the inverse as in \eqref{eq:inversespec}, we perform the integral by closing the contour in the appropriate region and summing over poles. By inspection of eq. \eqref{eq:bulktobulk}, we see that $G_\Delta$ as a function of $\Delta$ has poles at $\Delta = - n$, with $n$ a non-negative integer, coming from the $\Gamma(\Delta)$ in the prefactor (the poles of the hypergeometric function at $\Delta -\frac{d}{2} + 1 = -n$ are canceled by the zeros of the prefactor). As a consequence, the first term in eq. \eqref{eq:eigen} has poles in $\nu$ located at $\nu = i (\tfrac{d}{2} + n)$, while the second term has poles at $\nu = -i (\tfrac{d}{2} + n)$. The leading behavior of the propagator at large $\Delta$ with $\zeta$ fixed is $\sim\left(\sqrt{\zeta/4+1}-\sqrt{\zeta/4}\right)^\Delta$, implying that the for the first term we can only close the contour in the lower-half $\nu$ plane, and for the second only in the upper-half $\nu$ plane.\footnote{We are using that $\sqrt{\zeta/4+1}-\sqrt{\zeta/4} < 1$ for any $\zeta > 0$. The asymptotic behavior of the propagator can be inferred from that of the hypergeometric function ${}_2F_1(a,a,2a,z)$ for large $a$ and $z$ fixed. The latter is discussed for instance in appendix D of \cite{Rattazzi:2008pe}.} As a result, both terms only contribute through either one of the two poles of the kernel $(\nu^2 + \left(\Delta -\frac{d}{2}\right)^2)^{-1}$, and it is easy to see that each contribution gives $\frac 12 G_\Delta$, thus proving the formula \eqref{eq:srprop}.

Consider now the bulk-to-bulk propagator for the alternative choice of boundary condition, corresponding to $\Delta = \Delta_- < \frac{d}{2}$. In this case, the propagator does not define a square-integrable function in the measure $d\sigma (\sinh(\sigma))^d$, so a priori we do not how to define a spectral representation. However we can still try and proceed similarly to the previous example, and prove an integral representation by evaluating the integral via Cauchy's theorem. In this case we need the first term in eq. \eqref{eq:eigen} to contribute through a pole in the upper-half plane, at $\nu = i(\frac{d}{2} - \Delta)$, and viceversa the second term should contribute through a pole in the lower-half plane, at $\nu = - i(\frac{d}{2} - \Delta)$. This can be easily achieved by modifying the contour
\begin{equation}
G_{\Delta = \Delta^-}(x_1,x_2) = \frac{1}{L^{d-1}}\int_{\mathbb{R} + C_u+C_l} d\nu \,\frac{1}{\nu^2 + \left(\Delta -\frac{d}{2}\right)^2} \, \Omega_\nu(x_1,x_2)~,\label{eq:srpropmin}
\end{equation}
where $C_u$ is a clockwise circle around $\nu = i (\frac{d}{2}-\Delta)$, and $C_l$ is a counterclockwise circle around $\nu = -i (\frac{d}{2}-\Delta)$. Another way to derive this result is to note in eq. \eqref{eq:bulktobulk} that there is no discontinuity in the propagator as we go from $\Delta > \frac{d}{2}$ to $\Delta < \frac{d}{2}$. Therefore, we can start from the integral representation \eqref{eq:srprop} and change continuosly the parameter $\Delta$. This requires us to keep the contribution of the poles as they cross the contour when $\Delta$ crosses $\frac{d}{2}$, leading to the modified contour in eq. \eqref{eq:srpropmin}. We see that as a consequence of the lack of square-integrability, the function $G_{\Delta = \Delta^-}$ does not have a standard spectral representation with a $\nu$ integral on the real line.

Using equations \eqref{eq:Green}, \eqref{eq:boxeigen} and \eqref{eq:srprop} we obtain the spectral representation of the delta function 
\begin{equation}
\delta^{d+1}(x_1,x_2) = \frac{1}{L^{d+1}} \int^{+\infty}_{-\infty} d\nu \, \Omega_\nu(x_1,x_2)~.
\end{equation}

\subsection{Spectral Representation of a Power-Law $\zeta^{-\rho}$}\label{app:power}

We will now show how to compute the spectral representation of a power of the chordal distance in AdS$_{d+1}$, namely 
\begin{equation}
F_\rho(\sigma) = \left(\sinh\left(\frac{\sigma}{2}\right)\right)^{-2\rho} = \left(\frac{\zeta}{4 L^2}\right)^{-\rho}~. 
\end{equation}
This is relevant to probe the limit in which the two points in AdS$_{d+1}$ are very close to each other. We will borrow the method used in \cite{Hogervorst:2017sfd} to compute the Jacobi transform in terms of a Mellin-Barnes integral. Note that the above power-law only defines a square-integral function in the measure $d \sigma (\sinh(\sigma))^d$ for $d<1$ and $\frac d2 < \rho < \frac{d+1}{4}$. On the other hand, the integral that defines the spectral representation is convergent for the larger range $\frac d2 < \rho < \frac{d+1}{2}$, and any $d$. For more general values, one needs to perform some regularization, e.g. analytic continuation from the region in which it is well-defined. Similar considerations are familiar in the case of the Fourier transform of power-laws in flat space.

Using eq.s \eqref{eq:inversespec}-\eqref{eq:kern2} we have
\begin{align}
\frac{\tilde{F_\rho}(\nu)}{\mathrm{Vol}(S^d)} & =  \int_0^\infty d \sigma (\sinh(\sigma))^d  \left(\sinh\left(\frac{\sigma}{2}\right)\right)^{-2\rho}{}_2F_1\left(\tfrac d2 + i \nu , \tfrac{d}{2}-i \nu , \tfrac{d+1}{2},- \sinh\left(\tfrac{\sigma}{2}\right)^2\right) \nonumber\\
& = 2^d \int_0^1 dx \, x^{-\tfrac{d+3}{2}} \left(\frac{1}{x}-1\right)^\frac{d-1-2\rho}{2} {}_2F_1\left(\tfrac d2 + i \nu , \tfrac{d}{2}-i \nu , \tfrac{d+1}{2},\tfrac{x-1}{x}\right)~.\label{eq:trpowint}
\end{align}
In the second line we changed variable to $x = \cosh\left(\frac{\sigma}{2}\right)^{-2}$. We now use the following Mellin-Barnes representation of the ${}_2 F_1$
\begin{equation}
{}_2 F_1(a,b,c,z)= \frac{\Gamma(c)}{\Gamma(a)\Gamma(b)} \int[ds]\frac{\Gamma(-s)\Gamma(a+s)\Gamma(b+s)}{\Gamma(c+s)} (-z)^s~,
\end{equation}
where $\int[ds]\equiv \int_{-i\infty}^{i\infty} \frac{ds}{2\pi i}$. We plug this in \eqref{eq:trpowint}, and exchange the $s$ and $x$ integrals. The $x$ integral can be easily performed in terms of gamma functions, hence we are left with
\begin{equation}
\frac{\tilde{F_\rho}(\nu)}{\mathrm{Vol}(S^d)} \!=\!\frac{2^d \Gamma(\tfrac{d+1}{2})}{\Gamma(\tfrac{d}{2}\pm i \nu) \Gamma(\tfrac{1-d}{2})} \!\int \![ds] \frac{\Gamma(\tfrac d2 \pm i \nu + s)\Gamma(\tfrac{d+1}{2}-\rho+s)\Gamma(-d\!+\!\rho\!-\!s)\Gamma(-s)}{\Gamma(\tfrac{d+1}{2} +s)}.\!\!\!
\end{equation}
This Mellin-Barnes integral can be also evaluated in terms of a product of gamma functions using the second Barnes lemma
\begin{align}
& \int[ds]\frac{\Gamma(\alpha
+s)\Gamma(\beta+s)\Gamma(\gamma+s)\Gamma(1-\delta-s)\Gamma(-s)}{\Gamma(\epsilon+s)} \nonumber \\& = \frac{\Gamma(\alpha)\Gamma(\beta)\Gamma(\gamma)\Gamma(1-\delta +\alpha)\Gamma(1-\delta +\beta)\Gamma(1-\delta +\gamma)}{\Gamma(\epsilon-\alpha)\Gamma(\epsilon-\beta)\Gamma(\epsilon-\gamma)}~, 
\end{align}
valid when $\epsilon = \alpha+\beta+\gamma-\delta+1$. Using this formula we finally obtain
\begin{equation}
\tilde{F_\rho}(\nu) =  (4\pi)^{\tfrac{d+1}{2}}\frac{\Gamma(\tfrac{d+1}{2}-\rho)\Gamma(-\tfrac d2+\rho\pm i \nu )}{\Gamma(\rho)\Gamma(\tfrac 12 \pm i \nu)}~.\label{eq:trpow}
\end{equation}
As a check, let us take the flat-space limit. $F_\rho(\sigma)$ approaches a power of the distance $r$ from the origin in flat space, namely $F^{\rm flat}_\rho(r)=4^{\rho}L^{2\rho}r^{-2\rho}$. The Fourier transform in $\mathbb{R}^{d+1}$ of this function is
\begin{equation}
\tilde{F}^{\rm flat}_\rho(|p|) = L^{2\rho} (4\pi)^{\tfrac{d+1}{2}} \frac{\Gamma(\tfrac{d+1}{2}-\rho)}{\Gamma(\rho)} |p|^{2\rho-d-1}~.
\end{equation}
Comparing this formula with the limit $L\to \infty$ of eq. \eqref{eq:trpow} evaluated at $\nu = L|p|$, we find agreement with \eqref{eq:specfsl}. 

\subsection{Spectral Representation of a Single-Fermion Loop}\label{apsubsec:femionloop}
Here we derive the spectral representation for the trace of the propagator of the fermion, ${\rm tr}\left[\frac{1}{\gamma\cdot \nabla +M}\right]$. For this purpose, we use the representation of the fermion propagator\footnote{Note that the propagator defined in \cite{Kawano:1999au} is $-\frac{1}{\gamma\cdot \nabla +M}$.} in \cite{Kawano:1999au},
\begin{align}
\begin{aligned}
\frac{1}{\gamma\cdot \nabla +M}=&\frac{2}{\sqrt{z_1 z_2}}\left[\{(x_1\cdot \gamma) \Pi_{-}-\Pi_{+}(x_2\cdot \gamma)\}\frac{dG_{\frac{d}{2}+M-\frac{1}{2}}(x_1,x_2)}{d\zeta }\right.\\
&\left.+\{(x_1\cdot \gamma) \Pi_{+}-\Pi_{-}(x_2\cdot \gamma)\}\frac{dG_{\frac{d}{2}+M+\frac{1}{2}}(x_1,x_2)}{d\zeta}\right]\,,
\end{aligned}
\end{align}
where $\zeta$ is the chordal distance in AdS \eqref{eq:chordaldistance} and $\Pi_{\pm}$ are the chiral projection operator $\frac{1\pm \gamma_0}{2}$. After taking the trace of the spinor indices and taking the coincident point limit, we arrive at 
\begin{align}
{\rm tr}\left[\frac{1}{\gamma\cdot \nabla +M}\right]=2 c_{d+1} \left(\left.\frac{dG_{\frac{d}{2}+M+\frac{1}{2}}(x_1,x_2)}{d\zeta}\right|_{x_2\to x_1}-\left.\frac{dG_{\frac{d}{2}+M-\frac{1}{2}}(x_1,x_2)}{d\zeta}\right|_{x_2\to x_1}\right)\,.
\end{align}
Using the spectral representation of the scalar bulk-to-bulk propagator $G_{\Delta}$, we can express $dG_{\Delta}/d\zeta$ as
\begin{align}
\left.\frac{dG_{\Delta}(x_1,x_2)}{d\zeta}\right|_{x_2\to x_1}=-\int_{-\infty}^{\infty} d\nu \frac{1}{\nu^2+(\Delta-\frac{d}{2})^2}\frac{\Gamma(\tfrac{d}{2})\Gamma(1+\tfrac{d}{2}\pm i\nu)}{8\pi^{\frac{d}{2}+1}(d+1)\Gamma(d)\Gamma(\pm i\nu)}\,.
\end{align} 
As a result we get
\begin{align}
\begin{aligned}
&{\rm tr}\left[\frac{1}{\gamma\cdot \nabla + M}\right]=\\
&c_{d+1}\int_{-\infty}^{\infty} d\nu \left[\frac{1}{\nu^2+(M-\tfrac{1}{2})^2}-\frac{1}{\nu^2+(M+\tfrac{1}{2})^2}\right]\frac{\Gamma(\tfrac{d}{2})\Gamma(1+\tfrac{d}{2}\pm i\nu)}{4 \pi^{\frac{d}{2}+1}(d+1)\Gamma(d)\Gamma(\pm i\nu)}\,.
\end{aligned}
\end{align}

\section{Split Representation and Conformal Integral}\label{ap:split}
In this appendix, we derive the split representation of the harmonic function $\Omega_{\nu}$ and use it to compute the integral \eqref{eq:integralweneedtocompute}.

The bulk-to-bulk propagator is known to admit the so-called spectral representation (see for instance \cite{Fitzpatrick:2011ia}). In terms of the embedding coordinates, it reads
\begin{align}
G_{\Delta}(X_1,X_2)=\sqrt{\mathcal{C}_{\frac{d}{2}+i\nu}\mathcal{C}_{\frac{d}{2}-i\nu}}\int\frac{d\nu}{2\pi}\frac{2\nu^2}{\nu^2+(\Delta-\frac{d}{2})^2}\int dP_0\, K_{\frac{d}{2}+i\nu}(P_0,X_1)K_{\frac{d}{2}-i\nu}(P_0,X_2)\,,
\end{align} 
where $P$ is a boundary point and $\mathcal{C}_{x}$ is a constant defined in \eqref{eq:CDeltaDef}.  Comparing this expression with the spectral representation of the bulk-to-bulk propagator \eqref{eq:srprop}, we conclude that the harmonic function $\Omega_{\nu}$ can be expressed as
\begin{align}
\Omega_{\nu}(X_1,X_2)=\frac{\nu^2\sqrt{\mathcal{C}_{\frac{d}{2}+i\nu}\mathcal{C}_{\frac{d}{2}-i\nu}}}{\pi}\int dP_0\, K_{\frac{d}{2}+i\nu}(P_0,X_1)K_{\frac{d}{2}-i\nu}(P_0,X_2)\,.
\end{align}

\begin{figure}
\centering
\includegraphics[clip,height=2.5cm]{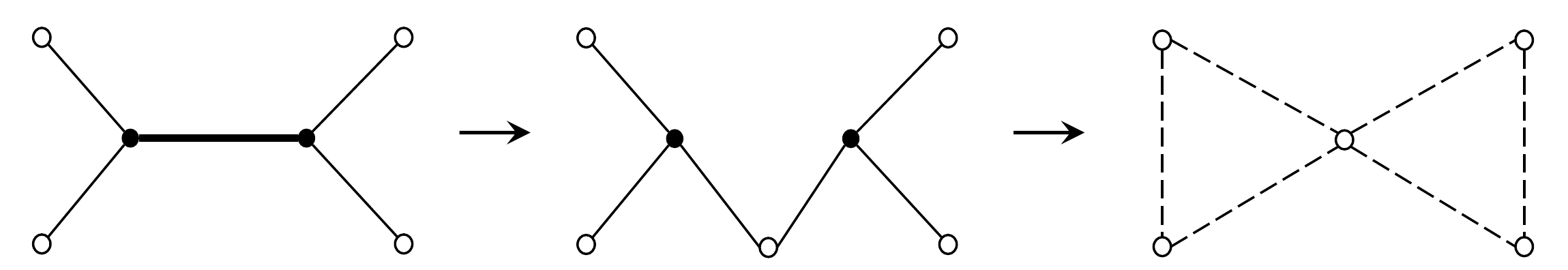}
\caption{Schematic explanation for the evaluation of the integral. The black and white dots represent the points in AdS and the points on the boundary respectively. The thick black line is the bulk-to-bulk propagator while the thin black lines denote the bulk-to-boundary propagators. To evaluate the integral \eqref{eq:integralweneedtocompute} (depicted in the left figure), we first use the split representation and convert the bulk-to-bulk propagator into a product of bulk-to-boundary propagators (see the middle figure). We can then perform the integral of the bulk points to arrive at the integral defined purely on the boundary (the right figure). The resulting integral only involves the propagators on the boundary (depicted as dashed lines) and can be identified with the shadow representation of the conformal partial wave. \label{fig:split}}
\end{figure}
Let us now use this representation to compute the integral \eqref{eq:integralweneedtocompute}. Substituting the expression into the integral, we get (see figure \ref{fig:split})
\begin{align}
\begin{aligned}
\eqref{eq:integralweneedtocompute}=&\frac{\nu^2\sqrt{\mathcal{C}_{\frac{d}{2}+i\nu}\mathcal{C}_{\frac{d}{2}-i\nu}}}{\pi}\int dP_0\int dX_1 \, K_{\Delta}(P_1,X_1)K_{\Delta}(P_2,X_1)K_{\frac{d}{2}+i\nu}(P_0,X_1)\\
&\times \int dX_2  K_{\Delta}(P_3,X_2)K_{\Delta}(P_4,X_2)K_{\frac{d}{2}-i\nu}(P_0,X_2)
\end{aligned}
\end{align}
We can then perform the integrals of $X_1$ and $X_2$ since they are precisely the integrals that appear in the computation of the three-point function \cite{Muck:1998rr, Freedman:1998tz},
\begin{align}
\int dX K_{a} (P_1, X) K_{b}(P_2,X)K_{c} (P_3,X)=\frac{B_{a,b,c}}{(P_{12})^{\frac{a+b-c}{2}}(P_{23})^{\frac{b+c-a}{2}}(P_{31})^{\frac{c+a-b}{2}}}\,,
\end{align} 
with
\begin{align}
B_{a,b,c}=\frac{\pi^{\frac{d}{2}}}{2}\Gamma_{-\frac{d}{2}+\frac{a+b+c}{2}}\sqrt{\mathcal{C}_a\mathcal{C}_b\mathcal{C}_c}\frac{\Gamma_{\frac{a+b-c}{2}}\Gamma_{\frac{a-b+c}{2}}\Gamma_{\frac{-a+b+c}{2}}}{\Gamma_{a}\Gamma_{b}\Gamma_{c}}\,.
\end{align}
Here we used the abbreviation $\Gamma_{x}\equiv \Gamma(x)$.
Using this formula, we obtain
\begin{align}
\begin{aligned}
\eqref{eq:integralweneedtocompute}=&\frac{\nu^2\mathcal{C}_{\frac{d}{2}+i\nu}\mathcal{C}_{\frac{d}{2}-i\nu} }{16\pi}\frac{\Gamma_{\Delta-\frac{d}{4}+\frac{i \nu}{2}}^2\Gamma_{\Delta-\frac{d}{4}-\frac{i \nu}{2}}^2\Gamma_{\frac{d}{4}+\frac{i\nu}{2}}^2\Gamma_{\frac{d}{4}-\frac{i\nu}{2}}^2}{\Gamma_{\Delta}^2\Gamma_{\Delta-\frac{d}{2}+1}^2\Gamma_{\frac{d}{2}+i\nu}\Gamma_{\frac{d}{2}-i\nu}}\\
&\times\frac{1}{(P_{12})^{\Delta-\frac{d}{4}-\frac{i\nu}{2}}(P_{34})^{\Delta-\frac{d}{4}+\frac{i\nu}{2}}}\int \frac{dP_0}{(P_{10})^{\frac{d}{4}+\frac{i\nu}{2}}(P_{20})^{\frac{d}{4}+\frac{i\nu}{2}}(P_{30})^{\frac{d}{4}-\frac{i\nu}{2}}(P_{40})^{\frac{d}{4}-\frac{i\nu}{2}}}\,.
\end{aligned}
\end{align}
Now the remaining task is to perform the integral of $P_0$. This integral turns out to be equivalent to the shadow representation of the conformal partial wave (see for instance \cite{Simmons-Duffin:2017nub}),
\begin{align}
\begin{aligned}
&\frac{1}{(P_{12})^{\Delta-\frac{d}{4}-\frac{i\nu}{2}}(P_{34})^{\Delta-\frac{d}{4}+\frac{i\nu}{2}}}\int \frac{dP_0}{(P_{10})^{\frac{d}{4}+\frac{i\nu}{2}}(P_{20})^{\frac{d}{4}+\frac{i\nu}{2}}(P_{30})^{\frac{d}{4}-\frac{i\nu}{2}}(P_{40})^{\frac{d}{4}-\frac{i\nu}{2}}}=\\
&\frac{1}{(P_{12})^{\Delta}(P_{34})^{\Delta}}\left(k_{\frac{d}{2}-i\nu}\mathcal{K}_{\frac{d}{2}+i\nu}(z,\bar{z})+k_{\frac{d}{2}+i\nu}\mathcal{K}_{\frac{d}{2}-i\nu}(z,\bar{z})\right)\,,
\end{aligned}
\end{align}
where $\mathcal{K}_{\Delta}$ is a $d$-dimensional scalar conformal block with dimension $\Delta$, $z$ and $\bar{z}$ are the conformal cross ratio given in \eqref{eq:crossratiodef}, and $k_{a}$ is given by
\begin{align}
k_{a}=\frac{\pi^{\frac{d}{2}}\Gamma_{a-\frac{d}{2}}\Gamma_{\frac{d}{2}-\frac{a}{2}}^2}{\Gamma_{d-a}\Gamma_{\frac{a}{2}}^2}\,.
\end{align}
Therefore we finally obtain the result
\begin{align}
\begin{aligned}
\eqref{eq:integralweneedtocompute}=&\frac{1}{(P_{12})^{\Delta}(P_{34})^{\Delta}}\frac{\Gamma_{\Delta-\frac{d}{4}-\frac{i\nu}{2}}^{2}\Gamma_{\Delta-\frac{d}{4}+\frac{i\nu}{2}}^{2}}{64\pi^{\frac{d}{2}+1}\Gamma_{\Delta}^2\Gamma_{1-\frac{d}{2}+\Delta}^2}\left[\frac{\Gamma_{\frac{d}{4}+\frac{i\nu}{2}}^{4}\mathcal{K}_{\frac{d}{2}+i\nu}(z,\bar{z})}{\Gamma_{\frac{d}{2}+i\nu}\Gamma_{i\nu}}+\frac{\Gamma_{\frac{d}{4}-\frac{i\nu}{2}}^{4}\mathcal{K}_{\frac{d}{2}-i\nu}(z,\bar{z})}{\Gamma_{\frac{d}{2}-i\nu}\Gamma_{-i\nu}}\right]\,.\nonumber
\end{aligned}
\end{align}

\section{OPE Coefficients in Fermionic Mean-Field Theory}
\label{app:fermgff}

We compute here the squared OPE coefficients for double-trace operators in the OPE of two fermionic generalized free fields, in $d=2$ and $d=1$. Let us start with the $d=2$ case.
The mean-field theory 4-point function for fermions with equal scaling dimensions $(h, \bar h) = (\frac{\D+\frac{1}{2}}{2},\frac{\D-\frac{1}{2}}{2})$ is:
\begin{align}\label{eq:correlatormeanfermion}
&\langle  \bar{\psi}^{i}(z_1,\bar{z}_1) \psi^{j}(z_2,\bar{z}_2)  \bar{\psi}^{k}(z_3,\bar{z}_3) \psi^{l}(z_4,\bar{z}_4)  \rangle  \nonumber\\ &=  
\frac{\d^{ij}\d^{kl}}{(z_{12}z_{34})^{2h}(\bar{z}_{12}\bar{z}_{34})^{2\bar h}} -\frac{\d^{il}\d^{jk}}{(z_{14}z_{23})^{2h}(\bar{z}_{14}\bar{z}_{23})^{2\bar h}}
\nn\\
&=\frac{1}{(z_{12}z_{34})^{2h}(\bar{z}_{12}\bar{z}_{34})^{2\bar h}} \left[ \d^{ij}\d^{kl} - \d^{il}\d^{jk}\left(\frac{z}{1-z}\right)^{2h} \left(\frac{\bar{z}}{1-\bar{z}}\right)^{2\bar h} \right]~.
\end{align} 
In the last line we used the cross-ratio $z\equiv\frac{z_{12}z_{34}}{z_{13}z_{24}}$.

To read off the coefficients $c_n^2$ used in \eqref{eq:GFFOPEfermion}, we equate this to the conformal block expansion\footnote{Note that here we omitted an extra overall minus sign present in the full four-point function \eqref{eq:correlatormeanfermion} since the coefficients in \eqref{eq:GFFOPEfermion} is defined without the minus sign.}
\bea
\label{eq:g313}
\sum_{n, \bar n}c^2_{h_n, \bar{h}_{\bar n}}\mm{K}_{h_n, \bar{h}_{\bar n}}(z, \bar z) = \left(\frac{z}{1-z}\right)^{2h} \left(\frac{\bar{z}}{1-\bar{z}}\right)^{2\bar h}~, 
\eea 
where the exchanged double trace operators have $(h_n, \bar h_{\bar n}) = ( \D+n+\frac{1}{2},\D+\bar n-\frac{1}{2})$, and the conformal blocks are
\bea
\mm{K}_{h_n, \bar{h}_{\bar n}}(z, \bar z) = z^{h_n}\bar{z}^{\bar{h}_{\bar n}}\ _2F_1 (h_n,h_n,2h_n,z)\ _2F_1 (\bar{h}_{\bar n},\bar{h}_{\bar n},2\bar{h}_{\bar n},\bar z)~.
\eea
The solution to eq.~\eqref{eq:g313} is
\bea
c^2_{n,\bar{n}} = \frac{ (4\D^2-1)(\D+\frac{1}{2})_{\bar{n}-1}(\D+\frac{3}{2})_{n-1}(2\D-1)_{\bar{n}-1}(2\D+1)_{n-1}}{4^{n+\bar{n}}\,\,n!\bar{n}!(\D)_{\bar{n}-1}(\D+1)_{n-1}}~.
\eea 
For an exchange of a scalar double trace operators we have $h_n=h_{\bar{n}}$, and thus $\bar{n} = n+1$. This gives
\bea
 \label{eq:ope2}
c^2_n \equiv c^2_{n,n+1} = \frac{\pi 2^{2-4n-4\D}\G^2(\D+n+\frac{1}{2}) \G(2\D+n-1)\G(2\D+n)}{(n+1)!n!\G^2(\D-\frac{1}{2})\G^2(\D+\frac{1}{2})\G^2(\D+n)}~.
\eea 

We now repeat the process in $d=1$. The mean-field theory four-point function is
\bea
\langle  \bar{\psi}^{i}_1(x_1) \psi^{j}_2(x_2)   \bar{\psi}^{k}_3(x_3) \psi^{l}_4(x_4)  \rangle   =  
\frac{\d^{ij}\d^{kl}}{(x_{12}x_{34})^{2\Delta} } -\frac{\d^{il}\d^{jk}}{(x_{14}x_{23})^{2\Delta} }
\nn\\
=- \frac{1}{(x_{12}x_{34})^{2\Delta} } \left[ \d^{ij}\d^{kl}- \d^{il}\d^{jk}\left(\frac{z}{1-z}\right)^{2\Delta} \right]~.
\eea 
Equating this to the conformal block expansion
\bea
\label{eq:g3132}
\sum_{n }\bar{c}^2_{h_n  }\mm{K}_{h_n }(z ) = \left(\frac{z}{1-z}\right)^{2\Delta} ~,
\eea 
where the exchanged double trace operators have $h_n = 2\D+n $, and the conformal blocks are
\bea
\mm{K}_{h_n}(z) = z^{h_n}  {}_2F_1 (h_n,h_n,2h_n,z)~,
\eea
we find
 \bea
\bar{c}^2_{h_n} =\frac{4^{1-n}\D (4\D)_{n-1}(2\D+1)_{n-1}}{n!(2\D+\frac{1}{2})_{n-1}}~.
\eea 
Note that $c_{n}^{2}$ defined in \eqref{eq:GFFOPEfermion} corresponds to $\bar{c}_{h_{2n+1}}^2$ in the notation here and reads
\begin{align}\label{eq:ope1}
c_{n}^2=\frac{\D (4\D)_{2n}(2\D+1)_{2n}}{4^{2n}\,\,(2n+1)!(2\D+\frac{1}{2})_{2n}}\,.
\end{align}

\section{Pauli-Villars Regularization of GN Model on $\mathbb{R}^3$}
\label{app:PV}

A regulator of the GN model that does not break the discrete symmetry consists in adding $N/2$ bosonic Dirac fields $\chi_+$ that for $\Sigma = 0$ have large positive mass $M_{PV}>0$ and $N/2$ bosonic Dirac fields $\chi_-$ with large negative mass $-M_{PV}$. We will refer to this regulator as parity-preserving Pauli-Villars (PV), and has to be contrasted with the parity-violating version that consists in adding $N$ bosonic Dirac fields of mass $M_{PV}$. The discrete symmetry flips the sign of the mass, and in order for the regulator to preserve it we need to accompany this with the exchange $\chi_+\leftrightarrow \chi_-$. These fields need to have exactly the same couplings as the physical Dirac fermions $\Psi$, so they also will couple to $\sigma$ and in the massive phase $\Sigma = M \neq 0$ their masses get shifted to $M_\pm = M \pm M_{PV}$. We will now show how to use this regulator to compute the finite, scheme-independent two-point function of $\sigma$ for the case of flat, three-dimensional background, i.e. on $\mathbb{R}^3$, since this is a useful comparison for the AdS calculation. We will follow the appendix of \cite{Rosenstein:1990nm}. Of course in flat-space it is not strictly necessary to adopt the PV regulator, and there are many other possible choices, but PV is the only one that we were able to straightforwardly import to AdS$_3$.

In flat space it is convenient to first compute the diagrams with a hard cutoff, and then see that the cutoff cancels in the PV-regularized expression. The fermionic bubble with a hard cutoff is
 \begin{align}
\tilde{B}_F(p^2,M)\vert_\Lambda& = \int^\Lambda \frac{d^3 q}{(2\pi)^3} \Tr\left[\frac{i\slashed{p} + i\slashed{q} + M}{(p+q)^2 + M^2} \frac{ i\slashed{q} + M}{q^2 + M^2}\right] \nonumber \\& = (p^2 + 4 M^2)\tilde{B}(p^2,M^2) -  \int^\Lambda \frac{d^3 q}{(2\pi)^3}\frac{1}{(p+q)^2 + M^2} -  \int^\Lambda \frac{d^3 q}{(2\pi)^3}\frac{1}{q^2 + M^2}\nonumber\\
& =(p^2 + 4 M^2)\tilde{B}(p^2,M^2)- \frac{1}{\pi^2} \Lambda + \frac{1}{2\pi}|M| + \dots~,
\end{align}
where $\dots$ are terms that are suppressed at large $\Lambda$ and $\tilde{B}(p^2,M^2)$ is the finite scalar bubble
\begin{equation}
\tilde{B}(p^2,M^2) = \int \frac{d^3 q}{(2\pi)^3} \frac{1}{(p+q)^2 + M^2}\frac{1}{q^2 + M^2} = \frac{1}{4\pi |p|}{\rm atan}\left(\frac{|p|}{2|M|}\right)~.
\end{equation}
The PV regulated fermionic bubble then is
 \begin{align}
\tilde{B}_F(p^2,M)\vert_{PV} & =\tilde{B}_F(p^2,M)\vert_\Lambda - \frac 12 \tilde{B}_F(p^2,M_+)\vert_\Lambda - \frac 12 \tilde{B}_F(p^2,M_-)\vert_\Lambda \nonumber\\
& =(p^2 + 4 M^2)\tilde{B}(p^2,M^2) - \frac 12 (p^2 + 4 M_+^2)\tilde{B}(p^2,M_+^2) - \frac 12 (p^2 + 4 M_-^2)\tilde{B}(p^2,M_-^2) \nonumber\\ & ~~+ \frac{1}{2\pi}|M| - \frac{1}{4\pi}|M_+| - \frac{1}{4\pi}|M_-| \nonumber\\
& = (p^2 + 4 M^2)\tilde{B}(p^2,M^2) + \frac{1}{2\pi}|M| - \frac{1}{2\pi}|M_+| - \frac{1}{2\pi}|M_-| + \dots\nonumber\\
& = (p^2 + 4 M^2)\tilde{B}(p^2,M^2) + \frac{1}{2\pi}|M| - \frac{1}{\pi}M_{PV} + \dots ~,\label{eq:flatPVbubble}
\end{align}
where $\dots$ are terms that are suppressed at large $M_{PV}$, and going from the second to the third equality we replaced the functions $\tilde{B}(p^2,M_\pm^2)$ with their large-$M_{PV}$ limit. Using a parity-breaking PV regulator for the fermionic bubble
 \begin{align}
\tilde{B}_F(p^2,M)\vert_{PV, \slashed{P}} & =\tilde{B}_F(p^2,M)\vert_\Lambda - \tilde{B}_F(p^2,M_{PV})\vert_\Lambda \nonumber\\
& =(p^2 + 4 M^2)\tilde{B}(p^2,M^2) - (p^2 + 4 M_{PV}^2)\tilde{B}(p^2,M_{PV}^2) + \frac{1}{2\pi}|M| -  \frac{1}{2\pi}M_{PV}\nonumber\\
& = (p^2 + 4 M^2)\tilde{B}(p^2,M^2) + \frac{1}{2\pi}|M| -  \frac{1}{\pi}M_{PV} + \dots~.
\end{align}
we find precisely the same result.

Next, we consider the trace of the fermionic propagator, that appears on the right-hand-side of the gap equation \eqref{eq:gap}, and that can equivalently be thought of as $(-)$the one-point function of $\bar{
\Psi}\Psi$. With a hard cutoff, this gives
\begin{align}
\Tr\int^\Lambda \frac{d^3 q}{(2\pi)^3}\frac{1}{-i\slashed{q}+M}\nonumber& =2 M \int^\Lambda \frac{d^3 q}{(2\pi)^3}\frac{1}{q^2+M^2}  = M\left( \frac{1}{\pi^2}\Lambda  -\frac{1}{2\pi} |M|\right)~.
\end{align}
Therefore the PV-regulated answer is
\begin{align}
& \Tr\int^\Lambda \frac{d^3 q}{(2\pi)^3}\frac{1}{-i\slashed{q}+M}- \frac 12\Tr\int^\Lambda \frac{d^3 q}{(2\pi)^3}\frac{1}{-i\slashed{q}+M_+} - \frac 12\Tr\int^\Lambda \frac{d^3 q}{(2\pi)^3}\frac{1}{-i\slashed{q}+M_-}\nonumber\\
& =  M\left( \frac{1}{\pi^2}\Lambda  -\frac{1}{2\pi} |M|\right)-\frac 12 M_+\left( \frac{1}{\pi^2}\Lambda  -\frac{1}{2\pi} |M_+|\right) -\frac 12 M_-\left( \frac{1}{\pi^2}\Lambda  -\frac{1}{2\pi} |M_-|\right)\nonumber\\
& = -\frac{1}{2\pi} M  |M| + \frac{1}{4\pi}(M^2_+ - M^2_-)\nonumber\\
& = -M\left(\frac{1}{2\pi}|M| - \frac{1}{\pi} M_{PV}\right)~.\label{eq:1pointMPV}
\end{align}
Note that with a parity-breaking PV regulator we would have instead
\begin{align}
&-\Tr\int^\Lambda \frac{d^3 q}{(2\pi)^3}\frac{1}{-i\slashed{q}+M}+\Tr\int^\Lambda \frac{d^3 q}{(2\pi)^3}\frac{1}{-i\slashed{q}+M_{PV}}\nonumber \\&~~~~~~~~~~~~~~~~~~~~~~~~=  -M\left( \frac{1}{\pi^2}\Lambda  -\frac{1}{2\pi} |M|\right)+M_{PV}\left( \frac{1}{\pi^2}\Lambda  -\frac{1}{2\pi} |M|\right)~,
\end{align}
and the dependence on $\Lambda$ would fail to cancel, meaning that this is not a good regulator. For this reason we are forced to resort to the parity-preserving version above.

Recalling that the gap equation \eqref{eq:gap} equates the trace \eqref{eq:1pointMPV} to $-M g^{-1}$, and comparing this to \eqref{eq:flatPVbubble}, we find that indeed there is a cancelation and the two-point function of $\delta\sigma$ is finite
\begin{equation}
\langle \delta \sigma(p) \delta \sigma(-p)\rangle = -\frac{1}{(g^{-1})_{PV}-(\tilde{B}_F(p^2,M^2))_{PV}} =\frac{1}{(p^2 + 4 M^2)\tilde{B}(p^2,M^2) }~.
\end{equation}

With the result \eqref{eq:1pointMPV} we can also check that the relation between the two-point function and the one-point function of $\bar{\Psi}{\Psi}$ in free-fermion theory ---that we used in AdS to fix the constant shift in the bubble diagram--- is also valid in flat space. Indeed, in flat-space the integrated, connected two-point function of $\bar{\Psi}{\Psi}$ is nothing but $(-)$the bubble $\tilde{B}_F$ evaluated at $p=0$, which gives
\begin{equation}
-\tilde{B}_F(p^2=0,M) = -\frac{1}{\pi}(|M| - M_{PV})~.
\end{equation}
On the other hand, the one-point function is $(-)$the result of \eqref{eq:1pointMPV}, and we see that indeed the two are related precisely by $(-)$a derivative w.r.t. $M$.

\bibliographystyle{JHEP}
\bibliography{referencesOPE}

\end{document}